\documentclass[aps,prd,twocolumn,groupedaddress, nofootinbib, longbibliography,superscriptaddress,amsmath,amssymb,amsfonts]{revtex4-2}
\usepackage[dvipsnames]{xcolor}
\usepackage{hyperref}

\definecolor{mydarkblue}{RGB}{46, 48, 146}

\hypersetup{colorlinks=true,allcolors=mydarkblue}
\usepackage{multirow}
\usepackage{graphicx}
\usepackage{adjustbox}
\usepackage[utf8]{inputenc} 
\usepackage[T1]{fontenc}    
\usepackage{microtype}      
\usepackage{tensor}
\usepackage{orcidlink}
\usepackage{float}
\usepackage[all]{hypcap}

\DeclareUnicodeCharacter{03C3}{$\sigma$}
\DeclareUnicodeCharacter{03B1}{$\alpha$}

\global\long\def\UPCBRACE#1#2{\overset{{\scriptstyle #2}}{\overbrace{#1}}}

\newcommand{\aap}{A\&A}
\usepackage{tikz}
\usetikzlibrary{decorations,calligraphy}
\NewDocumentCommand{\evalat}{sO{\big}mm}{%
  \IfBooleanTF{#1}
   {\mleft. #3 \mright|_{#4}}
   {#3#2|_{#4}}%
}
\usepackage[capitalise]{cleveref}

\begin{document}

\title{Cosmology-informed neural networks to solve the background dynamics of the Universe}

\author{\mbox{Augusto T. Chantada}\orcidlink{0000-0002-4480-9595}}
 \email{augustochantada01@gmail.com}
 \affiliation{Departamento de Física, Facultad de Ciencias Exactas y Naturales, Universidad de Buenos Aires, Av. Intendente Cantilo S/N 1428 Ciudad Autónoma de Buenos Aires, Argentina}
\author{\mbox{Susana J. Landau}\orcidlink{0000-0003-2645-9197}}
 \affiliation{Departamento de Física, Facultad de Ciencias Exactas y Naturales, Universidad de Buenos Aires, Av. Intendente Cantilo S/N 1428 Ciudad Autónoma de Buenos Aires, Argentina}
 \affiliation{IFIBA - CONICET - UBA, Av. Intendente Cantilo S/N 1428 Ciudad Autónoma de Buenos Aires, Argentina}
\author{\mbox{Pavlos Protopapas}\orcidlink{0000-0002-8178-8463}}
 \affiliation{John A. Paulson School of Engineering and Applied Sciences, Harvard University, Cambridge, Massachusetts 02138, USA}
\author{\mbox{Claudia G. Scóccola}\orcidlink{0000-0002-3565-4771}}
 \affiliation{Consejo Nacional de Investigaciones Científicas y Técnicas (CONICET), Godoy Cruz 2290, 1425, Ciudad Autónoma de Buenos Aires, Argentina}
 \affiliation{Facultad de Ciencias Astronómicas y Geofísicas,
Universidad Nacional de La Plata, Observatorio Astronómico, Paseo del Bosque,
B1900FWA La Plata, Argentina}
\author{\mbox{Cecilia Garraffo}\orcidlink{0000-0002-8791-6286}}
 \affiliation{Center for Astrophysics | Harvard \& Smithsonian, 60 Garden Street, Cambridge, Massachusetts 02138, USA}
 \affiliation{Institute for Applied Computational Science, Harvard University, 33 Oxford St., Cambridge, Massachusetts 02138, USA}

\date{\today}

\begin{abstract}
The field of machine learning has drawn increasing interest from various other fields due to the success of its methods at solving a plethora of different problems. An application of these has been to train artificial neural networks to solve differential equations without the need of a numerical solver. This particular application offers an alternative to conventional numerical methods, with advantages such as lower memory required to store solutions, parallelization, and, in some cases, a lower overall computational cost than its numerical counterparts. In this work, we train artificial neural networks to represent a bundle of solutions of the differential equations that govern the background dynamics of the Universe for four different models. The models we have chosen are $\Lambda \mathrm{CDM}$, the Chevallier-Polarski-Linder parametric dark energy model, a quintessence model with an exponential potential, and the Hu-Sawicki $f(R)$ model. We use the solutions that the networks provide to perform statistical analyses to estimate the values of each model's parameters with observational data; namely, estimates of the Hubble parameter from cosmic
chronometers, type Ia supernovae data from the Pantheon compilation, and measurements from baryon
acoustic oscillations. The results we obtain for all models match similar estimations done in the literature using numerical solvers. In addition, we estimate the error of the solutions that the trained networks provide by comparing them with the analytical solution when there is one, or to a high-precision numerical solution when there is not. Through these estimations we find that the error of the solutions is at most $\sim1\%$ in the region of the parameter space that concerns the $95\%$ confidence regions that we find using the data, for all models and all statistical analyses performed in this work. Some of these results are made possible by improvements to the method of solving differential equations with artificial neural networks conceived in this work.
\end{abstract}

\maketitle

\section{Introduction}
To improve our understanding of the Universe, new models are constantly being proposed to better describe its different aspects.
All models need to be continuously tested against observational data to determine which models best describe nature. Besides, often the theoretical predictions are calculated from a system of differential equations that need to be solved first to test the model's predictions against the data. The usual approach to solving these equations is to use numerical methods, which usually provide the solution in the form of an array of points matching a discretization of the domain. In addition, any statistical analysis which aims to constrain the parameters of a model requires exploring the parameter space of such a model. For this, the numerical solver needs to integrate each time the parameters of the differential system are changed by any amount, which, in sufficiently complex systems, can create a bottleneck in the computation time required to make the analysis statistically robust.
The past few years have seen an increased interest in the use of artificial neural networks (ANNs) as new tools for analyzing both data and physical models.
For example, the so called physics-informed neural networks are able to interpolate the physical behavior of a given data set, provided the underlying physical laws are assumed \cite{pinns}. Some approaches in the field of cosmology include using ANNs to reconstruct cosmological data, like in Ref.~\cite{ANN_cosmo_reconstruction} where this was done for both cosmic expansion and large-scale structure data, and creating artificial data extending the range of current data sets as in Ref.~\cite{RNN+BNN_trained_on_Pantheon_to_generate_dl_at_high_z}. 
Additionally, ANNs have been used as a replacement for numerical methods to serve as solutions to physical models. For example, in Ref.~\cite{Cosmopower} ANNs were trained through supervised learning on the outputs of a numerical method to constrain the Lambda cold dark matter ($\Lambda \mathrm{CDM}$) model with  cosmic microwave background (CMB) and large-scale structure data. Also, ANNs trained through unsupervised learning were used to compute the tunneling profiles of cosmological phase transitions in Ref.~\cite{early_universe_model_lagaris_method} with a modified version of the method introduced in Ref.~\cite{ANN_diff_eqs}. 
Contrary to the numerical method, the solutions provided by ANNs are continuous and fully differentiable \cite{goodfellow2016deep}.\footnote{A discussion concerning this affirmation can be found in \cref{NN_details}.} Furthermore, typically the storage for ANNs is tiny, and especially compared to numerical methods. In this way, scientists and engineers can share the solutions used in their research with a wider community, easing the process of reproducibility.

An extension of the unsupervised method for solving differential equations with ANNs was proposed in Ref.~\cite{bundlesolutions}, which introduces the possibility of training ANNs whose outcome is a bundle of solutions corresponding to a continuous range for the values of the parameters of the differential system, which can include initial and boundary conditions.
The main advantage of this method can be summarized as follows: the integration required to obtain the solutions is performed once, during training, after which the solutions can be used indefinitely to perform parameter inference, requiring in some cases (mainly those where the differential system is not easy to integrate) less computational time than with traditional numerical solvers. In this work, we adapt the bundle ANN method introduced in Ref.~\cite{bundlesolutions}, to obtain solutions of physical models in a cosmological context and perform statistical analyses to obtain constraints on the model's parameters. For this we implement the \href{https://github.com/NeuroDiffGym/neurodiffeq}{NeuroDiffEq} \textsc{python} library \cite{outdated_neurodiff_ref} to solve the background dynamics of four cosmological models. Also, to adapt the method to the cosmological background equations, some improvements to the method are developed.

The physical models considered in this paper are part of an active area of research in cosmology, which constitutes modeling the late-time accelerated expansion of the Universe as first evidenced by measurements of type Ia supernovae \cite{Riess_1998, Perlmutter}. Currently, the favored model to explain this behavior is the $\Lambda \mathrm{CDM}$ model, where a positive cosmological constant $\Lambda$ is introduced to the Einstein-Hilbert action of general relativity to reproduce the late-time acceleration in the expansion of the Universe. Although this model represents the simplest solution, it has some serious issues, like the lack of a theoretical explanation for the observed value of $\Lambda$. Therefore, the physical mechanism that drives the accelerated expansion is still unknown. Another problem is the so-called Hubble tension. This tension concerns the apparent difference in the current value of the Hubble parameter $H_0$ between the estimate obtained with model-independent observations and those inferred from indirect measurements which therefore also require the assumption of a cosmological model. The most prominent example of this is the $5 \sigma$ tension between the value reported by direct measurements of the SH0ES Collaboration \cite{riess2022comprehensive} and the one inferred from CMB data by the \textit{Planck} Collaboration \cite{Planck}.
While this tension could be attributed to $\Lambda \mathrm{CDM}$ not being the correct cosmological model, there is still the possibility that the tension arises from systematics in the direct measurements. For recent reviews on this topic, we  refer the reader to Refs.~\cite{H_0_tension_review, cosmo_white_paper}.

A myriad of alternative models have been considered to explain the current accelerated expansion of the Universe. Such models can fall into one of three categories. A first alternative is to add a fluid to the energy-momentum tensor with an equation of state that can eventually change with time. These kinds of models are referred to in the literature as parametric dark energy models. The second one is to, instead, include a new field that is minimally coupled to gravity or matter in the energy-momentum tensor. The simplest examples of such models are those called as quintessence models, where the new field is a self-interacting scalar field. Last, another choice is to consider alternative theories to general relativity to describe the gravitational interaction, such as for example $f(R)$ theories, where a modification to the Einstein-Hilbert action in the form of a function of the Ricci scalar $R$ is introduced. In this work, we focus on one model from each of these categories in addition to $\Lambda \mathrm{CDM}$ and compute the expansion rate of the Universe using ANNs instead of numerical solvers. We consider i) the Chevallier-Polarski-Linder (CPL) parametric dark energy model \cite{darkenergy1,darkenergy2}, ii) 
a quintessence model with an exponential potential \cite{quintessence_exp_pot1,quintessence_exp_pot2}, and iii) the Hu-Sawicki $f(R)$ model \cite{Hu-Sawicki}. For both the $\Lambda \mathrm{CDM}$ and CPL models, the solutions to the differential equations can be found analytically, while the differential equations of quintessence and a cosmological model in $f(R)$ gravity do not have analytical solutions. Then, to conclude, we use the trained ANNs together with observational data from cosmic chronometers (CC) \cite{CC1,CC2,CC3,CC4,CC5,CC6}, type Ia supernovae (SNIa) \cite{SnIa_pantheon} and baryon acoustic oscillations (BAO) \cite{SDSS_DR7,WiggleZ,SDSS-IV_quasars,DES_Y1,BOSS,SDSS_IV_LRG_anisotropicCF,SDSS_IV_QSO_anisotropicPS,SDSS-III_La_forests,La_forests_quasars_cross} to do Bayesian inference on the parameters of all of these models.

We start by explaining the details of each model in Sec.~\ref{theo}, underlining the similarities between them. In addition, we describe the variables that we use in each model to have dimensionless quantities as the dependent variables of the differential equations. Next, in Sec.~\ref{methods}, we explain in further detail the ANN method and how it can be applied to an inference pipeline. We also describe the improvements made to the bundle method to implement it in the cosmological scenario. In Sec.~\ref{implementation}, we explain how we apply the ANN method to the cosmological models, and showcase the accuracy of our cosmology-informed neural networks in the parameter space of each model. Then, in Sec.~\ref{obs_data}, we give a description of the different data sets we use, the physical quantities that are derived from them, and their corresponding likelihoods, which are used to test the models' predictions in the statistical analysis. Later on, in Sec.~\ref{results}, we present the inferred constraints for each model. In Sec.~\ref{comp_eff} we analyze how the neural network method compares to a commonly used numerical method regarding computational efficiency. Finally, we share our conclusions and give a summary of our work in Sec.~\ref{conclusions}.

The codes used to train the ANNs in this work can be found in the following repository: \href{https://github.com/at-chantada/cosmo-nets}{https://github.com/at-chantada/cosmo-nets}.
\section{Theoretical models}\label{theo}
In this section, we explain the theoretical background of the models tested in this work. While each of the models has distinct characteristics, they still share some common underlying assumptions.

First, let us recall the cosmological principle, according to which the geometry of the spacial part of space-time is homogeneous and isotropic. Such space-time can be described by the Friedmann–Lemaître–Robertson–Walker (FLRW) metric. In reduced circumference polar coordinates, this metric sets the line element as\footnote{Throughout this paper we use $c=1$.}

\begin{equation}
\label{FLRW}
    ds^2=-dt^2 + a\left(t\right)^2\left[\dfrac{dr^2}{1-kr^2} + r^2\left(d\theta^2 + \sin^2\theta d\phi^2\right)\right].
\end{equation}
In this work, we use $k=0$, representing a spatially flat universe. We recall the definition of the ``Hubble parameter,'' $H=\dot{a}/a$, where the dot denotes the derivative with respect to the cosmological time and $a\left(t\right)$ is the scale factor 
introduced in \cref{FLRW}.

Second, it is usually assumed that the Universe is comprised of a perfect fluid and therefore the energy-momentum tensor is $T\indices{^\mu_\nu}=\text{diag}\left(-\rho,p,p,p\right)$,
where $\rho$ denotes the energy density and $p$ is the pressure of the fluid. Moreover, due to the cosmological principle, both $\rho$ and $p$ are independent of the position.
Then, the continuity equation (which in turn follows from the conservation of the energy-momentum tensor) reads
\begin{equation}
\label{continuity}
    \dot{\rho} + 3H\left(\rho + p \right)=0.
\end{equation}
Every model considered in this paper includes nonrelativistic matter that is assumed to be mostly dust. From this, and \cref{continuity}, an analytical solution for the nonrelativistic matter density $\rho_m$ as a function of the redshift, $z=\dfrac{1}{a}-1$, can be found,
\begin{equation}
\label{matter_sol}
    \rho_m\left(z\right)=\rho_{m,0}\left(1+z\right)^3,
\end{equation}
where $\rho_{m,0}$ denotes the matter density at the present time.
Last, because our analysis is done for values of redshift $z\in\left[0,10\right]$, the effects of radiation can be ignored.

\subsection{\texorpdfstring{$\boldsymbol{\Lambda \mathrm{CDM}}$}{LambdaCDM}}\label{LambdaCDM_theo}

In this model, the background cosmological evolution considering an energy-momentum tensor with only nonrelativistic matter, is described by the Friedmann equations as follows:
\begin{subequations}
\label{Friedmann_eqs}
\begin{align}
\label{friedmann_eq_1}
    H^2=&\dfrac{\kappa \rho_m+\Lambda}{3},\quad\\
\label{friedmann_eq_2}
    2\dot{H}+3H^2=&\Lambda,
\end{align}
\end{subequations}
where $\Lambda$ is the cosmological constant, $\kappa=8\pi G$ and $G$ is Newton's gravitational constant. Next,
$H_0$ is defined as the value of $H\left(z\right)$ at $z=0$, and $\Omega_{m,0}=\kappa\rho_{m,0}/3 H^2_0$. Therefore, using \cref{matter_sol} in \cref{friedmann_eq_1} we get
\begin{equation}
    H\left(z\right)=H_0\sqrt{\Omega_{m,0}\left(1+z\right)^3+\dfrac{\Lambda}{3H_0^2}},
\end{equation}
which can be evaluated at $z=0$ to obtain $\Lambda=3H_0^2\left(1-\Omega_{m,0}\right)$. With this result, we can express $H\left(z\right)$ as
\begin{equation}
\label{H_Lambda}
H\left(z\right)=H_0\sqrt{\Omega_{m,0}\left(1+z\right)^3+1-\Omega_{m,0}}.
\end{equation}

Given the fact that the only differential equation that needs to be solved in this model is \cref{continuity} for $\rho_m$ with $p_m=0$, we define the variable $x=\kappa\rho_m/3 H^2_0$ to get the following differential system:
\begin{equation}
\begin{aligned}
\label{x_matter}
    \dfrac{dx}{dz}=\dfrac{3x}{1+z}, &&
    \evalat{x\left(z\right)}{z=0}=\dfrac{\kappa\rho_{m,0}}{ 3H^2_0}=\Omega_{m,0}.
\end{aligned}
\end{equation}

The Hubble parameter obtained from the solution of \cref{x_matter} is
\begin{equation}
    H\left(z\right)=H_0\sqrt{x\left(z\right)+1-\Omega_{m,0}}.
\end{equation}

\subsection{Parametric dark energy}
As described in the Introduction, there are both theoretical and observational reasons to consider alternative cosmological models. A simple alternative to $\Lambda \mathrm{CDM}$, introduced phenomenologically, is to consider a new component of the energy-momentum tensor, whose equation of state is that of a fluid and a function of redshift. For this model, the equations that describe the expansion of the Universe replace \cref{friedmann_eq_1,friedmann_eq_2} to become

\begin{subequations}
\begin{align}
\label{friedmann_eqs_DE_1}
    H^2=&\dfrac{\kappa}{3}\left(\rho_m+\rho_{\rm DE}\right),\quad\\
\label{friedmann_eqs_DE_2}
    2\dot{H}+3H^2=&-\kappa p_{\rm DE},
\end{align}
\end{subequations}
where $\rho_{\rm DE}$ and $p_{\rm DE}$ represent the density and pressure of the new fluid component, respectively. This new fluid is often referred to as ``dark energy'' and it is characterized by the following equation of state:
\begin{equation}
\label{eq_state_DE}
    p_{\rm DE}=\omega\left(z\right)\rho_{\rm DE},
\end{equation}
where $\omega\left(z\right)$ is the model's parametrization of the equation of state.
 In this work, we use the CPL parametric dark energy model introduced in Refs.~\cite{darkenergy1, darkenergy2}, with the equation of state
\begin{equation}
    \omega\left(z\right)=\omega_0 + \dfrac{\omega_1 z}{1+z},
\end{equation}
where $\omega_0$ and $\omega_1$ are free parameters.
Applying the CPL parametrization in \cref{eq_state_DE} and using \cref{continuity}, the differential equation of the model is
\begin{equation}
\label{diff_DE}
\dfrac{d\rho_{\rm DE}}{dz}=\dfrac{3\rho_{\rm DE}}{1+z}\left(1+\omega_0 + \dfrac{\omega_1 z}{1+z}\right).
\end{equation}
This differential equation has the analytical solution
\begin{equation}
\label{DE_sol}
    \rho_{\rm DE}\left(z\right)=\rho_{\mathrm{DE},0}\left(1+z\right)^{3\left(1+\omega_0+\omega_1\right)}\exp\left({-\dfrac{3\omega_1z}{1+z}}\right),
\end{equation}
where $\rho_{\mathrm{DE},0}$ is the value of the dark energy density at $z=0$.
Using \cref{matter_sol,DE_sol} in \cref{friedmann_eqs_DE_1}, along with the definition of $H_0$ and $\Omega_{m,0}$, the expression for the Hubble parameter is
\begin{widetext}
\begin{equation}
    H\left(z\right)=H_0\sqrt{\Omega_{m,0}\left(1+z\right)^3+\left(1-\Omega_{m,0}\right)\left(1+z\right)^{3\left(1+\omega_0+\omega_1\right)}\exp\left({-\dfrac{3\omega_1z}{1+z}}\right)}.
\end{equation}
\end{widetext}

Doing an analogous change of variable as the one done in $\Lambda \mathrm{CDM}$, we define $x=\kappa\rho_{\rm DE}/3 H^2_0$, and highlighting the fact that $\kappa\rho_{\mathrm{DE},0}/3 H^2_0=1-\Omega_{m,0}$, the differential equation and initial condition of the problem can be expressed as
\begin{equation}
\begin{aligned}
\label{x_DE}
    \dfrac{dx}{dz}=&\dfrac{3x}{1+z}\left(1+\omega_0 + \dfrac{\omega_1 z}{1+z}\right),\quad\\
    \evalat{x\left(z\right)}{z=0}=&\dfrac{\kappa\rho_{\mathrm{DE},0}}{3 H^2_0}=1-\Omega_{m,0}.
\end{aligned}
\end{equation}
Considering this change of variable, the expression of the  Hubble parameter is
\begin{equation}
    H\left(z\right)=H_0\sqrt{\Omega_{m,0}\left(1+z\right)^3+x\left(z\right)}.
    \label{H_DE}
\end{equation}
\subsection{Quintessence}\label{quintessence}
An alternative proposal to explain the present accelerated expansion of the Universe, is to consider a scalar field $\phi$ minimally coupled to gravity with an appropriate potential $V\left(\phi\right)$. These kinds of cosmological models assume general relativity as the underlying gravitational theory and are usually referred to in the literature as quintessence models \cite{quintessence1,quintessence2,quintessence3}. Let us recall that the energy density and pressure of a homogeneous scalar field in a flat FLRW background are
given by $\rho_\phi=\dot{\phi}^2/2+V\left(\phi\right)$ and $p_\phi=\dot{\phi}^2/2-V\left(\phi\right)$. Besides, the expansion of the Universe can be described by the following Friedmann equations:
\begin{subequations}
\label{friedmann_eqs_q}
\begin{align}
\label{fried_q1}
H^2=&\dfrac{\kappa}{3}\left(\rho_m+\dot{\phi}^2/2+V\left(\phi\right)\right),\\
\label{fried_q2}
2\dot{H}+3H^2=&-\kappa \left(\dot{\phi}^2/2-V\left(\phi\right)\right).
\end{align}
\end{subequations}
 In addition, applying \cref{continuity} to $\rho_\phi$ and $p_\phi$, the equation of motion of the scalar field can be obtained:
\begin{equation}
\label{phi_dynamics}
    \ddot{\phi}+3H\dot{\phi}+\dfrac{dV}{d\phi}=0.
\end{equation}
The quintessence model considered in this work corresponds to a model characterized by an exponential potential $V\left(\phi\right)=V_0\exp{\left(-\lambda \sqrt{\kappa}\phi\right)}$ as proposed in Refs.~\cite{quintessence_exp_pot1,quintessence_exp_pot2}, where $\lambda$ is a free parameter, and $V_0$ is set by the initial conditions.
Equations \eqref{friedmann_eqs_q} and \eqref{phi_dynamics}, along with \cref{continuity} applied to the matter density, provide the necessary equations to solve the dynamics of the problem given the initial conditions.

In order to solve \cref{fried_q1,fried_q2}, we use the change of variables introduced in Ref.~\cite{var_change_quintessence} such that
\begin{equation}
\begin{aligned}
\label{var_phi}
x=\dfrac{\sqrt{\kappa}\dot{\phi}}{\sqrt{6}H},  && y =\dfrac{\sqrt\kappa V\left(\phi\right)}{\sqrt{3}H}.
\end{aligned}
\end{equation}
Using these new variables along with \cref{friedmann_eqs_q,phi_dynamics}, plus defining $N=-\ln\left(1+z\right)$, the differential equations to solve for $x$ and $y$ are
\begin{equation}
\left\{ \begin{aligned}\dfrac{dx}{dN}= & -3x+\dfrac{\sqrt{6}}{2}\lambda y^{2}+\dfrac{3}{2}x\left(1+x^{2}-y^{2}\right)\\
\dfrac{dy}{dN}= & -\dfrac{\sqrt{6}}{2}xy\lambda+\dfrac{3}{2}y\left(1+x^{2}-y^{2}\right).
\end{aligned}
\right.\label{diff_q}
\end{equation}
To set the initial conditions of $x$ and $y$, we assume that $H\left(z\right)$ should match 
the corresponding $\Lambda \mathrm{CDM}$ solution at $z=z_0$. In this work, we choose $z_0=10$ and set $N_0$ accordingly.
In this way, we can write
\begin{equation}
\begin{aligned}
\label{cond_xy}
x_0=0,  && y_0 =\sqrt{\dfrac{1-\Omega_{m,0}^{\Lambda}}{\Omega_{m,0}^{\Lambda}\left(1+z_{0}\right)^{3}+1-\Omega_{m,0}^{\Lambda}}},
\end{aligned}
\end{equation}
where the superscript $\Lambda$ refers to the quantities defined in the $\Lambda \mathrm{CDM}$ model. Therefore, $\Omega_{m,0}^{\Lambda}$ is the matter density parameter in the $\Lambda \mathrm{CDM}$ model which should be distinguished from $\Omega_{m,0}$ the matter density parameter defined in the quintessence model. Both quantities are related as follows:
\begin{equation}
\Omega_{m,0} H_0^2 = \Omega_{m,0}^{\Lambda} \left(H_0^{\Lambda}\right)^2.
\label{LCDMtoAlt}
\end{equation}
Once the solutions for $x$ and $y$ for the system of equations \eqref{diff_q} with initial conditions described by \cref{cond_xy} are obtained, the Hubble parameter is computed from the following expression:
\begin{equation}
    H=H^{\Lambda}_0\sqrt{\dfrac{\Omega_{m,0}^{\Lambda}\left(1+z\right)^3}{1-x^2-y^2}}.
    \label{H_quintessence}
\end{equation}
Here the Hubble factor that describes the expansion of the Universe in the quintessence model is expressed in terms of the Hubble constant and matter density of the $\Lambda \mathrm{CDM}$ model. However, we will report the results of the parameter inference (see Sec.~\ref{results}) in terms of the parameters defined in the quintessence model ($H_0$ and $\Omega_{m,0}$).
\subsection{\texorpdfstring{$f(R)$}{f(R)} gravity}
\label{fR_gravity}
Another possibility to explain the physical mechanism responsible for the accelerated expansion of the Universe is to assume an alternative theory to general relativity to describe the gravitational interaction \cite{ReviewMOG}.
In this paper, we focus on a particular class of theories [referred to in the literature as $f(R)$ theories], namely, those in which the action is written as \cite{first_f_R_paper}
\begin{equation}
\label{S_fR}
    S = \dfrac{1}{2\kappa}\int d^4x\sqrt{-g}f(R) + S_m.
\end{equation}
Here $R$ is the Ricci scalar and $f(R)$ is an arbitrary function of the latter.
Applying the modified Einstein equations derived from \cref{S_fR}, along with the assumptions listed at the start of this section, results in the modified Friedmann equations
\begin{subequations}
\label{friedmod}
\begin{align}
\label{friedmod1}
H^{2}=&\dfrac{1}{3f_{R}}\left[\kappa \rho_{m}+\dfrac{Rf_{R}-f}{2}-3H\dot{R}f_{RR}\right],\\
\begin{split}
    2\dot{H}+3H^{2}=&\dfrac{-1}{f_{R}}\left[-\dfrac{Rf_{R}-f}{2}+f_{RRR}\dot{R}^{2}\right.\\
    &\left.+\left(\ddot{R}+2H\dot{R}\right)f_{RR}\right],
\label{eq:friedmod2}
\end{split}
\end{align}
\end{subequations}
where the notation $f_R=df/dR$, $f_{RR}=d^2f/dR^2$, etc., is used.

It has been shown that only $f(R)$ models that show a ``chameleon-like'' effect are able to explain observational bounds from Solar System and equivalence principle tests \cite{Brax2008,Hui2009}. For this reason, we focus on one of the few $f(R)$ models that have not been ruled out by the aforementioned local gravity tests, namely, the Hu-Sawicki model \cite{Hu-Sawicki} where \footnote{In the Hu-Sawicki model, usually $f(R)=R-\dfrac{m^{2}c_{1}\left(R/m^{2}\right)^{n}}{c_{2}\left(R/m^{2}\right)^{n}+1} $. In this paper, we focus on the case $n=1$ and we write $f(R)$ as a function of $b=2/c_1$ while $c_2$ and $m^2$ are fixed such that in the high-curvature regime, the $\Lambda \mathrm{CDM}$ behavior is reproduced.}
\begin{equation}
f(R)=R-2\Lambda\left[1+\dfrac{1}{\dfrac{R}{\Lambda b}+1}\right].\label{f_R_HS_b}
\end{equation}
As \cref{f_R_HS_b} shows, when $b\xrightarrow{}0$, the usual general relativity with the cosmological constant is recovered. Therefore, $b$ quantifies the deviation from general relativity.

To finally set the parameters of the problem, we express the cosmological constant in terms of the parameters in $\Lambda \mathrm{CDM}$ as $\Lambda=3\left(H^\Lambda_{0}\right)^{2}\left(1-\Omega^\Lambda_{m,0}\right)$. 
In this way, the parameters of the problem are $\left(b,\Omega^\Lambda_{m,0},H^\Lambda_{0}\right)$. However, as explained for the quintessence model, results for the parameter inference will be shown in terms of the parameters $\Omega_{m,0}$ and $H_0$ defined in the $f(R)$ models and related to the $\Lambda \mathrm{CDM}$ parameters through \cref{LCDMtoAlt}.

To solve the modified Friedmann equations, we use the variables introduced in Ref.~\cite{f_R_var}
\begin{equation}
\label{var_f_R}
\begin{aligned}\begin{aligned}x= & \dfrac{\dot{R}f_{RR}}{Hf_{R}},\\
v= & \dfrac{R}{6H^{2}},
\end{aligned}
 & \begin{aligned}\qquad y= & \dfrac{f}{6H^{2}f_{R}},\\
\qquad\Omega= & \dfrac{\kappa\rho_{m}}{3H^{2}f_{R}}.
\end{aligned}
\end{aligned}
\end{equation}
Using these variables, \cref{friedmod1} gives the relation
\begin{equation}
\label{eq_1}
    1=\Omega + v - x - y.
\end{equation}
The differential system that follows from the derivatives of the variables in \cref{var_f_R} with respect to $z$, has the problem of including a function $\Gamma=f_R/Rf_{RR}$. For some simple $f(R)$ models, it is possible to express $\Gamma$ in terms of $v$ and $y$, but in the case of the Hu-Sawicki model this is not the case. A way to solve this problem is to add an extra variable $r=R/\Lambda$ to \cref{var_f_R} to be able to write $\Gamma$ in terms of it. With this new addition, the differential system is

\begin{equation}
\left\{ \begin{aligned}\dfrac{dx}{dz} & =\dfrac{1}{1+z}\left(-\Omega-2v+x+4y+xv+x^{2}\right)\\
\dfrac{dy}{dz} & =\dfrac{-1}{1+z}\left(vx\Gamma-xy+4y-2yv\right)\\
\dfrac{dv}{dz} & =\dfrac{-v}{1+z}\left(x\Gamma+4-2v\right)\\
\dfrac{d\Omega}{dz} & =\dfrac{\Omega}{1+z}\left(-1+2v+x\right)\\
\dfrac{dr}{dz} & =-\dfrac{r\Gamma x}{1+z},
\end{aligned}
\right.\label{diff_f_R_2}
\end{equation}
where the $\Gamma$ corresponding to the $f(R)$ in \cref{f_R_HS_b} can be expressed as follows:
\begin{equation}
\label{eq:Gamma}
\Gamma\left(r\right)=\dfrac{\left(r+b\right)\left[\left(r+b\right)^{2}-2b\right]}{4br}.
\end{equation}

Details of the calculation of the differential equation for $r$ can be found in \cref{models_calcs}. As we did for the quintessence model, we choose the initial conditions so that the behavior of the model matches that of the $\Lambda\mathrm{CDM}$ model at $z=z_0$. Furthermore, the relation between $H\left(z\right)$ and the Ricci scalar is given by $R=6\left(2H^2+\dot{H}\right)$, and therefore the initial conditions of \cref{diff_f_R_2} can be written as
\begin{subequations}
\label{f_R_ci}
\begin{align}
\label{x_0}
    x_0&=0,\\
\label{y_0}
    y_{0}&=\dfrac{\Omega_{m,0}^{\Lambda}\left(1+z_{0}\right)^{3}+2\left(1-\Omega_{m,0}^{\Lambda}\right)}{2\left[\Omega_{m,0}^{\Lambda}\left(1+z_{0}\right)^{3}+1-\Omega_{m,0}^{\Lambda}\right]},\\
\label{v_0}
    v_{0}&=\dfrac{\Omega_{m,0}^{\Lambda}\left(1+z_{0}\right)^{3}+4\left(1-\Omega_{m,0}^{\Lambda}\right)}{2\left[\Omega_{m,0}^{\Lambda}\left(1+z_{0}\right)^{3}+1-\Omega_{m,0}^{\Lambda}\right]},\\
\label{Omega_0}
    \Omega_{0}&=\dfrac{\Omega_{m,0}^{\Lambda}\left(1+z_{0}\right)^{3}}{\Omega_{m,0}^{\Lambda}\left(1+z_{0}\right)^{3}+1-\Omega_{m,0}^{\Lambda}},\\
\label{r_0}
    r_{0}&=\dfrac{\Omega_{m,0}^{\Lambda}\left(1+z_{0}\right)^{3}+4\left(1-\Omega_{m,0}^{\Lambda}\right)}{1-\Omega_{m,0}^{\Lambda}}.
\end{align}
\end{subequations}
After solving the differential equations, one can obtain the solution for $H$ using the following relation:
\begin{equation}
    H=H^\Lambda_{0}\sqrt{\dfrac{r}{2v}\left(1-\Omega^\Lambda_{m,0}\right)}.
    \label{H_fR}
\end{equation}
The final change of variables that we make is due to the nature of $r$. This variable reaches high values and grows rapidly with $z$. Such behavior presents a challenge to the ANN (see discussion in Sec.~\ref{methods_1}). Therefore, we propose to use $r^\prime=\ln r$, so now the variable that we solve for does not present such a challenge to the ANN.

It is necessary to mention that this last variable change restricts the values that $R$ can take to just $R>0$. However, doing an exploration of the parameter space $\left(z, b, \Omega^{\Lambda}_{m,0}\right)\in[0, 10]\times[0, 5]\times[0.1, 0.4]$ using a numerical solver on \cref{diff_f_R_2} without the variable change, we find that $R>0$ in that region, so this is not a concern.
\section{Methods}\label{methods}
As mentioned in the Introduction, we employ ANNs to solve differential equations. In particular, we use the \href{https://github.com/NeuroDiffGym/neurodiffeq}{NeuroDiffEq} \textsc{python} library \cite{outdated_neurodiff_ref}, which provides the tools to solve differential equations using ANNs.
We explain some aspects of the method that are specific to this application and refer the reader to the original paper, documentation and library for more details. We explain the neural network approach for ordinary differential equations (ODEs) or systems of ODEs, which are the relevant ones in this paper, but note that the library also has the capability of solving partial differential equations (PDEs) and systems of PDEs.

\subsection{Neural networks that solve differential equations}
\label{methods_1}
The problem of solving the equations using ANNs is framed as an optimization problem, where the task is to find the set of internal parameters of the network (see \cref{NN_details} for details) that minimize a cost (or loss) function $L$ that quantifies how well the output of the ANN satisfies the differential equation. The ANN solution also has to satisfy the initial and/or boundary conditions of the system. We use the simplest form for the loss function:
\begin{equation}
\label{loss_res}
L\left(\boldsymbol{u}_{\mathcal{N}},t\right)=\sum_{i}^{M}\mathcal{R}_{i}\left(\boldsymbol{u}_{\mathcal{N}},t\right)^{2},
\end{equation}
where $\boldsymbol{u}_{\mathcal{N}}=\boldsymbol{u}_{\mathcal{N}}\left(t\right)=\left(u_{\mathcal{N}}^{1}\left(t\right),u_{\mathcal{N}}^{2}\left(t\right),\dots,u_{\mathcal{N}}^{M}\left(t\right)\right)$ is the vector comprised of the outputs of the ANN for each of the $M$ dependent variables of the problem at $t$ (the independent variable) and the superscript constitutes a label for each dependent variable, while the subscript $\mathcal{N}$ denotes that $u_{\mathcal{N}}$ is the output of the ANN. Besides, $\mathcal{R}_i$ represents the residual of the $i$-th equation in the differential system with $M$ equations. 
While the true solution satisfies exactly the differential equation, if we evaluate the
$i$-th differential equation using the ANN solution ($\boldsymbol{u}_{\mathcal{N}}$), we obtain something different from zero, which we call the residual of the $i$-th equation. 
In particular, we employ an architecture in which we have one ANN for each dependent variable of the problem. This way, each $u_{\mathcal{N}}^{j}\left(t\right)$ is the only output of a single ANN with a single input $t$. 
We remark that the analytical or numerical solutions are not used in the training of the ANNs [which can be inferred from \cref{loss_res}], stressing the unsupervised nature of the method used in this work.

The criteria for determining when a residual is acceptable is that its scale must be several orders of magnitude lower than the scale of the variables of the problem. Therefore, rescaled variables of order one are preferred for this method. For example, in the case of the Hu-Sawicki $f(R)$ model, one of the variables of the problem reaches high values and therefore we implement a change of variables so that this variable is of order one (see discussion at the end of Sec.~\ref{fR_gravity}). Besides, the relation between the ANN solution errors and the residuals has not been analyzed yet for the cosmological context. However, a recent work \cite{liu2022evaluating} studied this issue for some types of differential equations, and its application to cosmology is left for future work.

While \cref{loss_res} is a good representation for how well the ANNs solve the equations of a given differential system, the information for the initial/boundary conditions is missing. To impose them, a reparametrization $\tilde{\boldsymbol{u}}\left(\boldsymbol{u}_{\mathcal{N}},t\right)$ of $\boldsymbol{u}_{\mathcal{N}}$ can be defined. For example, in the case of a system of first-order ODEs with initial conditions as $\evalat{\boldsymbol{u}\left(t\right)}{t=t_0}=\boldsymbol{u}_0$, the reparametrization must satisfy $\tilde{\boldsymbol{u}}\left(\boldsymbol{u}_{\mathcal{N}}, t_0\right)=\boldsymbol{u}_0$. An example of a good reparametrization choice for this case introduced in Ref.~\cite{marios_exp_reparam} is
\begin{equation}
    \tilde{\boldsymbol{u}}\left(t\right)=\boldsymbol{u}_0+\left(1-e^{-\left(t-t_0\right)}\right)\boldsymbol{u}_{\mathcal{N}}\left(t\right).
    \label{default_reparam}
\end{equation}
This way, to solve ODEs, the loss function to optimize for is
\begin{equation}
    L\left(\tilde{\boldsymbol{u}},t\right)=\sum_{i}^{M}\mathcal{R}_{i}\left(\tilde{\boldsymbol{u}},t\right)^{2}.
    \label{loss_res_reparam}
\end{equation}
The loss displayed in \cref{loss_res_reparam} takes as input all of the reparametrized outputs of the ANNs, and outputs a scalar that serves as a measurement for the accuracy of the ANNs at solving a given ODE system at $t$.

\subsection{Bundle solution}\label{bundles}
An extension of the regular ANN method described before can be made, as suggested in Ref.~\cite{bundlesolutions}, for the trained neural networks to represent a set (or bundle) of solutions of a given differential system for a continuous range of its $S$ parameters $\boldsymbol{\theta}=\left(\theta^{1},\theta^{2},\dots,\theta^{S}\right)$. To accomplish this, $\boldsymbol{u}_{\mathcal{N}}$ must now have $S$ extra inputs for the parameters of the system. Therefore, if $t_0$ is not part of the $S$ parameters, now \cref{default_reparam} turns into
\begin{equation}
    \tilde{\boldsymbol{u}}\left(t, \boldsymbol{\theta}\right)=\boldsymbol{u}_0\left(\boldsymbol{\theta}\right)+\left(1-e^{-\left(t-t_0\right)}\right)\boldsymbol{u}_{\mathcal{N}}\left(t, \boldsymbol{\theta}\right),
    \label{default_reparam_bundle}
\end{equation}
where, because the parameters entering in the initial conditions are also parameters of the system, $\boldsymbol{u}_0=\boldsymbol{u}_0\left(\boldsymbol{\theta}\right)$. In addition, \cref{loss_res_reparam} turns into
\begin{equation}
    L\left(\tilde{\boldsymbol{u}},t,\boldsymbol{\theta}\right)=\sum_{i}^{M}\mathcal{R}_{i}\left(\tilde{\boldsymbol{u}},t,\boldsymbol{\theta}\right)^{2},
    \label{loss_res_reparam_bundle}
\end{equation}
representing the loss function for the bundle solution at time $t$, for the differential system defined by the parameters $\boldsymbol{\theta}$. For a concrete example, the details of the implementation of this method to the cosmological background equation of the $\Lambda\mathrm{CDM}$ model can be found in \cref{ex_lcdm}.

A natural next step, is to integrate this method into a pipeline whose goal is to perform a statistical analysis to test the theoretical models using observational data.

\begin{figure*}
    \centering
    \includegraphics{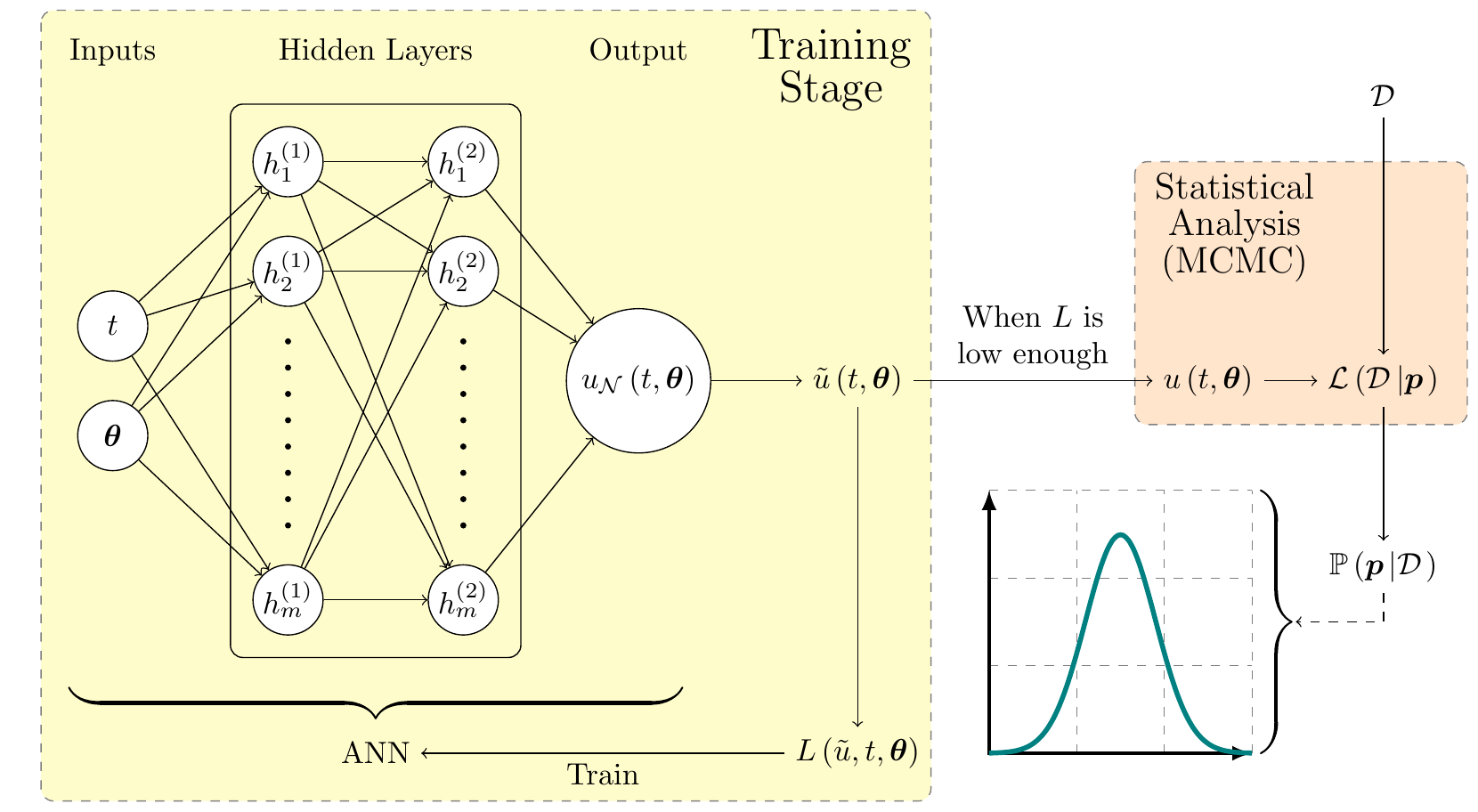}
        \caption{Illustration of an example implementation of ANN-based bundle solutions to a differential system of a model $\mathcal{M}\left(\boldsymbol{p}\right)$ in a statistical analysis with a data set $\mathcal{D}$ to obtain samples from the posterior $\mathbb{P}\left(\boldsymbol{p}\left|\mathcal{D}\right.\right)$.}
        \label{fig:workflow}
\end{figure*}

Let us say that we have a data set $\mathcal{D}$ and a model $\mathcal{M}\left(\boldsymbol{p}\right)$, where $\boldsymbol{p}$ is a vector comprised of all the parameters and observables of the model. Our objective is to draw conclusions on the possible values of $\boldsymbol{p}$ using $\mathcal{D}$. In general, to achieve this outcome, one has to solve a differential system characterized by its differential equations, with initial and/or boundary conditions.
We can describe the equations of the differential system by its $M$ residuals $\mathcal{R}_i\left(\boldsymbol{u},t, \boldsymbol{\theta}\right)$, where $\boldsymbol{u}=\left(u^1,u^2,\dots,u^M\right)$ are the dependent variables of the system. Here, it is important to clarify that $\boldsymbol{\theta}$ is the vector that contains only the parameters of the differential equations and those that are necessary to write the initial conditions, such as $\Omega_{m,0}$ in \cref{x_matter}. This implies that $\boldsymbol{\theta}$ is contained within $\boldsymbol{p}$, and there can be parameters of the model that do not change the differential system itself but are included in the observable quantities that are considered in the statistical analysis. For example, $H_0^\Lambda$ is included in the expression of the observable $H\left(z\right)$ in all models analyzed in this paper [see Eqs.~\eqref{H_Lambda}, \eqref{H_DE}, \eqref{H_quintessence} and \eqref{H_fR}] but not in their respective differential equations or initial conditions. Now that the problem is defined, the next step is to set up a loss and reparametrization as was described at the start of this subsection, with the goal of training ANNs to solve the given bundle problem. Once the training is finished, $\tilde{\boldsymbol{u}}\left(t, \boldsymbol{\theta}\right) $ becomes the bundle solution ${\boldsymbol{u}}\left(t, \boldsymbol{\theta}\right)$. Then, the latter can be used together with the data in a statistical analysis to infer bounds on the parameters $\boldsymbol{p}$.
The method described before is illustrated in \cref{fig:workflow} for the case where the differential system is a single ODE.

\subsection{Case-specific extensions}

The regular and bundle methods described so far can be used to tackle any differential system. 
When applying the bundle method to the cosmological background scenario, we improve the method in Ref.~\cite{bundlesolutions} by proposing two new reparametrizations of the ANN solutions. These improvements of the ANN bundle method are applied to the CPL, quintessence, and Hu-Sawicki $f(R)$ cosmological models and are an important achievement of this work.

\subsubsection{Integrating factor case}
\label{integrating_factor}

To describe the first reparametrization, let us consider a homogeneous linear first-order ODE.
We can take advantage of the fact that for the general case 
\begin{equation}
    \dfrac{du}{dt}=uf\left(t\right),
\end{equation}
where $f$ is an arbitrary function of the independent variable $t$, the general solution to the problem with an initial condition $u_0$ at $t_0$ is
\begin{equation}
    \label{integrating_factor_diff}
    u=u_0\exp\left[\int^t_{t_0}f\left(t^\prime\right)dt^\prime\right].
\end{equation}
Now, we take this one step further into the bundle problem
\begin{equation}
    \label{bundle_linear}
    \dfrac{du}{dt}=uf\left(t,\boldsymbol{\theta}\right),
\end{equation}
with the initial condition $u_0\left(\boldsymbol{\theta}\right)$. If we consider the special case where the function $f\left( t, \boldsymbol{\theta}\right)$ can be expressed as $f\left(t,\boldsymbol{\theta}\right)=\sum_{i}^{Q} h_i\left(\boldsymbol{\theta}\right) g_i\left(t\right)$,\footnote{While there can be multiple ways of expressing a given function $f\left(t,\boldsymbol{\theta}\right)$ as the required sum, one should strive to choose the one that corresponds to the smallest $Q$ possible to avoid unnecessary complexity.} the general solution to \cref{bundle_linear} is 
\begin{equation}
    u=u_0\left(\boldsymbol{\theta}\right)\exp\left[\sum_{i}^{Q} h_i\left(\boldsymbol{\theta}\right) \int^t_{t_0}g_i\left(t^\prime\right)dt^\prime\right].
\end{equation}

We can now take this knowledge and come up with an ANN architecture and reparametrization $\tilde{u}$ that takes advantage of it. Choosing an ANN with $Q$ outputs $i$ and one input $t$ we propose
\begin{equation}
    \label{bundle_reparam_linear}
    \tilde{u}\left(t, \boldsymbol{\theta}\right)=u_0\left(\boldsymbol{\theta}\right)\exp\left\{\sum^Q_i h_i\left(\boldsymbol{\theta}\right)\left[u_{\mathcal{N},i}\left(t\right) - u_{\mathcal{N},i}\left(t_0\right)\right]\right\},
\end{equation}
where $u_{\mathcal{N},i}$ is the $i$-th output of the neural network. The main advantage of this reparametrization is that it provides a way of getting a bundle solution where the neural network does not need to have the parameters of the differential system as an input. Also, it is important to note that there is only one ANN involved in this specific implementation, which corresponds to the single dependent variable of \cref{bundle_linear}. The $Q$ different outputs of this ANN are necessary for the viability of this specific implementation. This reparametrization is implemented for the bundle method in the CPL model. The details are shown in \cref{ex_cpl}.

\subsubsection{Perturbative reparametrization}\label{pert_reparam}

To make the job of minimizing the loss $L$ in \cref{loss_res_reparam} easier for the ANNs, we propose to modify the reparametrization in \cref{default_reparam} by providing it with an approximate solution $\hat{\boldsymbol{u}}\left(t\right)$ to the problem , which also satisfies $\hat{\boldsymbol{u}}\left(t=t_0\right)=\boldsymbol{u}_0$. This way the reparametrization becomes
\begin{equation}
    \tilde{\boldsymbol{u}}\left(t\right)=\hat{\boldsymbol{u}}\left(t\right)+\left(1-e^{-\left(t-t_0\right)}\right)\boldsymbol{u}_{\mathcal{N}}\left(t\right).
    \label{perturbative_reparam}
\end{equation}
This new reparametrization can be thought of as a perturbative solution to the problem, where the job of the second term of \cref{perturbative_reparam} is to correct the approximate solution.

This approach can be specialized further for the bundle case, by choosing the approximate solution as the true solution for a fixed value of one or more of the $S$ parameters. Here, the approximate solution plays the role of a boundary condition in the bundle solution. In this way, if $\hat{\boldsymbol{u}}\left(t,\theta^{2},\dots,\theta^{S}\right)=\boldsymbol{u}\left(t,\theta^{1}=\theta_0^{1},\theta^{2},\dots,\theta^{S}\right)$ denotes the true solution when the parameter $\theta^{1}=\theta_0^{1}$ we can recast \cref{default_reparam_bundle} as

\begin{equation}
\begin{split}
    \tilde{\boldsymbol{u}}\left(t, \boldsymbol{\theta}\right)=&\hat{\boldsymbol{u}}\left(t,\theta^{2},\dots,\theta^{S}\right)\\
    &+\left(1-e^{-\left(t-t_0\right)}\right)\left(1-e^{-\left(\theta^{1}-\theta^{1}_0\right)}\right)\boldsymbol{u}_{\mathcal{N}}\left(t, \boldsymbol{\theta}\right),
    \label{perturbative_reparam_bundle}
\end{split}
\end{equation}
where this can be naturally extended when more than one parameter needs to be fixed to use a true solution as the basis for the reparametrization.

\paragraph*{Example:}
We show a simple example to illustrate the method with the following first-order ODE:
\begin{equation}
\begin{aligned}
\label{pert_example_ODE}
\dfrac{dx}{dt}=x\left(\alpha + \beta \sin x\right),  && \evalat{x\left(t\right)}{t=t_0}=x_0.
\end{aligned}
\end{equation}
For this ODE, we know that the solution when $\beta=0$ is $x=x_0\exp\left[{\alpha\left(t-t_0\right)}\right]$, which we can call $\hat{x}\left(t,x_0, \alpha\right)$. Therefore, we can use $\hat{x}\left(t,x_0,\alpha\right)$ as a boundary condition in the bundle reparametrization when $\beta=0$. So the perturbative reparametrization would be
\begin{equation}
\begin{split}
    \tilde{x}\left(t,x_0,\alpha,\beta\right)=& \hat{x}\left(t,x_0, \alpha\right)\\
    &+ \left(1-e^{-\left(t-t_0\right)}\right)\left(1-e^{-\beta}\right)\\
    &\times x_{\mathcal{N}}\left(t,x_0,\alpha, \beta\right),
\end{split}
\end{equation}
where $x_{\mathcal{N}}$ is the output of the ANN. Then, the loss function that the ANN must minimize to solve the problem is
\begin{equation}
    L\left(\tilde{x},t,x_0,\alpha,\beta\right)=\left[\dfrac{d\tilde{x}}{dt}-\tilde{x}\left(\alpha + \beta \sin\tilde{x}\right)\right]^2.
\end{equation}
The perturbative reparametrization is implemented for the quintessence and $f(R)$ Hu-Sawicki models. For the latter, this allows us to tackle the problem introduced by a singularity in the system of equations (see discussion in Sec.~\ref{f_R_loss}).
\section{Implementation specifics}\label{implementation}
In this section, we describe how we implement the methods described in Sec.~\ref{methods} in the quintessence and Hu-Sawicki $f(R)$ models.\footnote{The  specifics of the implementations for the $\Lambda\mathrm{CDM}$ and CPL models are detailed in \cref{app_1}.} Some of the details of each implementation are omitted in this section for the sake of readability. We elaborate on these details in \cref{details_imp_models}. To close out this section, we share an estimation of the errors of our cosmology-informed neural networks that we obtain for each model in their respective parameter space.

\subsection{Quintessence implementation}
The perturbative reparametrization described in the last section can be implemented to successfully solve the background equations of the quintessence model. We show in \Cref{lambda=0} that when $\lambda=0$ the solution of \cref{diff_q} is equivalent to the $\Lambda \mathrm{CDM}$ solution, which implies that
\begin{equation}
\begin{aligned}
\label{xy_lambda=0}
x\left(N\right)=0,  && y\left(N\right)=\sqrt{\dfrac{1-\Omega_{m,0}^{\Lambda}}{\Omega_{m,0}^{\Lambda}e^{-3N}+1-\Omega_{m,0}^{\Lambda}}}.
\end{aligned}
\end{equation}
Therefore, defining $\hat{x}\left(N, \Omega_{m,0}^{\Lambda}\right)=0$ and setting $\hat{y}\left(N, \Omega_{m,0}^{\Lambda}\right)$ equal to the $y\left(N\right)$ in \cref{xy_lambda=0}, we can apply the method described in Sec.~\ref{pert_reparam}, and set a boundary condition at $\lambda=0$. Thus, the reparametrizations of the outputs of the ANNs that solve the differential system in \cref{diff_q} are
\begin{subequations}
\label{quintessence_reparam}
\begin{align}
\label{x_reparam}
    \tilde{x}\left(N, \lambda, \Omega_{m,0}^{\Lambda}\right)=&\left(1-e^{-\left(N-N_0\right)}\right)\left(1-e^{-\lambda}\right)\\
    &\times x_{\mathcal{N}}\left(N, \lambda, \Omega_{m,0}^{\Lambda}\right),\nonumber\\
\begin{split}
\label{y_reparam}
    \tilde{y}\left(N, \lambda,\Omega_{m,0}^{\Lambda} \right)=&\hat{y}\left(N,\Omega_{m,0}^{\Lambda}\right)\\
    &+\left(1-e^{-\left(N-N_0\right)}\right)\left(1-e^{-\lambda}\right)\\
    &\times y_{\mathcal{N}}\left(N, \lambda, \Omega_{m,0}^{\Lambda}\right).
\end{split}
\end{align}
\end{subequations}
In this model, there are two individual ANNs, one for each dependent variable. Both of them have $N$, $\lambda$ and $\Omega^{\Lambda}_{m,0}$ as inputs, and the outputs are denoted as $x_{\mathcal{N}}\left(N, \lambda, \Omega_{m,0}^{\Lambda}\right)$ for the network assigned to $x$, and $y_{\mathcal{N}}\left(N, \lambda, \Omega_{m,0}^{\Lambda}\right)$ for the one that corresponds to $y$.

\subsection{\texorpdfstring{$f(R)$}{f(R)} gravity implementation}\label{f_R_loss}
In a similar vein to quintessence, the solution to the $f(R)$ problem matches $\Lambda \mathrm{CDM}$ when $b=0$\footnote{In the actual implementation we use $b=10^{-13}$, because at $b=0$ there is a singularity.}. The reparametrization of the variables is then
\begin{equation}
\begin{split}
    \tilde{\boldsymbol{u}}\left(z, b, \Omega_{m,0}^{\Lambda} \right)=&\hat{\boldsymbol{u}}\left(z,\Omega_{m,0}^{\Lambda}\right)\\
    &+\left(1-e^{-\left(z-z_0\right)}\right)\left(1-e^{-b}\right)\\
    &\times\boldsymbol{u}_{\mathcal{N}}\left(z, b, \Omega_{m,0}^{\Lambda} \right),
    \label{perturbative_reparam_bundle_f_R}
\end{split}
\end{equation}
where $\boldsymbol{u}_{\mathcal{N}}=\left(x_{\mathcal{N}},y_{\mathcal{N}},v_{\mathcal{N}},\Omega_{\mathcal{N}},r_{\mathcal{N}}\right)$ \footnote{We remind the reader that we use $r'_{\cal N}=\ln r_{\cal N}$ in order that the variables are of order 1. See Sec.~\ref{methods}.} is the vector comprised of each of the individual outputs of the five ANNs that correspond to each dependent variable. Also, $\hat{\boldsymbol{u}}=\left(\hat{x},\hat{y},\hat{v},\hat{\Omega},\hat{r}\right)$ is the vector with the exact solutions of each variable when $b=0$.\footnote{These solutions are the same as those that were used for the initial conditions in \cref{f_R_ci}, but with the difference that $z$ is not fixed at $z_0$.}

If we observe Eqs.~\eqref{diff_f_R_2} and \eqref{eq:Gamma}, we see that there is a singularity in the differential equations at $b=0$. 
Here, the perturbative reparametrization allows us to get values of $b$ that are arbitrarily close to zero without the need for more computational time, which is a clear benefit of the ANN method. This is because, as the value of $b$ tends to zero, the first term in \cref{perturbative_reparam_bundle_f_R} becomes dominant, and the ANN only provides a minimal correction. In contrast, the numerical approach can demand more operations in those situations, which in turn implies more computational time. This problem became so important that a perturbative analytic method was developed in Ref.~\cite{Basilakos} to obtain an approximation of $H\left(z\right)$ for the Hu-Sawicki and Starobinsky models to be used to perform parameter inference. For example, in Refs.~\cite{Updated_Matias_paper,D'agostino_Nunes,Farrugia2021} the output of the numerical solvers in the region close to $b=0$ is replaced by the solution provided by the aforementioned method.

For this model, in addition to the modification of the reparametrization, we add three additional terms to the loss $L$ that involve relationships between the dependent variables of \cref{diff_f_R_2}. The first one is \cref{eq_1}, while the other two are the following:
\begin{subequations}
\label{consv}
\begin{align}
\label{consv_Om_1}
\Omega=&\dfrac{2y\Omega^\Lambda_{m,0}\left(1+z\right)^3\left(r+b\right)}{r\left(1-\Omega^\Lambda_{m,0}\right)\left(r+b-2\right)},\\
\label{consv_Om_2}
\Omega=&\dfrac{2v\Omega^\Lambda_{m,0}\left(1+z\right)^{3}\left(r+b\right)^{2}}{r\left(1-\Omega^\Lambda_{m,0}\right)\left[\left(r+b\right)^{2}-2b\right]}.
\end{align}
\end{subequations}
Both of these equations are related to the conservation of mass through the use of \cref{matter_sol} (the detailed calculation can be found in \cref{apx:consv_matter}). 
To impose these equations, the extra terms added to the loss are the squares of the relative difference of each equation's right-hand side and left-hand side. By adding these three extra terms we get the following expression:
\begin{equation}
\begin{split}
L_\mathcal{C}\left(\tilde{\boldsymbol{u}}, z, \boldsymbol{\theta}\right)=&\left(\tilde{\Omega}+\tilde{v}-\tilde{x}-\tilde{y}-1\right)^2\\
&+ \left\{\dfrac{2\tilde{y}\Omega^\Lambda_{m,0}\left(1+z\right)^3\left(\tilde{r}+b\right)}{\tilde{r}\tilde{\Omega}\left(1-\Omega^\Lambda_{m,0}\right)\left(\tilde{r}+b-2\right)}-1\right\}^2\\
&+ \left\{\dfrac{2\tilde{v}\Omega^\Lambda_{m,0}\left(1+z\right)^{3}\left(\tilde{r}+b\right)^{2}}{\tilde{r}\tilde{\Omega}\left(1-\Omega^\Lambda_{m,0}\right)\left[\left(\tilde{r}+b\right)^{2}-2b\right]}-1\right\}^2.
\end{split}
\label{consv_term}
\end{equation}
Denoting the part of the loss that only regards the residuals as $L_{\mathcal{R}}$, the final loss for this model reads
\begin{equation} 
L=L_{\mathcal{R}}+L_{\mathcal{C}} . 
\end{equation}
While \cref{consv} should be approximately obeyed by the ANN-based solutions without the extra term $L_\mathcal{C}$ in the loss, we find that explicitly imposing these relations through the extra term yields more accurate solutions.

\subsection{Accuracy of solutions}
The loss function is the quantity that reflects how well the ANNs are at solving a given differential system.
The ANN method employed in this work lacks the capability of estimating the uncertainty of its solutions.\footnote{A recent work \cite{graf2021uncertainty} showed how to estimate uncertainties in the general framework. The application to the cosmological context is left for future work.} Therefore, we choose to quantify the accuracy of the solutions provided by our trained ANNs as the absolute value of the relative difference between the ANN solution and i) the analytical solution in the case of the $\Lambda \mathrm{CDM}$ and CPL models or ii) the numerical solution in the case of quintessence and $f(R)$ models. We  stress  that this is the only time that the analytical or numerical solutions of the cosmological models are used in this work, i.e., these solutions are never used to train the ANNs or in the inference pipeline. It is important to note that this analysis is mainly done to give a more appropriate quantification of the errors of the solutions over the parameter space of interest of each model, which can not be inferred from the value of the loss
function. Nevertheless, it is still necessary to see if the errors shown here are small enough to not affect in any meaningful way the results of an inference pipeline such as the one we illustrated in \cref{fig:workflow}. In the case of our solutions, this situation is tested later on in Sec.~\ref{results} by comparing the results of the inference we perform using our cosmology-informed neural networks against ones found in the literature that use either similar or the same data to test the same models, but using numerical methods. 

\begin{figure}
    \centering
    \includegraphics[width=\columnwidth]{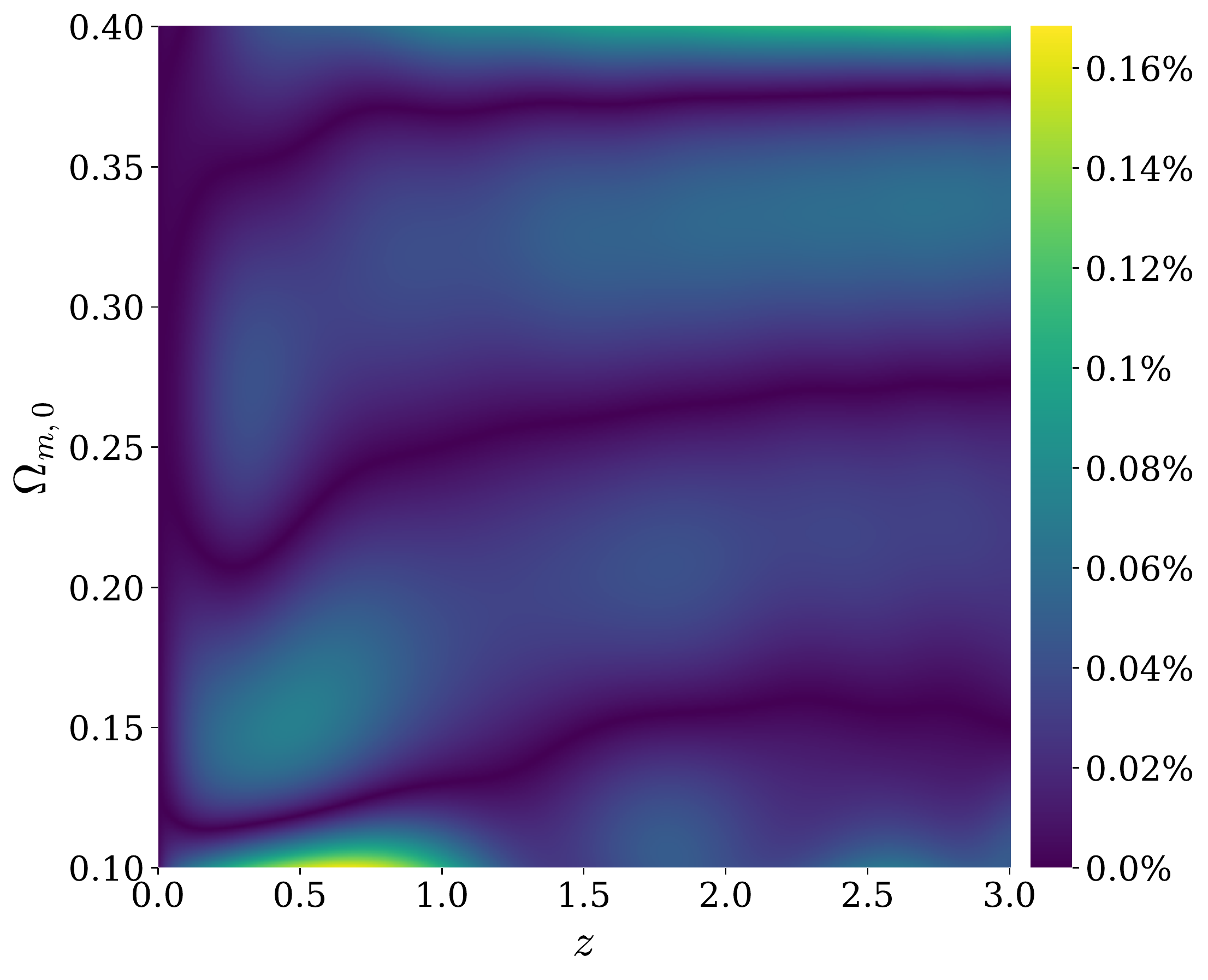}
    \caption{Percentage error of the neural network-based bundle solution to $\Lambda \mathrm{CDM}$ in its training range, through comparison to the analytical solution.}
    \label{percent_error_LCDM}
\end{figure}

Let us begin this analysis with the accuracy of the bundle solutions of \cref{x_matter}, which correspond to the $\Lambda \mathrm{CDM}$ model. 
\Cref{percent_error_LCDM} shows the percentage error
in the range of $z$ and $\Omega_{m,0}$ that the 
ANN was trained for. It follows that the maximum error in the whole training range is $\sim 0.16 \%$ and that the ANN solution is worse at the edges of its training range for the only parameter in the bundle ($\Omega_{m,0}$).
 
\begin{figure}
    \centering
    \includegraphics[width=\columnwidth]{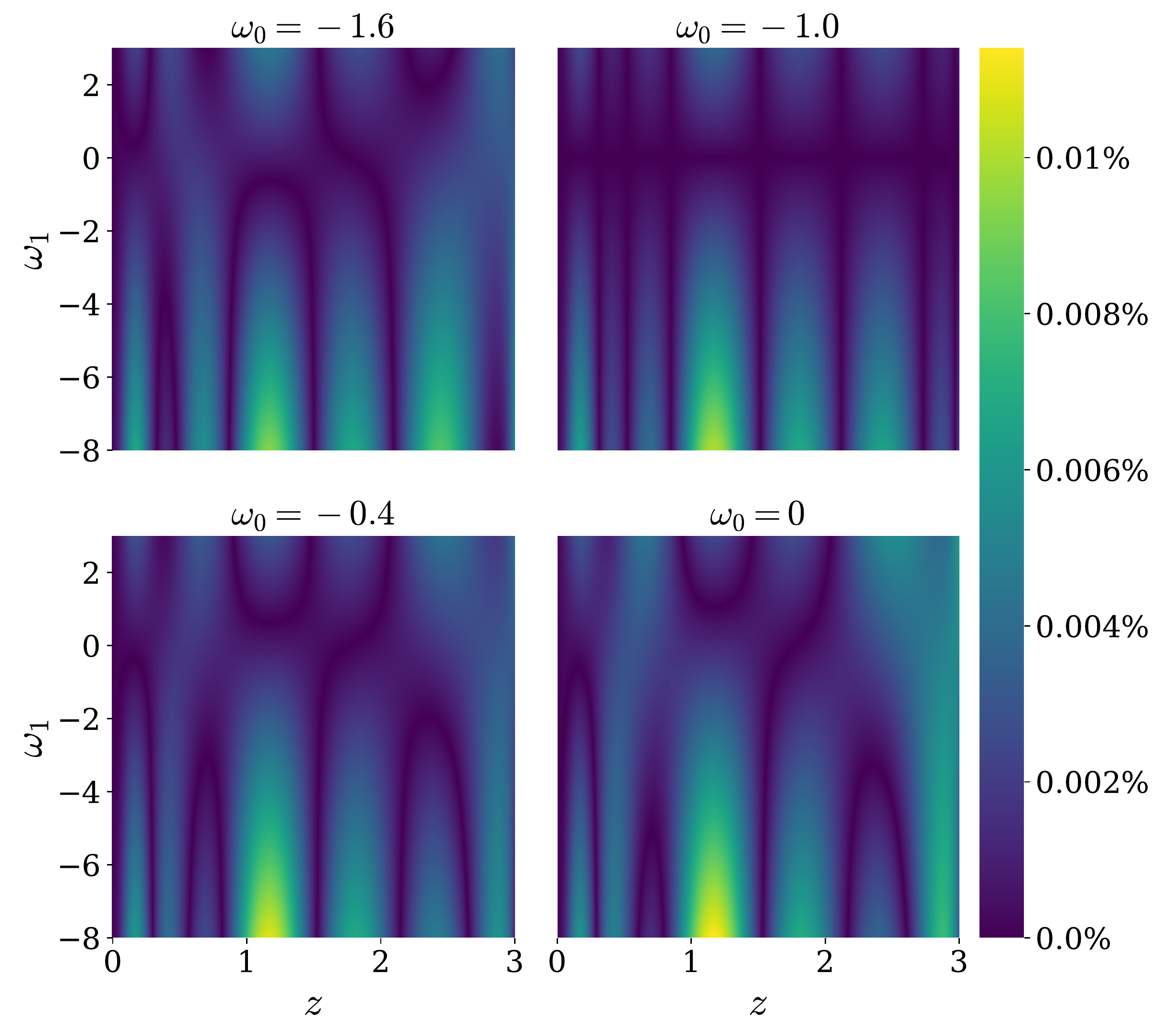}
    \caption{Percentage error of the neural network-based bundle solution to the CPL parametric dark energy model in its training range of $z$, a range of $\omega_1$ covering the range used in the statistical analysis, and four different values of $\omega_0$. The percentage error is calculated through comparison to the analytical solution.}
    \label{percent_err_CPL}
\end{figure}

Now we focus on the CPL parametric dark energy model. \Cref{percent_err_CPL} shows the percentage error for four different values of $\omega_0$ in a wide range of $\omega_1$ and the range of $z$ for which the solution was trained for. It follows from \cref{DE_sol} (analytical solution) and the reparametrization of the ANN solution [\cref{DE_reparam}] that the percentage error is independent of the initial condition. Besides, the maximum error is $\sim 0.01 \%$ which compared to the results obtained for the $\Lambda \mathrm{CDM}$ model, leads us to conclude that we have obtained a more accurate solution to a more general version of the problem. In particular, the plot corresponding to $\omega_0=0$ in \cref{percent_err_CPL} contains the solution to \cref{x_matter} at $\omega_1=0$. This improvement in accuracy is related to the method described in Sec.~\ref{integrating_factor}.

\begin{figure}
    \centering
    \includegraphics[width=\columnwidth]{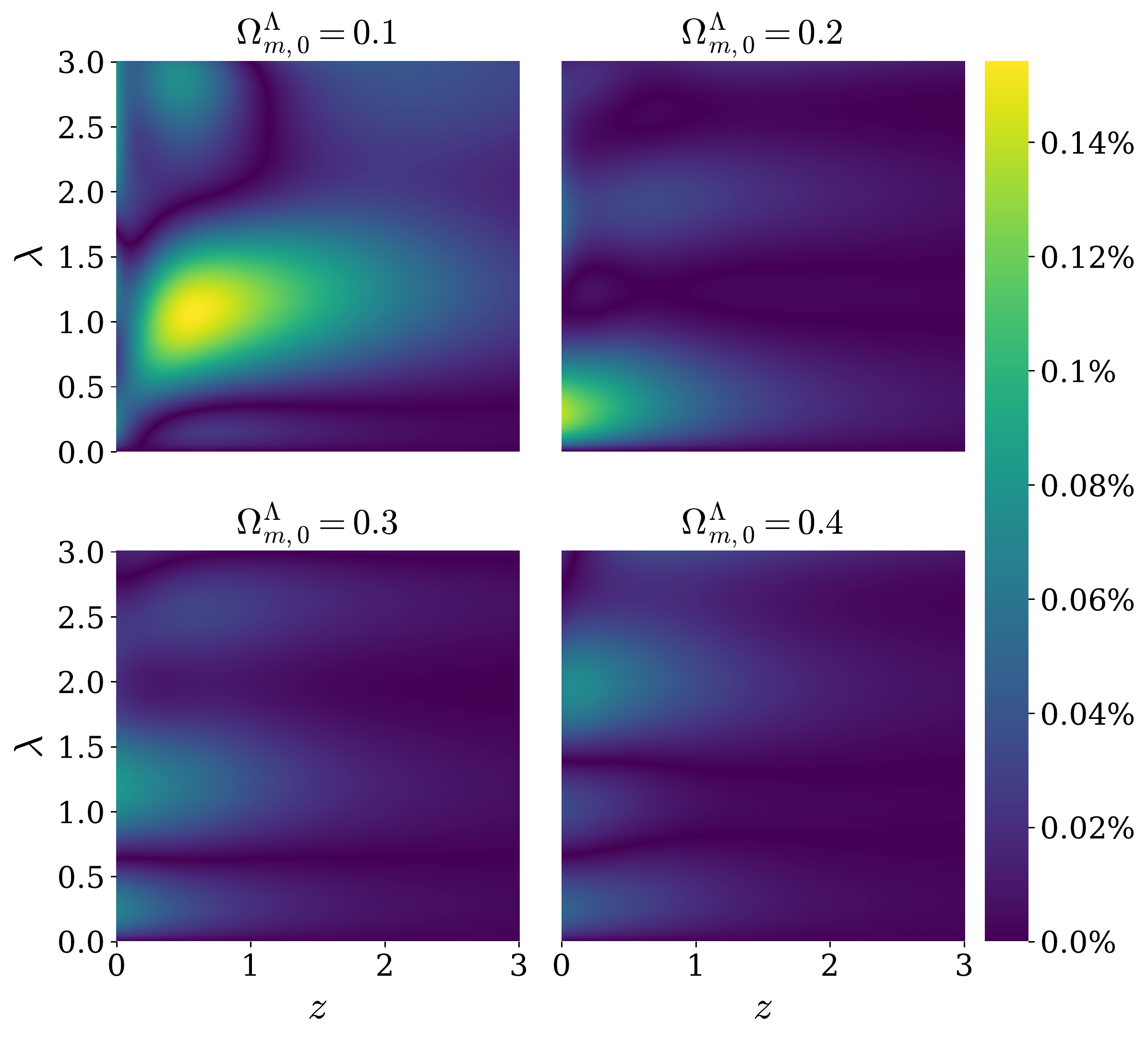}
    \caption{Percentage error of the neural network-based bundle solutions of $H\left(z\right)/H^{\Lambda}_0$ in the quintessence model, through comparison to numerical solutions. The range of the comparison goes through the training range of the parameters of the bundle.}
    \label{percent_err_quintessence}
\end{figure}

\Cref{percent_err_quintessence} shows the percentage error for the quintessence model for four different values of $\Omega^{\Lambda}_{m,0}$, and going through the whole training range of $\lambda$. In this case, since there is more than one dependent variable, the percentage error is calculated for $H\left(z\right)/H^{\Lambda}_0$. For the numerical solution, we use an adaptive explicit Runge-Kutta method of order 5(4), where the local estimated error $e$ for any dependent variable of the system $u^i$ can be bounded such that $\left|e\right|<a_{\rm tol} + r_{\rm tol}\left|u^i\right|$.\footnote{This is in fact a slight oversimplification. For a more detailed description we refer the reader to Ref.~\cite{supp_material}.} For the plot in \cref{percent_err_quintessence} we consider $r_{\rm tol}=10^{-10}$ and $a_{\rm tol}=10^{-13}$.

Although the ANNs are trained for $z\in\left[0,10\right]$, the plot is limited to the redshift range that is relevant for the data we use. Besides, in the region that is not shown in the plot, the value of the percentage error is lower or at most equal to the one in \cref{percent_err_quintessence} and it gets lower as $z$ gets closer to the initial condition $z_0$.

\begin{figure}
    \centering
    \includegraphics[width=\columnwidth]{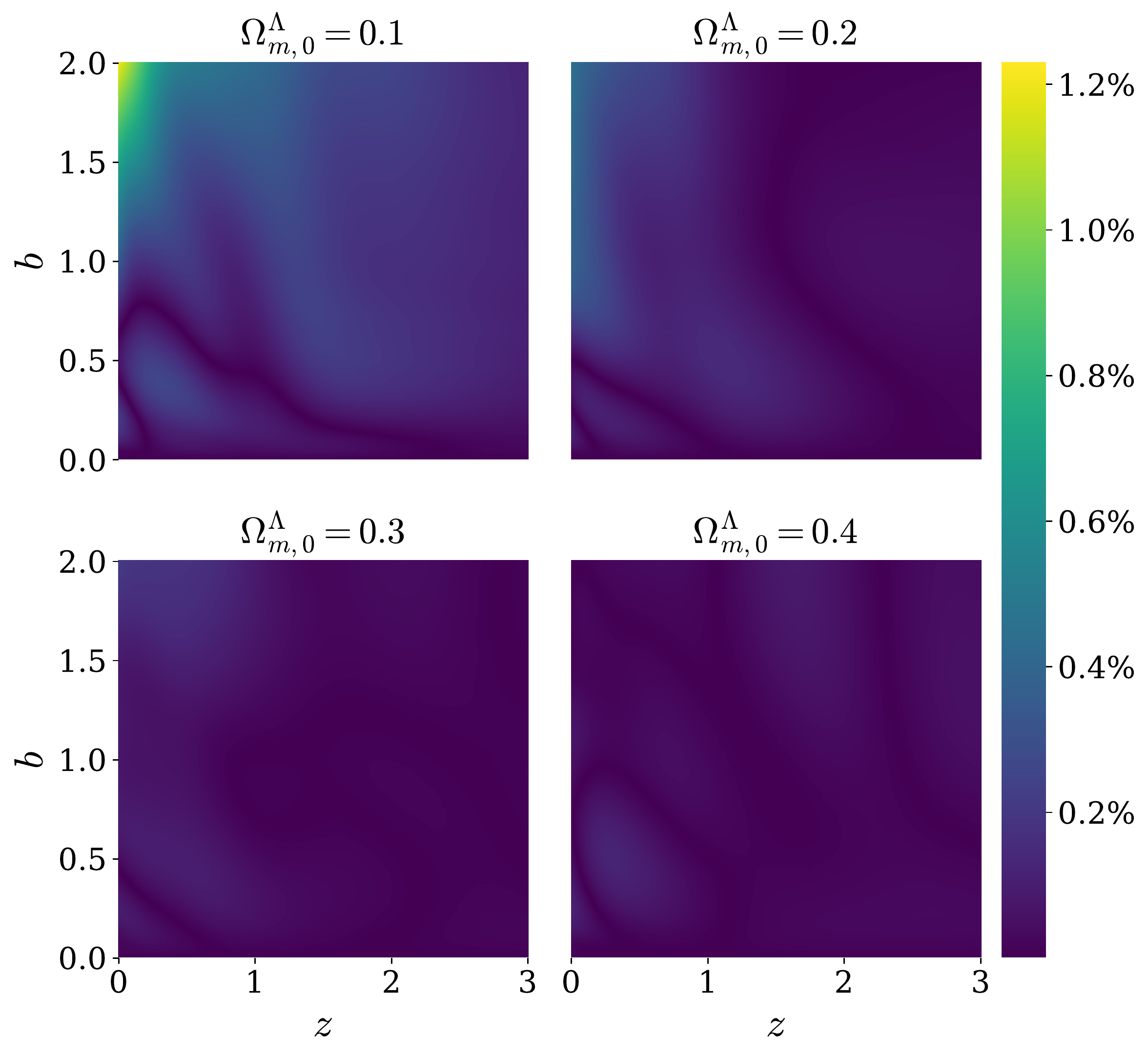}
    \caption{Percentage error of the neural network-based bundle solution of $H\left(z\right)/H^{\Lambda}_0$ in the Hu-Sawicki model, through comparison to numerical solutions. The range of the comparison goes through a section of the training range of the parameters of the bundle.}
    \label{percent_err_f_R}
\end{figure}

Finally, \cref{percent_err_f_R} shows the percentage error for $H\left(z\right)/H^{\Lambda}_0$ in the $f(R)$ Hu-Sawicki model; here the values of the error tolerance for the numerical solver are set as follows: $r_{\rm tol}=10^{-11}$ and $a_{\rm tol}=10^{-16}$. We also restrict the range of $b$ in the plot to the region that is more relevant for the cosmological inference.
However, it is important to report that there is a small region with an error $\sim 5\%$, which is located at $\left(b, \Omega^{\Lambda}_{m,0}\right)\in\left[2,5\right]\times0.1$ far away from the obtained $2 \sigma$ region of the parameter space obtained later on in Sec.~\ref{results} with both data sets (see \cref{contours:f_R}).

\section{Observational data}\label{obs_data}
In this section, we describe the different data sets that we use to do inference on the parameters of the models presented in Sec.~\ref{theo}, along with their associated likelihood functions.
\subsection{Cosmic chronometers}
\label{sect:CC}
The method of cosmic chronometers allows to obtain measurements of $H\left(z\right)$ at different values of $z$ through the measurement of the difference in age $\Delta t$ between passively evolving galaxies whose light is observed at different redshifts, which differ by $\Delta z$ \cite{CC_technique}.\footnote{We denote passively evolving galaxies as those galaxies with no current star formation or interaction with other galaxies.} Using the relation between the scale factor, $a$, and the cosmological redshift, $z$, we can obtain the equation
\begin{equation}
\label{dz_dt}
    H=\dfrac{\dot{a}}{a}=\dfrac{-1}{1+z}\dfrac{dz}{dt},
\end{equation}
from which it is possible to determine $H\left(z\right)$, if the following approximation holds: $\Delta z/\Delta t\simeq dz/dt$.
A key advantage of this method is that it relies on the measurement of relative ages $dt$, which has smaller systematic errors than the determination of absolute ages. Besides, the measurement of $dt$ depends only on atomic physics and therefore it is model independent.
We consider 30 measurements of $H\left(z\right)$ obtained with this technique \cite{CC1,CC2,CC3,CC4,CC5,CC6} within 
a range of redshift $z\in\left[0.07,1.965\right]$ which we summarize in \Cref{table_CC} of \Cref{datasets}. 
The likelihood function for this data set can be written as
\begin{equation}
{\cal L}_{\bf \rm CC} \propto \exp \left\{ -\dfrac{1}{2}\sum_{i=1}^{30} \left[\frac{H^{\rm obs}\left(z_i\right) - H^{\rm th}\left(z_i,\boldsymbol{p}\right)}{\sigma_{H^{\rm obs}\left(z_i\right)}}\right]^2 \right\},
\label{like_CC}
\end{equation}
where $H^{\rm obs}\left(z_i\right)$ refers to the observational measurements of $H\left(z\right)$ obtained with the CC method and $\sigma_{H^{\rm obs}\left(z_i\right)}$ refers 
to the corresponding error (see \cref{table_CC}). The model's prediction is $H^{\rm th}\left(z_i,\boldsymbol{p}\right)$ obtained with the ANN method where $\boldsymbol{p}$ is a vector that includes the free parameters
of the model. All the models share $\Omega_{m,0}^{\Lambda}$ and $H_0^{\Lambda}$ while the following extra parameters are added: 
$\omega_0$ and $\omega_1$ for the CPL model, $\lambda$ for the quintessence model and $b$ for the $f(R)$ Hu-Sawicki model; see Sec.~\ref{theo} for details.
\subsection{Type Ia supernovae}\label{sec:SNeIa}
Type Ia supernovae are powerful and extremely luminous stellar explosions. Due to the homogeneity of
their light curves and spectra, they are considered as standard candles, which allows to determine cosmological distances from their luminosities. In this work, we consider the 1048 measurements of supernovae luminosities comprised in the Pantheon compilation in the redshift range $z\in\left[0.01,2.3\right]$ \cite{SnIa_pantheon}. The observed quantity is the distance modulus $\mu$, which can be expressed by a modified version of the formula proposed by Tripp in Ref.~\cite{Tripp}:
\begin{equation}
    \label{mu_pantheon}
    \mu = m_b - M +\alpha x_1 -\beta c + \Delta_M + \Delta_B,
\end{equation}
where the color $c$, the light-curve shape parameter $x_1$ and the log of the overall flux normalization $m_b$ are provided by the compilation through light-curve fitting using SALT2 as presented in Ref.~\cite{SALT2} and implemented in SNANA~\cite{SNANA}. In addition, $\alpha$ is the coefficient of the relation
between luminosity and stretch, $\beta$ is the coefficient of the relation between luminosity
and color, and $M$ is the absolute B-band magnitude of a fiducial SNIa
with $x_1=0$ and $c=0$. The distance correction, $\Delta_B$, is based on predicted biases from simulations and $\Delta_M$ is a distance correction based on the host galaxy mass of the SNIa, which is determined by
\begin{equation}
    \Delta_M = \gamma\left[1+e^{\left(-\left(m-m_{\rm step}\right)/\tau\right)}\right],
\end{equation}
where $\gamma$ is a relative offset in luminosity, $m$ is the mass of the host galaxy, and $m_{\rm step}$ is the mass step for the split that separates the SNIa into ones that have a host galaxy with high mass and low mass. Last, $\tau$ is an exponential transition term in the Fermi function that describes the relative probability of masses being on one side or the other of the split. In this way, $\alpha$, $\beta$, $\gamma$ and $M$ are called nuisance parameters and they are usually estimated via a statistical analysis where a theoretical model for the dynamics of the Universe is assumed. However, as described before, the nature of these parameters is astrophysical and therefore their values should not depend on the cosmological model.

In this work, we use the mean values estimated by the Pantheon compilation \cite{SnIa_pantheon} ($\alpha=0.0154 \pm 0.06$, $\beta=3.02 \pm 0.06$, $\gamma=0.053\pm 0.009$) where a $\Lambda \mathrm{CDM}$ model is assumed. These values have been verified by two independent analyses with alternative cosmological models: the one in Ref.~\cite{Updated_Matias_paper} where the $f(R)$ Hu-Sawicki model was assumed and the one in Ref.~\cite{Negrelli_2020} performed in the context of the modified gravity theory. Moreover, $M$ will be considered as an additional free parameter for all models in the statistical analysis (meaning that it is included in $\boldsymbol{p}$).
To perform the statistical analysis, we compare the value of $\mu$ obtained from the data using \cref{mu_pantheon} against the theoretical value provided by
\begin{equation}
    \label{mu_theo}
    \mu\left(z\right) = 25 + 5\log_{10}\left(d_L\left(z\right)\right),
\end{equation}
where the luminosity distance $d_L\left(z\right)$ is related to $H\left(z\right)$ through
\begin{equation}
    d_L\left(z\right)=\left(1+z\right)\int^z_0\dfrac{dz^\prime}{H\left(z^\prime\right)}.
    \label{dl_theo}
\end{equation}
The likelihood function concerning this data set can be expressed as \cite{SnIa_pantheon}
\begin{equation}
{\cal L}_{\bf \rm SNIa} \propto \exp \left\{ -\dfrac{1}{2} \Delta \boldsymbol{\mu}^T \cdot\boldsymbol{C}^{-1}\cdot \Delta \boldsymbol{\mu}\right\},
\label{like_SN}
\end{equation}
where the vector $\Delta \boldsymbol{\mu} = \boldsymbol{\mu}^{\rm obs} - \boldsymbol{\mu}^{\rm th}\left(\boldsymbol{p}\right)$ contains the difference between the observed value and the theoretical prediction 
of the distance modulus [\cref{mu_theo,dl_theo}] for each of the 1048 measurements of the Pantheon compilation. Furthermore, the covariance matrix $\boldsymbol{C}$ is defined as follows:
\begin{equation}
\boldsymbol{C}=\boldsymbol{D}_{\rm stat} + \boldsymbol{C}_{\rm sys},
\end{equation}
where the statistical matrix $\boldsymbol{D}_{\rm stat}$ has only a diagonal component which includes measurement errors such as the photometric error of the SNIa distance, the uncertainty from the distance bias correction, the uncertainty from the peculiar velocity, the redshift measurement uncertainty (added in quadrature), and the uncertainty from stochastic gravitational lensing. Moreover, $\boldsymbol{C}_{\rm sys}$ is the systematic covariance matrix.
\subsection{Baryon acoustic oscillations}
\label{obs_bao}
Since the pioneering work in Refs.~\cite{BAO_theory_1, BAO_theory_2, BAO_theory_3} studying the connection between matter fluctuations and the anisotropies in the CMB, the physical mechanisms operating in the early Universe have now been well established. Before the recombination of hydrogen, baryons and photons interact through Thomson scattering, while baryons and dark matter interact gravitationally. The result is a photon-baryon fluid that suffers acoustic oscillations, thanks to the pull of gravity towards matter overdensities, and the push of photons due to the radiation pressure. When the temperature of the Universe becomes sufficiently low due to cosmic expansion, hydrogen atoms are formed, the density of free electrons dramatically drops, and Thomson scattering becomes inefficient. The subsequent decoupling between matter and radiation freezes the oscillations, leaving an imprint both in the CMB and in the distribution of galaxies thanks to the BAO mentioned above. The BAO offers a characteristic scale, namely, the sound horizon $r_d$ at recombination, which is the largest comoving distance a sound wave could have traveled before recombination, and it provides a very useful standard rod. Thus, we can use observations of galaxy clustering as a cosmological probe. The quantities related to the BAO data relevant for this paper are the Hubble parameter $H\left(z\right)$, the Hubble distance
\begin{equation}
    D_H\left(z\right)=\dfrac{1}{H\left(z\right)},
\end{equation}
the angular diameter distance
\begin{equation}
    D_A\left(z\right)=\dfrac{1}{1+z}\int^z_0 \dfrac{dz^\prime}{H\left(z^{\prime}\right)},
\end{equation}
the comoving angular diameter distance 
\begin{equation} 
    D_M\left(z\right)=\left(1+z\right)D_A\left(z\right),
\end{equation}
and the combination of $D_M$ and $D_H$, introduced in Ref.~\cite{D_V}
\begin{equation}
    D_V\left(z\right)=\left[D_A\left(z\right)^2\dfrac{z}{H\left(z\right)}\right]^{1/3}.
\end{equation}
To do inference on the BAO data that we use, we also need to calculate the sound horizon $r_d$. To do this, we follow the method employed in Ref.~\cite{r_s} which uses the fitting formula for the drag epoch redshift $z_d$ introduced in Ref.~\cite{z_drag_formula}. 
It is important to remark that the BAO data sets we use, which are listed in \Cref{table_BAO}, are independent. Thus, we define the likelihood function as follows:
\begin{equation}
{\cal L}_{\bf \rm BAO} \propto \exp \left\{ -\dfrac{1}{2}\sum_{i=1}^{20} \left[\frac{A_i^{\rm obs}\left(z_i\right) -
A_i^{\rm th}\left(z_i,\boldsymbol{p}\right)}{\sigma_{A_i^{\rm obs}\left(z_i\right)}}\right]^2 \right\},
\label{like_BAO}
\end{equation}
where the $A_i$ denotes the specific quantity that is measured (such as $H r_d$, $D_H/ r_d$, $D_A/ r_d$, and others; see the first column of \Cref{table_BAO}), $A_i^{\rm obs}\left(z_i\right)$ refers to the observed value and $\sigma_{A_i^{\rm obs}\left(z_i\right)}$ refers to the observational error.\footnote{For measurements with both statistical and systematic errors, we use the sum of both in quadrature.} Moreover, $A_i^{\rm th}\left(z_i,\boldsymbol{p}\right)$ indicates the model's prediction for the quantity $A_i$ at $z_i$.
\section{Results}\label{results}

In this section, we present the results of the statistical analyses performed with the data sets described in Sec.~\ref{obs_data}. We perform the statistical analyses using the likelihood functions through a Markov chain Monte Carlo (MCMC)
method. For this, we set the same flat priors for the parameters that all four models share: $H^\Lambda_{0}\in\left[50, 80\right]\mathrm{km}\;\mathrm{s}^{-1}\mathrm{Mpc}^{-1}$, $\Omega^\Lambda_{m,0}\in\left[0.1, 0.4\right]$ [for the CPL model with just CC and SNIa the prior is too restrictive for its parameter space as can be confirmed with the results of the statistical analysis (see \cref{table:DE}),\footnote{The first MCMC runs show that the obtained bounds on $\Omega_{m,0}$ matched the priors, therefore it was necessary to enlarge them.} so we use $\Omega^\Lambda_{m,0}\in\left[0, 0.6\right]$], and for the absolute magnitude $M\in\left[-22, -18\right]$ for all models. We recall that in both the $\Lambda \mathrm{CDM}$ and CPL models, $\Omega^\Lambda_{m,0}=\Omega_{m,0}$ and $H^\Lambda_{0}=H_{0}$ which is not necessarily the case in both the quintessence and $f(R)$ models. Therefore, for these models we carry out a post-processing of the chains obtained by MCMC, where for each chain we calculate $H_0=H\left(z=0\right)$ and use \cref{LCDMtoAlt} to obtain $\Omega_{m,0}$. For all models we report the values of $H_0$ and $\Omega_{m,0}$.
To implement MCMC, we use the emcee \cite{emcee} \textsc{python} library, which implements the samplers introduced in Ref.~\cite{Affine_invariance_sampler}. Also, to plot the contours displayed in this paper, we use the GetDist \cite{getdist} \textsc{python} library.
For each theoretical model considered in this paper we carry out a first analysis with only CC and SNIa data and a second analysis where the BAO data are added to the previous ones. In both cases the joint likelihood used is the product of all the individual likelihoods involved.

\begin{table*}
\centering
\begin{ruledtabular}
\begin{tabular}{lccccc}
Data                         &            & $\Omega_{m,0}$   & $H_{0}\left[\dfrac{\mathrm{km/s}}{\mathrm{Mpc}}\right]$ & $M$                  & ${\chi}_{\nu}^2$  \\ \hline
\multirow{3}{*}{CC+SNIa}     & $68\%$ C.L. & $[0.281, 0.322]$ & $[67.251, 70.959]$                                      & $[-19.434, -19.326]$ &                \\
                             & $95\%$ C.L. & $[0.26, 0.342]$ & $[65.331, 72.622]$                                      & $[-19.484, -19.271]$ &                \\
                             & Best fit   & 0.$301$            & $68.994$                                                  & $-19.38$              & $0.969$             \\ \hline
\multirow{3}{*}{CC+SNIa+BAO} & $68\%$ C.L. & $[0.284, 0.306]$ & $[67.861, 69.277]$                                      & $[-19.422, -19.347]$ &                \\
                             & $95\%$ C.L. & $[0.275, 0.317]$ & $[67.175, 69.944]$                                      & $[-19.444, -19.347]$ &                \\
                             & Best fit   & $0.296$            & $68.545$                                                  & $-19.396$              & $0.97$             \\
\end{tabular}
\end{ruledtabular}
\caption{Constraints on the parameters of the $\Lambda \mathrm{CDM}$ model. The table shows the $68\%$ and $95\%$ confidence intervals for each free parameter. The reduced $\chi^2_\nu$ is also shown for each analysis.}
\label{table:LCDM}
\end{table*}

Starting with $\Lambda \mathrm{CDM}$, we show the results of the statistical analysis for this model in \Cref{table:LCDM} and we observe that the constraints on the parameters are tighter when the BAO data are included. Furthermore, the obtained confidence intervals of the analysis with CC and SNIa are in agreement within $1\sigma$\footnote{This means that the $1\sigma$ confidence intervals obtained with one statistical analysis in a given work overlap with the $1\sigma$ confidence intervals calculated in another.} with the ones obtained in Ref.~\cite{D'agostino_Nunes}, where the same data sets were considered.

\begin{table*}
\begin{adjustbox}{center}
\begin{ruledtabular}
\begin{tabular}{lccccccc}
Data                         &            & $\omega_0$   & $\omega_1$       & $\Omega_{m,0}$   & $H_{0}\left[\dfrac{\mathrm{km/s}}{\mathrm{Mpc}}\right]$ & $M$                  & ${\chi}_{\nu}^2$  \\ \hline
\multirow{3}{*}{CC+SNIa}     & $68\%$ C.L. & $[-1.172, -0.822]$ & $[-2.089, 1.401]$     & $[0.296, 0.437]$ & $[66.36, 70.249]$                                      & $[-19.458, -19.342]$ &                \\
                             & $95\%$ C.L. & $[-1.363, -0.648]$ & $[-5.104, 1.757]$      & $[0.111, 0.47]$ & $[64.563, 72.216]$                                      & $[-19.514, -19.287]$ &                \\
                             & Best fit   & $-1.051$            & $-0.554$                & $0.343$            & $68.724$                                                  & $-19.39$              & $0.97$             \\ \hline
\multirow{3}{*}{CC+SNIa+BAO} & $68\%$ C.L. & $[-1.05, -0.889]$ & $[-0.295, 0.494]$ & $[0.279, 0.307]$ & $[66.219, 69.182]$                                      & $[-19.46, -19.37]$ &                \\
                             & $95\%$ C.L. & $[-1.12, -0.803]$ & $[-0.827, 0.734]$     & $[0.266, 0.32]$ & $[64.594, 70.444]$                                       & $[-19.509, -19.334]$ &                \\
                             & Best fit   & $-0.974$            & $0.088$            & $0.291$            & $67.441$                                                  & $-19.426$              & $0.971$             \\ 
\end{tabular}
\end{ruledtabular}
\end{adjustbox}
\caption{Constraints on the parameters of the CPL model. The table shows the $68\%$ and $95\%$ confidence intervals for each free parameter. The reduced $\chi^2_\nu$ is also shown for each analysis.}
\label{table:DE}
\end{table*}

Regarding the CPL model, the flat priors that we use for the parameters that are not shared with the other models are $\omega_{0}\in\left[-0.4, -1.6\right]$ and $\omega_{1}\in\left[-8, 3\right]$. Results of the statistical analysis for this model are shown in \Cref{table:DE}. As in the case of the $\Lambda \mathrm{CDM}$ model, we note that the constraints tighten when the BAO data are added, as this effect is more pronounced for the parameter $\omega_1$. Furthermore, the obtained confidence intervals are in agreement within $1\sigma$ with the ones obtained in Ref.~\cite{2021Univ....7..163M}, where only the CC data set were considered.
 
\begin{table*}
\centering
\begin{ruledtabular}
\begin{tabular}{lcccccc}
Data                         &          & $\lambda$              & $\Omega_{m,0}$   & $H_{0}\left[\dfrac{\mathrm{km/s}}{\mathrm{Mpc}}\right]$            & $M$    & ${\chi}_{\nu}^2$  \\ \hline
\multirow{3}{*}{CC+SNIa}     & $68\%$ C.L.   & $[0, 0.584]$   & $[0.27, 0.317]$ & $[67.097, 70.753]$ & $[-19.432, -19.324]$ &                \\
                             & $95\%$ C.L.   & $[0, 0.941]$   & $[0.245, 0.338]$ & $[65.382, 72.629]$  & $[-19.486, -19.273]$  &                \\
                             & Best fit & $0.008$              & $0.301$            & $68.995$             & $-19.38$              & $0.97$             \\ \hline
\multirow{3}{*}{CC+SNIa+BAO} & $68\%$ C.L.   & $[0.159, 0.744]$ & $[0.284, 0.306]$ & $[66.837, 68.856]$ & $[-19.442, -19.383]$ &                \\
                             & $95\%$ C.L.   & $[0, 0.893]$   & $[0.275, 0.317]$ & $[65.638, 69.677]$ & $[-19.474, -19.354]$ &                \\
                             & Best fit & $0.532$            & $0.295$            & $67.733$             & $-19.415$              & $0.97$             \\
\end{tabular}
\end{ruledtabular}
\caption{Constraints on the parameters of a quintessence model with an exponential potential. The table shows the $68\%$ and $95\%$ confidence intervals for each free parameter. The reduced $\chi^2_\nu$ is also shown for each analysis.}
\label{table:quintessence}
\end{table*}
\begin{figure}
    \centering
    \includegraphics[width=\columnwidth]{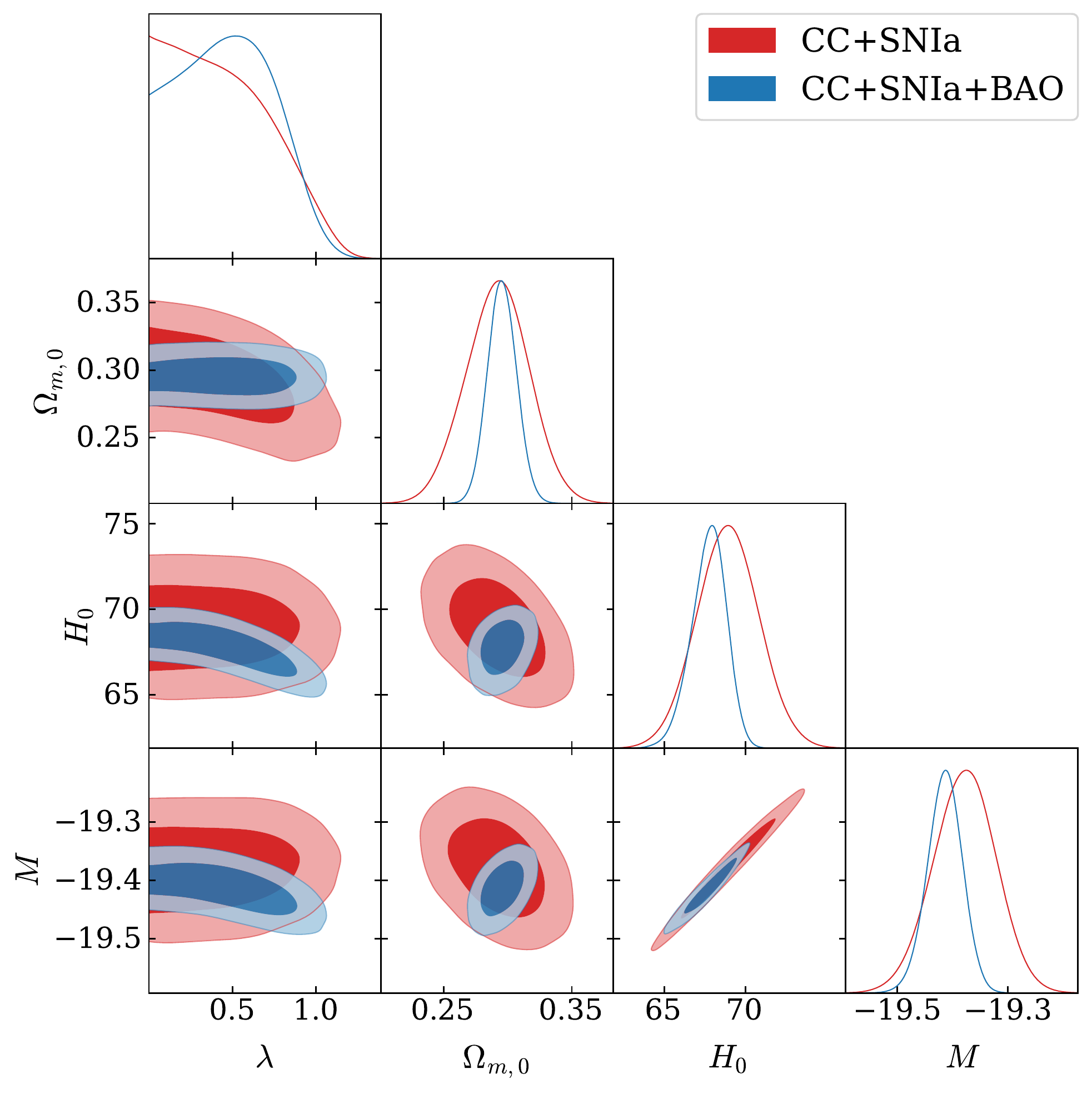}
    \caption{Results of the statistical analysis for the quintessence model with an exponential potential. The darker and brighter regions correspond to the 68\% and 95\% confidence regions, respectively. The plots on the diagonal show the posterior probability density for each of the free parameters of the model.}
    \label{contours:quintessence}
\end{figure}

\begin{table*}
\centering
\begin{ruledtabular}
\begin{tabular}{lcccc}
Data                                   &               & $\lambda$    & $\Omega_{m,0}$     & $H_0\left[\dfrac{\mathrm{km/s}}{\mathrm{Mpc}}\right]$  \\ 
\hline
SNIa+BAO+CMB+$H_0$                    & Akrami \textit{et al.} & $[0, 0.8]$    & \textit{N/A}              & \textit{N/A}                                                    \\
\multicolumn{1}{l}{CC+SNIa+BAO+$H_0$} & Tosone \textit{et al.} & $[0, 0.709]$  & $[0.257, 0.307$  & $[69.2, 73.9]$                                          \\
CC+SNIa+BAO                           & This work     & $[0, 0.941]$ & $[0.275, 0.317]$ & $[65.638, 69.677]$                                    
\end{tabular}
\end{ruledtabular}
\caption{Our results from the statistical analysis for the parameters of the quintessence model with an exponential potential, as well as similar analyses done with traditional numerical solvers. The table shows the $95\%$ confidence intervals for each free parameter.}
\label{table_comparison_quintessence}
\end{table*}

As regards the statistical analysis for the quintessence model considered in this work, we set a flat prior for $\lambda$, namely $\lambda\in\left[0,3\right]$. Results are shown in \Cref{table:quintessence} and \cref{contours:quintessence}. We note three changes in the results of the statistical analysis when the BAO data set is added: i) the constraints are tighter, ii) the correlation between $\Omega_{m,0}$ and $H_0$ changes sign, and iii) the model parameter $\lambda$ shows a small correlation with $H_0$, while its correlation with $\Omega_{m,0}$ noticeably decreases.
The comparison of our results with the ones obtained by other authors for the same model are summarized in \cref{table_comparison_quintessence}. In this table, we show the 95\% confidence intervals for each parameter instead of the 68\% ones because only the former were available in Ref.~\cite{2019ForPh..6700075A}.

The obtained 95 \% confidence interval for $\lambda$ is in agreement with the ones obtained in Ref.~\cite{2019ForPh..6700075A}. Besides, there is also agreement with the 95\% intervals of $\lambda$ and $\Omega_{m,0}$ reported in Ref.~\cite{2019PhRvD..99d3503T}, and a marginal agreement with the value of $H_0$. However, we note that the analysis performed in Ref.~\cite{2019PhRvD..99d3503T} included the current value of $H_0$ reported in Ref.~\cite{Riess_2018}, which affects the obtained results.

\begin{table*}
\centering
\begin{ruledtabular}
\begin{tabular}{lcccccc}
Data                         &          & $b$              & $\Omega_{m,0}$   & $H_{0}\left[\dfrac{\mathrm{km/s}}{\mathrm{Mpc}}\right]$            & $M$    & ${\chi}_{\nu}^2$  \\ \hline
\multirow{3}{*}{CC+SNIa}     & $68\%$ C.L.   & $[0, 0.649]$   & $[0.24, 0.302]$ & $[67.236, 70.834]$ & $[-19.425, -19.32]$ &                \\
                             & $95\%$ C.L.   & $[0, 1.313]$   & $[0.207, 0.324]$ & $[65.469, 72.547]$  & $[-19.474, -19.266]$  &                \\
                             & Best fit & $0$              & $0.301$            & $68.994$             & $-19.38$              & $0.97$             \\ \hline
\multirow{3}{*}{CC+SNIa+BAO} & $68\%$ C.L.   & $[0.031, 0.411]$ & $[0.28, 0.302]$ & $[65.897, 68.289]$ & $[-19.469, -19.399]$ &                \\
                             & $95\%$ C.L.   & $[0, 0.708]$   & $[0.27, 0.314]$ & $[64.611, 69.235]$ & $[-19.505, -19.369]$ &                \\
                             & Best fit & $0.218$            & $0.293$            & $67.349$             & $-19.426$              & $0.97$             \\
\end{tabular}
\end{ruledtabular}
\caption{Constraints on the parameters of a cosmological model in Hu-Sawicki $f(R)$ gravity with $n=1$. The table shows the $68\%$ and $95\%$ confidence intervals for each free parameter. The reduced $\chi^2_\nu$ is also shown for each analysis.}
\label{table:f_R}
\end{table*}
\begin{figure}
    \centering
    \includegraphics[width=\columnwidth]{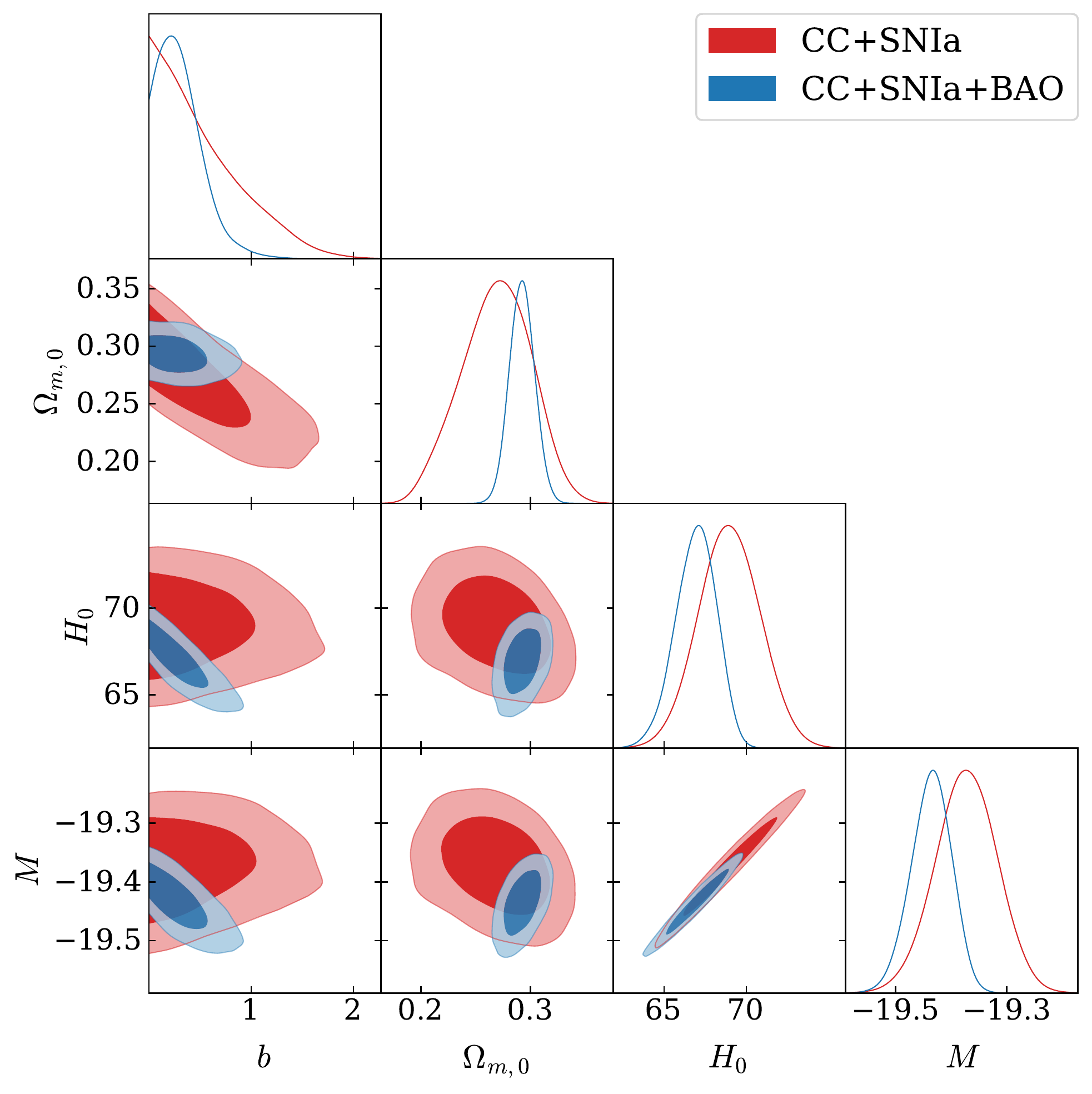}
    \caption{Results of the statistical analysis for the $f(R)$ Hu-Sawicki model. The darker and brighter regions correspond to the 68\% and 95\% confidence regions respectively. The plots on the diagonal show the posterior probability density for each of the free parameters of the model.}
    \label{contours:f_R}
\end{figure}

Concerning the statistical analysis of the $f(R)$ Hu-Sawicki model, we set a flat prior for $b$, namely $b\in\left[0,5\right]$ and show the results in \Cref{table:f_R} and \cref{contours:f_R}.
Similar to the analysis performed for the quintessence model, we note the following effects when the BAO data set is added to the statistical analysis: i) the constraints are tighter, ii) the correlation between $\Omega_{m,0}$ and $H_0$ changes sign and iii) the parameter $b$ shows an important correlation with $H_0$, while its correlation with $\Omega_{m,0}$ noticeably decreases.

\begin{table*}
\centering
\begin{ruledtabular}
\begin{tabular}{lcccc}
Data                          &                                                   & $b$              & $\Omega_{m,0}$     & $H_0\left[\dfrac{\mathrm{km/s}}{\mathrm{Mpc}}\right]$  \\ 
\hline
\multirow{3}{*}{CC+SNIa}     & Leizerovich \textit{et al.}                             & $[0, 0.587]$     & $[0.241, 0.303]$ & $[67.296, 70.701]$                                     \\
                              & D’Agostino \textit{et al.}                      & $[0.07, 0.49]$    & $[0.261, 0.314]$ & $[67.5, 71.5]$                                         \\
                              & This work                                         & $[0, 0.649]$     & $[0.24, 0.302]$  & $[67.236, 70.834]$                                     \\ 
\hline
CC+SNIa+H0LiCOW              & \multicolumn{1}{l}{D’Agostino \textit{et al.}} & $[0.03, 0.29]$    & $[0.244, 0.29]$   & $[71, 73.8]$                                            \\ 
\hline
\multirow{2}{*}{CC+SNIa+BAO} & Leizerovich \textit{et al.}                             & $[0.024, 0.382]$ & $[0.281, 0.304]$ & $[65.874, 68.313]$                                     \\
                              & This work                                         & $[0.031, 0.411]$ & $[0.28, 0.302]$  & $[65.897, 68.289]$                                     \\ 
\hline
CC+SNIa+BAO+RSD+CMB          & Farrugia \textit{et al.}                                   & $[0, 0.00005]$    & $[0.295, 0.307]$  & $[68.043, 69.007]$                                     
\end{tabular}
\end{ruledtabular}
\caption{Our results from the statistical analysis for the parameters of a cosmological model in Hu-Sawicki $f(R)$ gravity with $n=1$, as well as similar analyses done with traditional numerical solvers. The table shows the $68\%$ confidence intervals for each free parameter.}
\label{table:f_R_comparison}
\end{table*}

Let us now compare our results with similar ones published in the literature, which are summarized in \cref{table:f_R_comparison} (a similar comparison was shown in Fig.~3 of Ref.~\cite{Updated_Matias_paper}). We start with the ones performed in Refs.~\cite{D'agostino_Nunes,Updated_Matias_paper} where the same data sets of CC and SNIa were considered. First, we note that the estimates in Ref.~\cite{D'agostino_Nunes} for the $b$ parameter were smaller than the ones obtained in this work, while an excellent agreement is found with the results of Ref.~\cite{Updated_Matias_paper}. The reason for this is the following: due to the computational instabilities of \cref{diff_f_R_2} in the region $b \rightarrow 0$, in Ref.~\cite{D'agostino_Nunes} the expression for $H\left(z\right)$ was computed from a series expansion proposed in Ref.~\cite{Basilakos}. This approach allowed the authors to explore only a small region of the parameter space while there are no computational instabilities in the ANN method, and thus a wider region can be explored. On the other hand, in Ref.~\cite{Updated_Matias_paper}, the series expansion was only used for $b<0.15$ and a numerical integration was performed for other values of $b$ allowing for a complete exploration of the parameter space. We stress that the excellent agreement between the results of this work and the ones in Ref.~\cite{Updated_Matias_paper} where the very same data sets were considered, which shows that the errors of the ANN solutions are not significant
enough to affect the results of the statistical analysis.
Furthermore, we also note that our estimates for the $b$ parameter are in agreement within $1\sigma$ with the ones of another analysis performed in Ref.~\cite{D'agostino_Nunes}, where data from six systems of strongly lensed quasars analyzed by the H0LiCOW Collaboration \cite{HOLICOW2019} were added to the previous ones (CC+SNIa). On the other hand, another recent work \cite{Farrugia2021} also tested the $f(R)$ Hu-Sawicki model considering besides the CC and SNIa data, redshift-space distortions (RSD), BAO and CMB data. We note that the resulting values of the $b$ parameter were more restricted than in this work, but this can be explained since the former analysis considered more data sets than ours.

\begin{figure}
    \centering
    \includegraphics[width=\columnwidth]{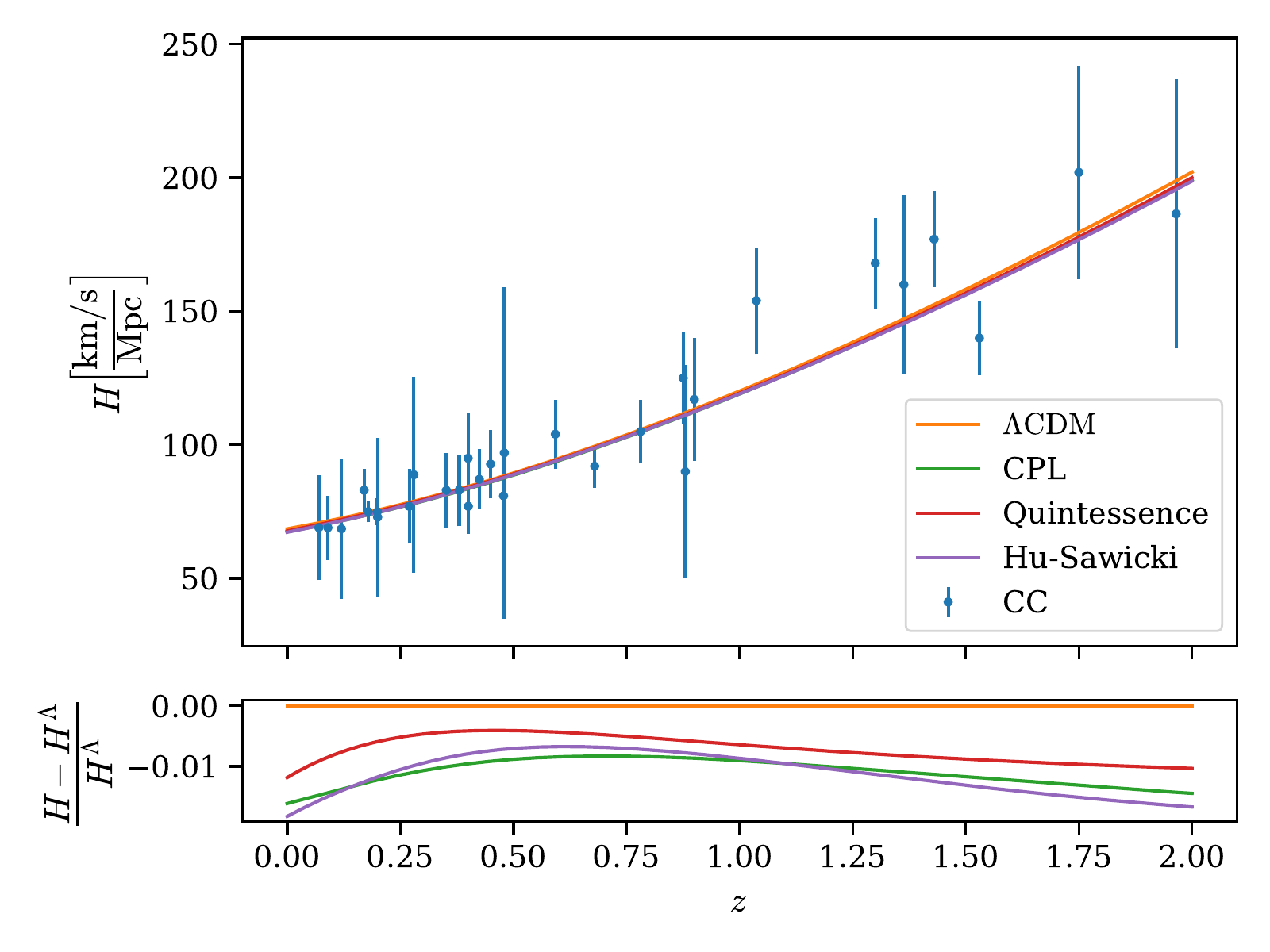}
    \caption{Top: $H\left(z\right)$ as a function of $z$ for all the models studied (using the parameters corresponding to the best fit obtained from the combination
CC+SNIa+BAO) together with CC data.
    Bottom: relative difference between the values of $H$ in the top panel for each model with respect to $\Lambda \mathrm{CDM}$.}
    \label{CC_plot}
\end{figure}

\begin{figure}
    \centering
    \includegraphics[width=\columnwidth]{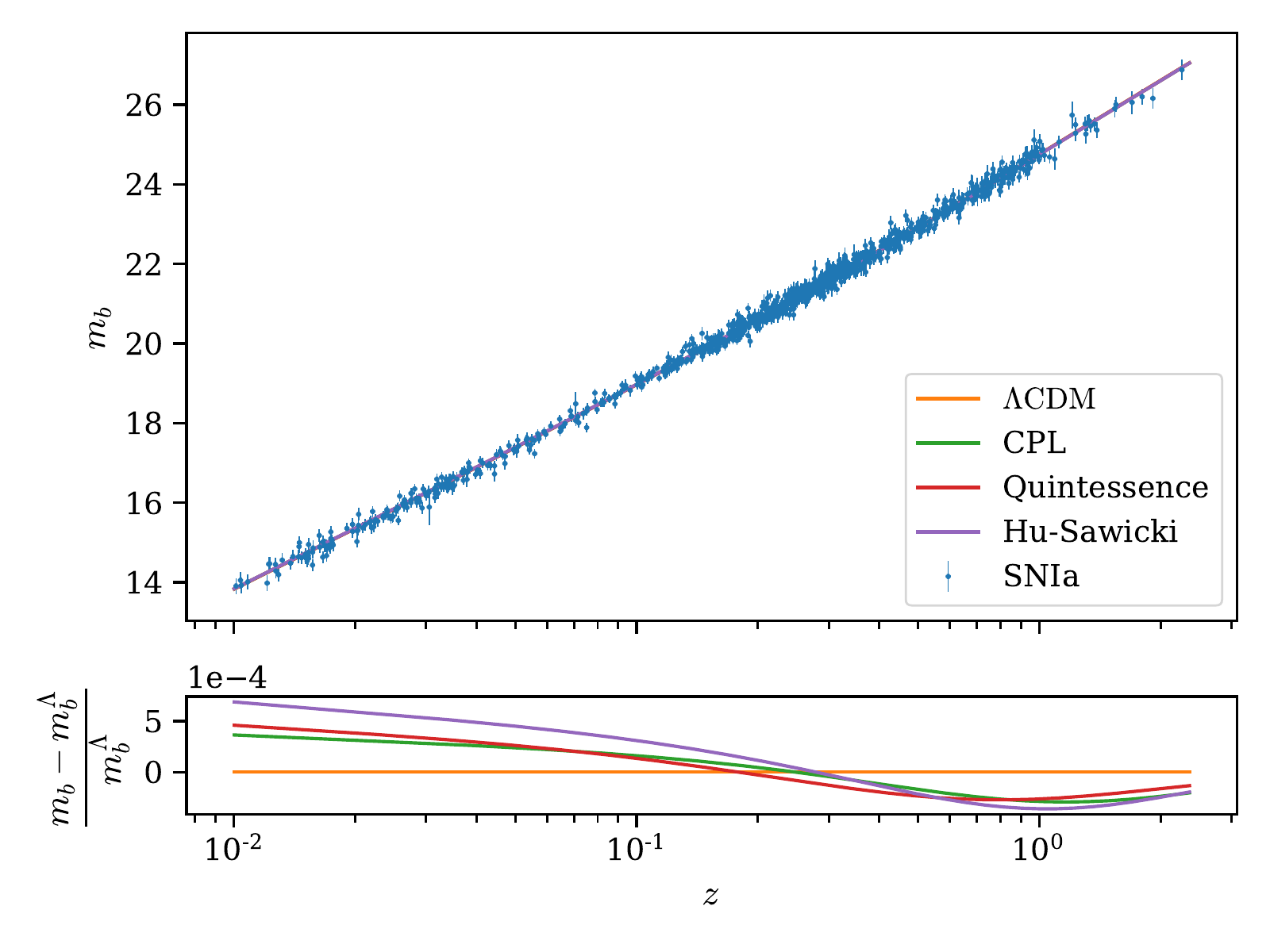}
    \caption{Top: apparent magnitude of a supernova, $m_b$, as a function of $z$ for all the models studied (using the parameters corresponding to the best fit obtained from the combination
CC+SNIa+BAO) together with SNIa data. The relation between $m_b$ and the Hubble parameter can be seen by looking at \cref{mu_pantheon,mu_theo,dl_theo}. Bottom: relative difference between the values of $m_b$ in the top panel for each model with respect to $\Lambda \mathrm{CDM}$.}
    \label{SnIa_plot}
\end{figure}

Finally, \cref{CC_plot,SnIa_plot} show the behavior of $H\left(z\right)$ and the SNIa apparent magnitude $m_b$ [related to $d_L\left(z\right)$] respectively for each of the cosmological models considered in this paper. For this,
we set the values of the free parameters at the best-fit values obtained by 
cosmological inference with CC+SNIa+BAO. 
The relative difference of each model with respect to $\Lambda \mathrm{CDM}$ is also shown in \cref{CC_plot,SnIa_plot}, and it allows us to conclude that the behavior of all these models concerning the CC and SNIa data is very similar.
\section{Computational efficiency}\label{comp_eff}

To discuss the topic of computational efficiency we perform an analysis on the amount of floating point operations (FLOPs) that it takes for a specific numerical method of common use [Runge-Kutta 5(4) with interpolation \cite{RK5(4)_formulae, RK_interpolation}] to solve the differential equations of a given initial value problem, as well as the equivalent for the ANN approach. In this section we only summarize the results of such an analysis, while the detailed calculation and discussion can be found in \cref{flops}.

Our analysis shows that if the number of points of the domain where the solution is evaluated is large or the problem is easy to integrate, as in the case of the quintessence model, the numerical solver is likely to outperform the ANNs in terms of computational cost, whereas in the case where the opposite is true (i.e few points to evaluate or a problem that is difficult to integrate), ANNs are most likely preferred. This is the case for the $f(R)$ Hu-Sawicki model, where the problem is rather more difficult to solve than in the quintessence model. This is confirmed in \cref{table:times} where the computational times for the MCMC runs performed for the numerical and ANN methods with different hardware are detailed. The main factor that makes the ANN method preferable in the context of problems that are hard to solve is the fact that once the ANNs are trained, the solutions can be used indefinitely. This is due to the process of solving the model’s differential equations being performed only during the training of the network, whereas in the numerical method, the integration
is done each time one needs to know an output of the
model for a given parameter combination.
This feature can be very important when using the obtained solution to perform statistical analyses like with MCMC. In the case of the $f(R)$ Hu-Sawicki model this is further amplified by the fact that, as we have discussed in Sec.~\ref{f_R_loss}, the ANNs require no additional computing time when there is a singularity in the differential system and an analytical solution at the singularity is available, which is not the case with traditional numerical solvers. Besides, we stress that we have left the optimization of the ANN method for future work, while the numerical method has been already tested and optimized. In fact, the use of parallelization can make the cost of both training and parameter inference much more efficient in terms of actual time spent executing the computations, thus making the scenario discussed before more favorable to ANNs.
\begin{table}
\begin{adjustbox}{center}
\begin{ruledtabular}
\begin{tabular}{lcc}
Method (hardware)                         &      $f(R)$      & Quintessence   \\ \hline 
Numerical method (CPU)  & 328 min & 120 min \\ 
ANN method (CPU) & 162 min & 375 min\\ 
ANN method (GPU) & 88 min & 240 min\\ 
\end{tabular}
\end{ruledtabular}
\end{adjustbox}
\caption{MCMC running times in minutes for different hardware and models using the CC+SNIa data set. For the CPU case, we use a ninth-generation Intel i7  (8 cores), while in the GPU case, we use a single Nvidia A100.}
\label{table:times}
\end{table}

\section{Conclusions}\label{conclusions}
In this work, we trained artificial neural networks through an unsupervised method to serve as bundle solutions to the differential equations of the background dynamics of the Universe for four different cosmological models. We then performed a statistical analysis using these solutions with observational data to draw conclusions on the possible values of each model's parameters. The novelty in this paper relies on the unsupervised nature of the training of the ANNs. This is in stark contrast to applications of neural networks in cosmology such as in Ref.~\cite{Cosmopower}, where the nature of training was supervised due to the use of solutions provided by numerical methods as training data.

We found our results from the statistical analysis to be compatible with the ones in the literature that used traditional numerical solvers. In addition, we provided estimations of the errors of our ANN-based solutions that show all errors to be either below or around $\sim 1 \%$ for the $95 \%$ confidence level region of the parameter space of all models for all the statistical analyses performed in this work.

Among the main achievements of this paper we stress the development of improvements to the ANN bundle method. One of these improvements was instrumental to making the method viable in the cosmological context by providing it with the capability to solve a special case of stiff\footnote{Here we use the term stiff in a broad sense to refer to problems where numerical solvers require extremely low step sizes to solve them.} problems without additional computational cost. We used this implementation in one of the models we solved, namely the differential equations of a cosmology in the $f(R)$ Hu-Sawicki gravity model, which has been shown to require significant computation time with numerical solvers for a specific region of its parameter space where the equations become stiff due to the presence of a singularity.

We showed in this paper that unsupervised ANN bundle solutions of differential equations can be effectively used to constrain physical models, and pose a viable alternative to traditional numerical methods. We also shared our codes to facilitate the implementation for future users in the following repository: \href{https://github.com/at-chantada/cosmo-nets}{https://github.com/at-chantada/cosmo-nets}.

We also performed an analysis of the computational efficiency of the ANN-based method against a common use numerical solver. We showed (see Sec.~\ref{comp_eff}) that once the training of the ANNs is concluded, the time required to perform parameter inference with MCMC is less than in the case of using traditional numerical methods when the differential system is not easy to integrate. In addition, the parallelization capabilities of ANNs can be exploited to make the use of these solutions even more computationally efficient. Nevertheless, we stress that the optimization of the ANN method to reduce computing times was not the main focus of this paper, and thus is left for future work.

While we showed an effective use of the ANN method, there are still many aspects of the method that can be improved which constitutes future work. Some of these future improvements include uncertainty quantification of the solutions provided by the method, model selection, a general tuning of the architecture (e.g., finding optimal activation functions, numbers of hidden layers and units, etc.), and improvements in the design of the training so as to achieve lower values for the loss, while also lowering the computational cost of training, namely the amount of iterations and batch sizes needed to obtain an optimal solution.

In summary, we used ANNs as the tool of choice to solve the differential equations of cosmological models, and used the solutions they provided to perform MCMC to constrain the models' parameters using data, without the use of numerical solvers in any step of the process.

Finally, our results show that ANNs provide a promising alternative to traditional numerical solvers for use in an inference process.

\begin{acknowledgments}
We thank Matías Leizerovich and Joy Parikh for valuable comments and discussions. The computations performed in this paper, which used GPUs, were run on the FASRC Cannon cluster supported by the FAS Division of Science Research Computing Group at Harvard University. S.L. is supported by grant PIP 11220200100729CO CONICET and grant 20020170100129BA UBACYT. C.G.S. is supported by grant PIP 11220200102876CO CONICET, and grants G157 and G175 from UNLP. C. G. was funded by NASA contract NAS8-03060 to the CXC and thanks the Director, Pat Slane, for advice and support.
\end{acknowledgments}

\appendix
\section{Implementation of the ANN method in specific models }
\label{app_1}
\subsection{Implementation in the \texorpdfstring{$\boldsymbol{\Lambda \mathrm{CDM}}$}{LambdaCDM} model} 
\label{ex_lcdm}
To solve the background equation of the $\Lambda\mathrm{CDM}$ model [\cref{x_matter}], using the bundle solution, one starts by assigning an ANN to each dependent variable. As described before, the inputs of the ANN are the independent variable, and the parameters of the bundle. Therefore, in this case, we have a single ANN whose output we label $x_{\mathcal{N}}\left(z,\Omega_{m,0}\right)$. Then, to enforce the initial condition of the system, the reparametrization in \cref{default_reparam_bundle} is implemented as follows:

\begin{equation}
    \tilde{x}\left(z, \Omega_{m,0}\right) = \Omega_{m,0}+\left(1-e^{-z}\right)x_\mathcal{N}\left(z, \Omega_{m,0}\right).
    \label{reparam_LCDM_example}
\end{equation}
We recall that $x=\kappa \rho_m(z)/3 H_0^2$, so $\tilde{x}$ in \cref{reparam_LCDM_example} satisfies the initial condition in \cref{x_matter}.
Next, the loss is constructed from the residual, $\mathcal{R}$, which is defined by the differential equation of the system,
\begin{equation}
    \mathcal{R}\left(\tilde{x}, z,\Omega_{m,0}\right)=\dfrac{d\tilde{x}}{dz}-\dfrac{3\tilde{x}}{1+z},
\end{equation}
which can be used in \cref{loss_res_reparam_bundle} to get the loss of the problem:
\begin{equation}
    L\left(\tilde{x},z,\Omega_{m,0}\right)=\left(\dfrac{d\tilde{x}}{dz}-\dfrac{3\tilde{x}}{1+z}\right)^2.
    \label{loss_LCDM_reparam}
\end{equation}
As the loss is defined as a function of $\tilde x$, this allows that the process of  training leads to a solution that meets the initial condition.

\begin{figure}
    \centering
    \includegraphics[width=\columnwidth]{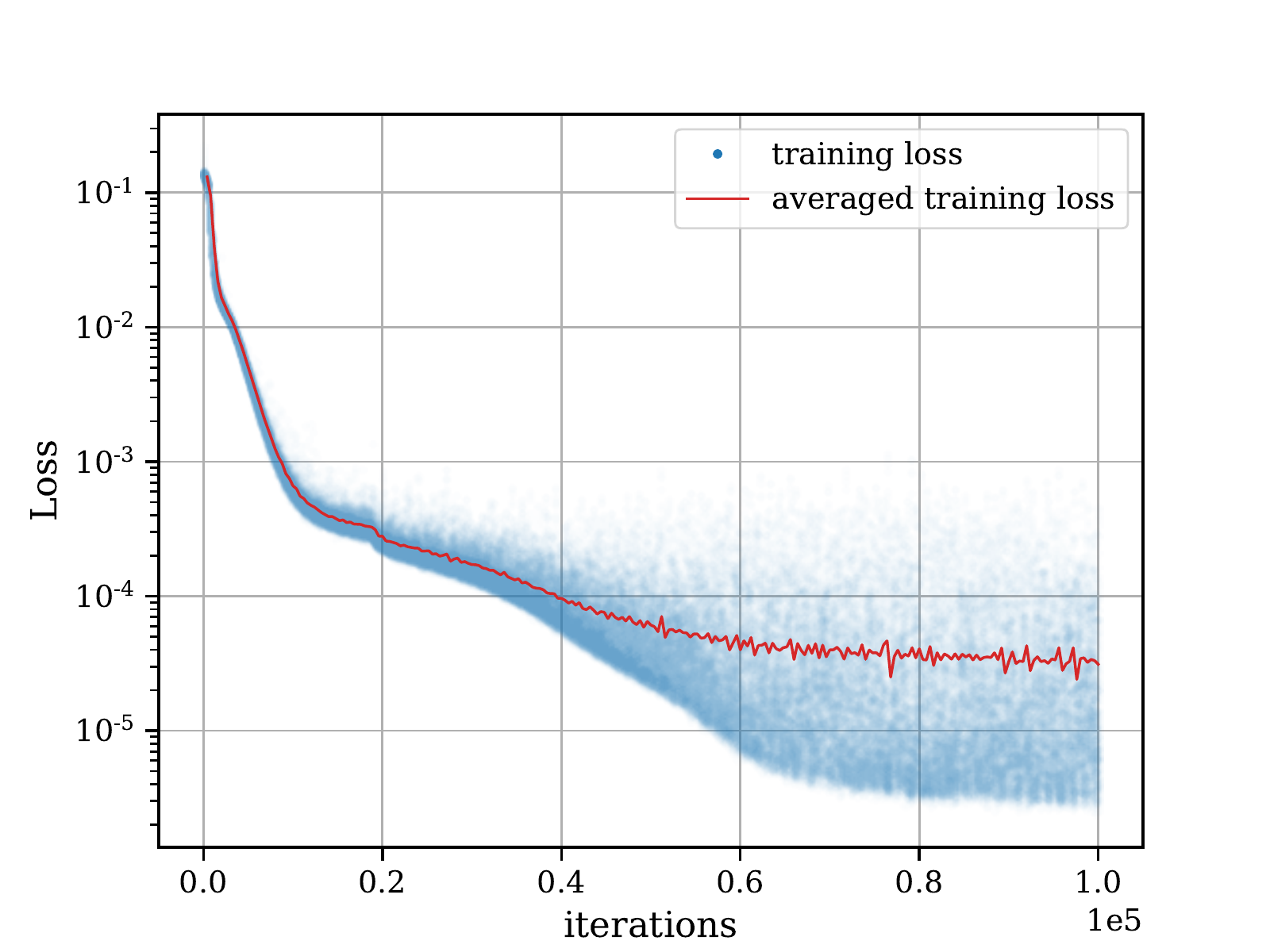}
    \caption{Example of a plot of the total loss per iteration of training. In blue are the values of the total loss at each iteration, and in red the average value of the total loss per $400$ iterations. This particular example comes from training an ANN to solve $\Lambda \mathrm{CDM}$ [\cref{x_matter}], defining the reparametrization in \cref{reparam_LCDM_example} and the loss in \cref{loss_LCDM_reparam}. }
    \label{loss_figure_LCDM}
\end{figure}
The training proceeds as follows. For a fixed value of the network's internal parameters [here we refer to the weights and biases detailed in \cref{uN_output}], $\tilde{x}\left(z, \Omega_{m,0}\right)$ is obtained and the respective loss [\cref{loss_LCDM_reparam}] is evaluated. Next, the training proceeds as an optimization problem performed with the gradient descent method where the network's internal parameters vary in order to minimize the loss function. 
\Cref{loss_figure_LCDM} shows one of the main indicators in the training of the ANN for the $\Lambda\mathrm{CDM}$ model, namely, the total loss per iteration.
Further details on how to train the ANNs were described in Ref.~\cite{ANN_diff_eqs}, and in Ref.~\cite{outdated_neurodiff_ref} on NeuroDiffEq's specific implementation.

The main takeaway from \cref{loss_figure_LCDM}, is that the value of the total loss decreases rapidly at the start of the training, and as the training continues this behavior starts to slow down. This is often the expected behavior of the loss during training. Therefore, the usual approach is to stop training when the relative change  in  the average value of the total loss per 400 iterations is of order $10^{-4}$.

\subsection{Implementation in the CPL model}
\label{ex_cpl}

To solve the background equation of the CPL dark energy model, we can take advantage of the reparametrization introduced in Sec.~\ref{integrating_factor}. We can identify $f(t,\boldsymbol{\theta})$ in \cref{bundle_linear} as $f(z,\omega_0,\omega_1)$ in the Friedmann equation of the CPL model [\cref{x_DE}]:
\begin{equation}
      f(z,\omega_0,\omega_1)=3(1+\omega_0)\dfrac{1}{1+z} + 3 \omega_1 \dfrac{z}{(1+z)^2}.
\end{equation}

Following \cref{bundle_reparam_linear}, the reparametrization $\tilde{x}$ for the outputs of the ANN in this model is

\begin{widetext}

\begin{equation}
\label{DE_reparam}
    \tilde{x}\left(z, \omega_0, \omega_1, \Omega_{m,0}\right)=\left(1 - \Omega_{m,0}\right)\exp\left\{3\left(1+\omega_0\right)\left[x_{\mathcal{N},1}\left(z\right) - x_{\mathcal{N},1}\left(z_0\right)\right] + 3\omega_1\left[x_{\mathcal{N},2}\left(z\right) - x_{\mathcal{N},2}\left(z_0\right)\right]\right\},
\end{equation}
\end{widetext}

where $x_{\mathcal{N},1}$ and $x_{\mathcal{N},2}$ represent the two outputs of the ANN and the respective loss function can be defined as follows:

\begin{equation}
  L (\tilde x,z,\omega_0,\omega_1)=\left[\dfrac{d\tilde{x}}{dz}-\dfrac{3\tilde{x}}{1+z}\left(1+\omega_0 + \dfrac{\omega_1 z}{1+z}\right)\right]^2.
\end{equation}

\section{Models' calculations}\label{models_calcs}
In this section we show some models' calculations that were avoided in the main text for the sake of readability.
\subsection{Dynamics of \texorpdfstring{$r$}{r}}
This subsection's goal is to show the calculation of the differential equation for $r$ that results in \cref{diff_f_R_2}, starting from the definition
\begin{equation}
    \dfrac{dr}{dz}=\dfrac{dt}{dz}\dfrac{dr}{dt}=\dfrac{1}{\Lambda}\dfrac{dt}{dz}\dfrac{dR}{dt}=\dfrac{-\dot{R}}{\Lambda H\left(1+z\right)}.
\end{equation}
Now using the definition of $x$ in \cref{var_f_R} as well as $\Gamma=Rf_R/f_{RR}$, we get
\begin{equation}
\begin{split}
    \dfrac{dr}{dz}&=\dfrac{-1}{1+z}\left(\dfrac{\dot{R}}{\Lambda H}\right)\\ &=\dfrac{-1}{1+z}\left(\UPCBRACE{\dfrac{R}{\Lambda}}{=r}\UPCBRACE{\dfrac{\dot{R}f_{RR}}{Hf_{R}}}{=x}\UPCBRACE{\dfrac{f_{R}}{Rf_{RR}}}{=\Gamma}\right)=-\dfrac{r\Gamma x}{1+z}.
\end{split}
\end{equation}
\subsection{\texorpdfstring{$\lambda=0$}{lambda=0}}\label{lambda=0}
Here we show that when $\lambda=0$ the behavior of the quintessence model shown in Sec.~\ref{quintessence}, is the very same as that of the $\Lambda\mathrm{CDM}$ model. Using $\lambda=0$, which results in $dV/d\phi=0$ we can now solve for $\dot{\phi}\left(z\right)$ in \cref{phi_dynamics} and get
\begin{equation}
    \dot{\phi}\left(z\right)=\dot{\phi}_0\left(\dfrac{1+z}{1+z_0}\right)^3.
    \label{phi_dot}
\end{equation}
Given the fact that the initial condition $x_0=0$ results in $\dot{\phi}_0=0$, using \cref{phi_dot} and the definition of $x$ in \cref{var_phi}, we can conclude that $x\left(N\right)=0\;\forall\,z$ when $\lambda=0$. Now, using the initial condition for $y_0$ we can obtain $V_0=\left(3/\kappa\right)\left(H_{0}^{\Lambda}\right)^{2}\left(1-\Omega_{m,0}^{\Lambda}\right)$, and using this result with \cref{fried_q1} one gets $H\left(z\right)=H^\Lambda \left(z\right)$. Last, by inserting the expression of $H^\Lambda\left(z\right)$ [\cref{H_Lambda}] into \cref{H_quintessence} we get
\begin{equation}
    y\left(N\right)=\sqrt{\dfrac{1-\Omega_{m,0}^{\Lambda}}{\Omega_{m,0}^{\Lambda}e^{-3N}+1-\Omega_{m,0}^{\Lambda}}}.
\end{equation}
\subsection{Extra loss terms in \texorpdfstring{$f(R)$}{f(R)} implementation}\label{apx:consv_matter}
To close out the models' extra calculations, here we show how to get \cref{consv}. Starting from the definition of $\Omega$ in \cref{var_f_R}, and using \cref{matter_sol}, we can obtain
\begin{equation}
\label{omega_eq}
    \Omega=\dfrac{\kappa \rho_{m,0}\left(1+z\right)^3}{3H^2f_R}.
\end{equation}
\subsubsection*{First equation}
To obtain \cref{consv_Om_1}, we start by multiplying and dividing \cref{omega_eq} by $2f$, and using the definition of $y$ in \cref{var_f_R}, the following expression is obtained:
\begin{equation}
\label{omega_eq_1}
    \Omega=2y\kappa \rho_{m,0}\left(1+z\right)^3f^{-1}.
\end{equation}
To develop this expression further, we first express \cref{f_R_HS_b} in a more compact form using $r=R/\Lambda$:
\begin{equation}
\begin{split}
    f\left(r\right)&=\Lambda\left\{r -2\left[1-\dfrac{1}{\dfrac{r}{b}+1}\right]\right\}=\Lambda\left\{ r-2\left[\dfrac{r}{r+b}\right]\right\}\\ &=3\left(H_0^{\Lambda}\right)^2\left(1-\Omega^{\Lambda}_{m,0}\right)r\left(\dfrac{r+b-2}{r+b}\right),
\end{split} 
\label{fr}
\end{equation}
where in the last step $\Lambda = 3\left(H_0^{\Lambda}\right)^2\left(1-\Omega^{\Lambda}_{m,0}\right)$ was used.
Using this last result with the expression for $\Omega$ in \cref{omega_eq_1}, while remembering the definition $\Omega^{\Lambda}_{m,0}=\kappa \rho_{m,0}/3\left(H_0^{\Lambda}\right)^2$, we obtain
\begin{equation}
\Omega=\dfrac{2y\Omega^\Lambda_{m,0}\left(1+z\right)^3\left(r+b\right)}{r\left(1-\Omega^\Lambda_{m,0}\right)\left(r+b-2\right)}.
\end{equation}
\subsubsection*{Second equation}
In a similar fashion to the first equation, we start by multiplying and dividing \cref{omega_eq} by $R/2$, and make use of the definition of $v$ in \cref{var_f_R}, which yields
\begin{equation}
\label{omega_eq_2}
    \Omega=\dfrac{2v\kappa \rho_{m,0}\left(1+z\right)^3}{Rf_R}.
\end{equation}
We now need an expression for $f_R$. Performing the derivative with respect to $R$ of $f$, and using \cref{f_R_HS_b} and $r=R/\Lambda$,
\begin{equation}
f_{R}=1-\dfrac{2b}{\left(\dfrac{R}{\Lambda}+b\right)^{2}}=\dfrac{\left(r+b\right)^{2}-2b}{\left(r+b\right)^{2}}.
\end{equation}
Replacing this expression for $f_R$ in \cref{omega_eq_2} while multiplying and dividing by $\Lambda$, we obtain
\begin{equation}
\label{almost_consv_2}
        \Omega=\dfrac{2v\kappa \rho_{m,0}\left(1+z\right)^3\left(r+b\right)^{2}}{\Lambda r\left[\left(r+b\right)^{2}-2b\right]}.
\end{equation}
Finally, using $\Lambda = 3\left(H_0^{\Lambda}\right)^2\left(1-\Omega^{\Lambda}_{m,0}\right)$ along with $\Omega^{\Lambda}_{m,0}=\kappa \rho_{m,0}/3\left(H_0^{\Lambda}\right)^2$ in \cref{almost_consv_2},
\begin{equation}
\Omega=\dfrac{2v\Omega^\Lambda_{m,0}\left(1+z\right)^{3}\left(r+b\right)^{2}}{r\left(1-\Omega^\Lambda_{m,0}\right)\left[\left(r+b\right)^{2}-2b\right]}.
\end{equation}
\section{Details of the ANN method}\label{NN_details}
In this appendix, we list a series of details that were omitted in the main explanation of the ANN method. 
In \cref{fig:workflow} we show a schematic of an example ANN. To outline how to calculate the outputs of the ANN, we start by translating what each element of that graph represents mathematically. Each circle represents a unit, and each one of these, except for the ones in the input layer, has an associated bias and activation function. For example, we can denote the bias of the $j$-th unit on the second hidden layer as $b_j^{\left(2\right)}$, and the activation function as $g_j^{\left(2\right)}$. Then, each line that connects a pair of units has an associated weight. In a similar vein to the previous example, we can denote the weight of the line that connects the $i$-th unit in the input layer to the $j$-th unit in the first hidden layer as $w_{ij}^{\left(1\right)}$. Let us consider an ANN architecture composed by an input layer with $n$ units, one hidden layer with $m$ units each and an output layer with $k$ units, this is an ANN of size $\left(n,m,k\right)$, and the outputs $u_{\mathcal{N},l}$ with inputs $x_i$\footnote{We recall that the inputs of the ANN are the independent variable and the parameters of the differential system ($\boldsymbol{\theta}$).} can be expressed as

\begin{equation}
    \label{ANN_calc}
    u_{\mathcal{N},l}=\sum_j^m w_{jl}^{\rm out}g_j^{\left(1\right)}\left(\sum_i^nw_{ij}^{\left(1\right)}x_i + b_j^{\left(1\right)}\right) + b_l^{\rm out}.
\end{equation}

We can dissect this last equation further by noticing that the outputs of the hidden layer are essentially

\begin{equation}
    \label{hid_out}
    h_j^{\left(1\right)}=g_j^{\left(1\right)}\left(\sum_i^nw_{ij}^{\left(1\right)}x_i + b_j^{\left(1\right)}\right).
\end{equation}

Therefore, if we wanted to calculate the outputs of an ANN of size $\left(n,m,m,k\right)$, (this is an ANN that includes two hidden layers with $m$ units each), it would be
\begin{equation}
        u_{\mathcal{N},l}=\sum_j^m w_{jl}^{\rm out}g_j^{\left(2\right)}\left(\sum_i^m w_{ij}^{\left(2\right)}h_i^{\left(1\right)} + b_j^{\left(2\right)}\right) + b_l^{\rm out},
        \label{uN_output}
\end{equation}

where $h_i^{\left(1\right)}$ is the same as in \cref{hid_out}.

It follows from the last equation that for a choice of continuous activation functions $g_i$, the continuity of $u_{\mathcal{N},l}$ is assured since the composition of continuous functions is also continuous. Besides, the last equation also shows that $u_{\mathcal{N},l}$ is differentiable with respect to the independent variable and the parameters of the differential system ($x_i$).

On the other hand, \cref{table:ANN_details} shows details that involve specifics of the training like the number of iterations for each model, the number of training points for each input considered in each iteration which is called the batch size, the training range of each parameter, the specific values of the ANN architecture used for the training of each cosmological model, the number of internal parameters of the networks (here we refer to the weights and biases described above) and the training time for each cosmological model.

For every network in this work we use the Adam \cite{Adam} optimizer with its default values and all of them use $\tanh$ as their only activation function. Regarding the sampling of the points from the independent variable and the parameter spaces, we use a uniform distribution for each one, where the limits correspond to their respective training range.
\begin{table*}
\begin{adjustbox}{center}
\begin{ruledtabular}
\begin{tabular}{lcccccc}
Model                                   & Iterations                 & Training range                                                   & Batch size             & \begin{tabular}[c]{@{}c@{}}ANN architecture\\$\left(\rm in,hid,hid,out\right)$\end{tabular} & \begin{tabular}[c]{@{}c@{}}Number of internal\\parameters of the ANNs\end{tabular} & Training time (hardware)           \\ 
\hline
\multirow{2}{*}{$\Lambda \mathrm{CDM}$} & \multirow{2}{*}{$100000$} & $z\in\left[0,3\right]$                                           & $64$                   & \multirow{2}{*}{$x:\left(2,32,32,1\right)$}                                               & \multirow{2}{*}{$1185$}                                                           & \multirow{2}{*}{20 minutes (CPU)}  \\
                                        &                            & $\Omega_{m,0}\in\left[0.1,0.4\right]$                              & $64$                   &                                                                                             &                                                                                   &                                    \\ 
\hline
CPL                                     & $50000$                   & $z\in\left[0,3\right]$                                           & $32$                   & $x:\left(1,32,32,2\right)$                                                               & $1186$                                                                            & 10 minutes (CPU)                   \\ 
\hline
\multirow{3}{*}{Quintessence}           & \multirow{3}{*}{$200000$} & $z\in\left[0,10\right]$                                          & $32$                   & $x:\left(3,32,32,1\right)$                                                                  & $1217$                                                                            & \multirow{3}{*}{17 hours (CPU)}    \\
                                        &                            & $\lambda\in\left[0,3\right]$                                     & $32$                   & \multirow{2}{*}{$y:\left(3,32,32,1\right)$}                                                 & \multirow{2}{*}{$1217$}                                                           &                                    \\
                                        &                            & $\Omega^\Lambda_{m,0}\in\left[0.1,0.4\right]$                      & $32$                   &                                                                                             &                                                                                   &                                    \\ 
\hline
\multirow{5}{*}{Hu-Sawicki}             & \multirow{5}{*}{$600000$} & \multirow{2}{*}{$z\in\left[0,10\right]$}                         & \multirow{2}{*}{$128$} & $x:\left(3,32,32,1\right)$                                                                  & $1217$                                                                            & \multirow{5}{*}{2 days (GPU)}      \\
                                        &                            &                                                                  &                        & $y:\left(3,32,32,1\right)$                                                                  & $1217$                                                                            &                                    \\
                                        &                            & $b\in\left[0,5\right]$                                           & $128$                  & $v:\left(3,32,32,1\right)$                                                                  & $1217$                                                                            &                                    \\
                                        &                            & \multirow{2}{*}{$\Omega^{\Lambda}_{m,0}\in\left[0.1,0.4\right]$} & \multirow{2}{*}{$64$}  & $\Omega:\left(3,32,32,1\right)$                                                             & $1217$                                                                            &                                    \\
                                        &                            &                                                                  &                        & $r^\prime:\left(3,32,32,1\right)$                                                           & $1217$                                                                            &                                   
\end{tabular}
\end{ruledtabular}
\end{adjustbox}
\caption{Training details of the models. We state the amount of training iterations employed, the training range for the parameters of each model, as well as their corresponding batch sizes used during training. In addition, we specify the architecture of the ANN that represents the dependent variables and the number of internal parameters of the network. In the last column the time that the training required is listed along with the hardware used to perform the training. In the case of a CPU, we used an Intel i7-6500U, while in the case of a GPU, we used a single Nvidia A100.}
\label{table:ANN_details}
\end{table*}
\section{Details of the implementation of the ANN method to cosmological models}
\label{details_imp_models}
In this appendix we describe minor changes to the reparametrization and loss function of some of the models analyzed in this paper, and variable changes to some inputs of the networks so that they do not reach outside of the interval $\left[0,1\right]$ are also discussed in this appendix.

\subsection{\texorpdfstring{$\boldsymbol{\Lambda \mathrm{CDM}}$}{LambdaCDM}}
This being the simplest model, we make only one modification to the default loss in \cref{loss_res_reparam_bundle} [or more explicitly \cref{loss_LCDM_reparam}], this being the addition of a weighting function as proposed in Ref.~\cite{bundlesolutions}. With this consideration we set the final loss used to solve \cref{x_matter}, for a bundle in $\Omega_{m,0}$ to be
\begin{equation}
        L\left(\tilde{x}_m,z,\Omega_{m,0}\right)=\mathcal{R}\left(\tilde{x}_m,z,\Omega_{m,0}\right)^{2}e^{-2\left(z-z_0\right)},
    \label{res_Lambda_exp}
\end{equation}
where $\tilde{x}_m$ is the reparametrization of the variable, that uses the default bundle reparametrization in \cref{default_reparam_bundle}.

\subsection{Parametric dark energy}
In spite of the fact that the network trained using \cref{DE_reparam} only has one input, the training is not strictly speaking only done in $z$. For each iteration of training of this ANN, the loss is calculated for two combinations of the pair $\left(\omega_0, \omega_1\right)$, namely, $\left(-1, -0.6\right)$ and $\left(-0.8, 0\right)$.
\subsection{Quintessence}
Due to the increasing complexity of the problem, and the fact that the inputs of the ANNs corresponding to $N$ and $\lambda$ are not bound between $0$ and $1$, we make two changes to \cref{diff_q}. Instead of using $N$ and $\lambda$, we define $\lambda^{\prime}=\lambda/3$ and $N^{\prime}=N/\left|N_0\right|+1$ (making $N_0^{\prime}$=0). This way, both of these inputs to the ANNs are bound as $N^{\prime}\times\lambda^{\prime}\in\left[0,1\right]\times\left[0,1\right]$. Making these changes, the differential system to solve is now
\begin{equation}
\left\{ \begin{aligned}\dfrac{1}{\left|N_0\right|}\dfrac{dx}{dN^{\prime}}= & -3x+3\dfrac{\sqrt{6}}{2}\lambda^{\prime} y^{2}+\dfrac{3}{2}x\left(1+x^{2}-y^{2}\right)\\
\dfrac{1}{\left|N_0\right|}\dfrac{dy}{dN^{\prime}}= & -3\dfrac{\sqrt{6}}{2}xy\lambda^{\prime}+\dfrac{3}{2}y\left(1+x^{2}-y^{2}\right).
\end{aligned}
\right.\label{diff_q_change}
\end{equation}
Similarly to $\Lambda \mathrm{CDM}$, we also add an exponential term to this model's loss. But, due to \cref{quintessence_reparam} having a boundary condition on top of an initial one, we choose to modify the weighting function to account for this. With this modification, the loss to solve \cref{diff_q_change} is
\begin{equation}
\begin{split}
        L\left(\tilde{x}, \tilde{y},N^{\prime},\lambda^{\prime},\Omega^{\Lambda}_{m,0}\right)=&\sum^2_i\mathcal{R}_i\left(\tilde{x}, \tilde{y},N^{\prime},\lambda^{\prime},\Omega^{\Lambda}_{m,0}\right)^{2}\\
        &\times e^{-2\lambda^{\prime} N^{\prime}}. 
\end{split}
\label{res_quint_exp}
\end{equation}
\subsection{\texorpdfstring{$f(R)$}{f(R)}}
For this model, $z$ is not really a suitable variable to use in conjunction with a reparametrization as in \cref{default_reparam}, due to the fact that $z_0$ is the maximum value of $z$ in the training range. Because of this, and for the same reasons as listed for quintessence, we define $z^{\prime}=1 - z/z_0$ and $b^{\prime}=b/5$. With these changes, the final differential system is now
\begin{equation}
\label{diff_f_R_3_change}
\left\{ \begin{aligned}\dfrac{dx}{dz^{\prime}} & =\dfrac{-z_{0}}{z_{0}\left(1-z^{\prime}\right)+1}\left(-\Omega-2v+x+4y+xv+x^{2}\right)\\
\dfrac{dy}{dz^{\prime}} & =\dfrac{z_{0}}{z_{0}\left(1-z^{\prime}\right)+1}\left(vx\Gamma\left(r^{\prime}\right)-xy+4y-2yv\right)\\
\dfrac{dv}{dz^{\prime}} & =\dfrac{z_{0}v}{z_{0}\left(1-z^{\prime}\right)+1}\left(x\Gamma\left(r^{\prime}\right)+4-2v\right)\\
\dfrac{d\Omega}{dz^{\prime}} & =\dfrac{-z_{0}\Omega}{z_{0}\left(1-z^{\prime}\right)+1}\left(-1+2v+x\right)\\
\dfrac{dr^{\prime}}{dz^{\prime}} & =\dfrac{z_{0}\Gamma\left(r^{\prime}\right)x}{z_{0}\left(1-z^{\prime}\right)+1},
\end{aligned}
\right.
\end{equation}
where $\Gamma$ is now
\begin{equation}
\Gamma\left(r^\prime\right)=\dfrac{\left(e^{r^\prime}+5b^{\prime}\right)\left[\left(e^{r^\prime}+5b^{\prime}\right)^{2}-10b^{\prime}\right]}{20b^{\prime}e^{r^\prime}}.
\end{equation}
In addition to this, we also perform a change to the loss analogous to the one in quintessence making the part of final loss $L=L_\mathcal{R} + L_\mathcal{C}$ to solve \cref{diff_f_R_3_change} corresponding to just the residuals $L_\mathcal{R}$ as
\begin{equation}
    \label{f_R_res_exp}
    L_\mathcal{R}\left(\tilde{\boldsymbol{u}},z^{\prime},b^{\prime}, \Omega^{\Lambda}_{m,0}\right)=\sum_{i}^{5}\mathcal{R}_{i}\left(\tilde{\boldsymbol{u}},z^{\prime},b^{\prime}, \Omega^{\Lambda}_{m,0}\right)^{2}e^{-2b^{\prime} z^{\prime}}.
\end{equation}
The final detail to note, is the addition of a hyperparameter $\alpha$ in the reparametrization expressed by \cref{perturbative_reparam_bundle_f_R}, that helps tuning how good the solution provided by the $\Lambda \mathrm{CDM}$ model approximates to the true solution of the Hu-Sawicki model. This change gives the following general expression for the reparametrization used in \cref{f_R_res_exp}:
\begin{equation}
\begin{split}
    \tilde{\boldsymbol{u}}\left(z^\prime, b^\prime, \Omega_{m,0}^{\Lambda} \right)=&\hat{\boldsymbol{u}}\left(z^\prime,\Omega_{m,0}^{\Lambda}\right)+\left(1-e^{-z^\prime}\right)\left(1-e^{-\alpha b^\prime}\right)\\
    &\times\boldsymbol{u}_{\mathcal{N}}\left(z^\prime, b^\prime, \Omega_{m,0}^{\Lambda} \right).
\end{split}
\label{perturbative_reparam_bundle_f_R_alpha}
\end{equation}
It is evident that $\alpha\sim0$ corresponds to a belief that the $\Lambda \mathrm{CDM}$ solution is close to the real one for all values of the parameters, whereas $\alpha\gg 1$ denotes that the real solution is far from the one provided by the $\Lambda \mathrm{CDM}$ model. We choose to use $\alpha=1/6$\footnote{This value was found by trial and error with the goal of obtaining the lowest value for the loss $L$.}.

\section{Data sets}\label{datasets}
In this appendix we show the values and associated errors of the CC and BAO data that we use to obtain the results of the statistical analysis. We list in \cref{table_CC} the measurements of the Hubble parameter $H$ from cosmic chronometers, and in \cref{table_BAO} the measurements of different physical quantities that use the scale provided by baryon acoustic oscillations.
\begin{table}[H]
    \centering
\begin{ruledtabular}
\begin{tabular}{lcc}

$z$ & $H\left(z\right)\pm\sigma_{H}\left[\dfrac{\mathrm{km/s}}{\mathrm{Mpc}}\right]$ & References
\tabularnewline

\hline 
$0.09$ & $69\pm12$ & \multirow{9}{*}{\cite{CC1}}\tabularnewline
$0.17$ & $83\pm8$ & \tabularnewline
$0.27$ & $77\pm14$ & \tabularnewline
$0.4$ & $95\pm17$ & \tabularnewline
$0.9$ & $117\pm23$ & \tabularnewline
$1.3$ & $168\pm17$ & \tabularnewline
$1.43$ & $177\pm18$ & \tabularnewline
$1.53$ & $140\pm14$ & \tabularnewline
$1.75$ & $202\pm40$ & \tabularnewline
\hline
$0.48$ & $97\pm62$ & \multirow{2}{*}{\cite{CC2}}\tabularnewline
$0.88$ & $90\pm40$ & \tabularnewline
\hline
$0.1791$ & $75\pm4$ & \multirow{8}{*}{\cite{CC3}}\tabularnewline
$0.1993$ & $75\pm5$ & \tabularnewline
$0.3519$ & $83\pm14$ & \tabularnewline
$0.5929$ & $104\pm13$ & \tabularnewline
$0.6797$ & $92\pm8$ & \tabularnewline
$0.7812$ & $105\pm12$ & \tabularnewline
$0.8754$ & $125\pm17$ & \tabularnewline
$1.037$ & $154\pm20$ & \tabularnewline
\hline
$0.07$ & $69\pm19.6$ & \multirow{4}{*}{\cite{CC4}}\tabularnewline
$0.12$ & $68.6\pm26.2$ & \tabularnewline
$0.2$ & $72.9\pm29.6$ & \tabularnewline
$0.28$ & $88.8\pm36.6$ & \tabularnewline
\hline
$1.363$ & $160\pm33.6$  & \multirow{2}{*}{\cite{CC5}}\tabularnewline
$1.965$ & $186.5\pm50.4$ & \tabularnewline
\hline
$0.3802$ & $83\pm13.5$ & \multirow{5}{*}{\cite{CC6}}\tabularnewline
$0.4004$ & $77\pm10.2$ & \tabularnewline
$0.4247$ & $87.1\pm11.2$ & \tabularnewline
$0.4497$ & $92.8\pm12.9$  & \tabularnewline
$0.4783$ & $80.9\pm9$ & \tabularnewline
\end{tabular}
\end{ruledtabular}
    \caption{Measurements of the Hubble parameter $H$ using the Cosmic Chronometers technique described in Sec.~\ref{sect:CC}.}
    \label{table_CC}
\end{table}
\begin{table}[H]
    \centering
\begin{ruledtabular}
\begin{tabular}{lccc}
Observable & $z_{\rm eff}$ & Value $\pm$ error & References
\tabularnewline
\hline 
\multirow{5}{*}{$D_{V}/r_{d}$} & $0.15$ & $4473\pm0159$ & \cite{SDSS_DR7}\tabularnewline
 & $0.44$ & $11.548\pm0.559$ & \multirow{3}{*}{\cite{WiggleZ}}\tabularnewline
 & $0.6$ & $14.946\pm0.680$ & \tabularnewline
 & $0.73$ & $16.931\pm0.579$ & \tabularnewline
 & $1.52$ & $26.005\pm0.995$ & \cite{SDSS-IV_quasars}\tabularnewline
\hline 
$D_{A}/r_{d}$ & $0.81$ & $10.75\pm0.43$ & \cite{DES_Y1}\tabularnewline
\hline 
\multirow{7}{*}{$D_{M}/r_{d}$} & $0.38$ & $10.272\pm0.135\pm0.074$ & \multirow{3}{*}{\cite{BOSS}}\tabularnewline
 & $0.51$ & $13.378\pm0.156\pm0.095$ & \tabularnewline
 & $0.61$ & $15.449\pm0.189\pm0.108$ & \tabularnewline
 & $0.698$ & $17.65\pm0.3$ & \cite{SDSS_IV_LRG_anisotropicCF}\tabularnewline
 & $1.48$ & $30.31\pm0.79$ & \cite{SDSS_IV_QSO_anisotropicPS}\tabularnewline
 & $2.3$ & $37.77\pm2.13$ & \cite{SDSS-III_La_forests}\tabularnewline
 & $2.4$ & $36.6\pm1.2$ & \cite{La_forests_quasars_cross}\tabularnewline
\hline 
\multirow{4}{*}{$D_{H}/r_{d}$} & $0.698$ & $19.77\pm0.47$ & \cite{SDSS_IV_LRG_anisotropicCF}\tabularnewline
 & $1.48$ & $13.23\pm0.47$ & \cite{SDSS_IV_QSO_anisotropicPS}\tabularnewline
 & $2.3$ & $9.07\pm0.31$ & \cite{SDSS-III_La_forests}\tabularnewline
 & $2.4$ & $8.94\pm0.22$ & \cite{La_forests_quasars_cross}\tabularnewline
\hline 

\multirow{3}{*}{$Hr_{d}\left[\mathrm{km/s}\right]$} & $0.38$ & $12044.07\pm251.226\pm133.002$ & \tabularnewline
 & $0.51$ & $13374.09\pm251.226\pm147.78$ & \cite{BOSS}\tabularnewline
 & $0.61$ & $14378.994\pm266.004\pm162.558$ & \tabularnewline
\end{tabular}
\end{ruledtabular}
    \caption{Measurements of different quantities reliant on the scale provided by Baryon Acoustic Oscillations that we used in this paper. For the values with two errors, the first corresponds to the
statistical uncertainty, while the second represents the systematic error. In those cases we used the sum of both errors in quadrature.}
    \label{table_BAO}

\end{table}

\section{Computational efficiency}\label{flops}

In this appendix, we explore how the ANN method stands against its numerical counterpart, and in which circumstances one might find either method more appealing. As mentioned in the Introduction, neural networks can provide an advantage over numerical methods when used to test a model whose predictions are obtained by solving a system of differential equations. To quantify this advantage we calculate an estimation of the amount of FLOPs that it takes for a numerical method to solve the differential equations of a given initial value problem, as well as the equivalent for the ANN approach. In addition, we remark on some other appealing features of the ANN method.
\subsection{Floating point operations estimation}
The general form of an initial value problem is as follows:
\begin{equation}
\begin{aligned}
\label{eq:ivp}
\dfrac{d\boldsymbol{u}}{dt}=f\left(t, \boldsymbol{u}\right),  && \evalat{\boldsymbol{u}\left(t\right)}{t=t_0}=\boldsymbol{u}_0,
\end{aligned}
\end{equation}
where $f\left(t, \boldsymbol{u}\right)$ is an arbitrary function, and the vector $\boldsymbol{u}=\left(u^1,u^2,\dots,u^M\right)$ contains the $M$ dependent variables of the system. With the general problem defined, we begin our analysis by analyzing the FLOPs of numerical methods, and we note that the computational cost can significantly vary depending on the specific method used. Because all the problems in this work are initial value problems, we choose to analyze the FLOPs of SciPy's \cite{scipy} implementation\footnote{The analysis for this implementation was done for SciPy 1.8.0, which was the latest release at the time of submission.} of the adaptive explicit Runge-Kutta method RK5(4) \cite{RK5(4)_formulae,RK_interpolation}. In this implementation the number of function evaluations of $f\left(t, \boldsymbol{u}\right)$ in \cref{eq:ivp} is independent from the amount of points that the algorithm is asked to output. Besides, due to the adaptive nature of the integrator, the number of function evaluations is highly dependent on both the differential system and the error tolerance for the solver. Taking these facts into account, we can define a rough expression for the number of FLOPs, $\mathcal{F}_{\rm RK5(4)}$, that it takes for the integrator to output the values of the $M$ dependent variables at $N_{\rm out}$ points for a given value of the parameters of the differential system ($\boldsymbol{\theta}$), as follows \cite{supp_material}:
\begin{equation}
    \label{FLOPs_num}
    \mathcal{F}_{\rm RK5(4)} \simeq 9 N_{\rm out}\left(M+1\right)+N_{\rm eval}\left(\mathcal{F}_{\rm eval}+20  M\right),
\end{equation}
where $N_{\rm eval}$ is the number of evaluations of $f\left(t, \boldsymbol{u}\right)$ and $\mathcal{F}_{\rm eval}$ is the amount of FLOPs that it takes to do such an evaluation. The first term in \cref{FLOPs_num} corresponds to the interpolation cost, and the second to the Runge-Kutta steps. We see from this expression, that the cost associated with this numerical method scales linearly with a slope of order $10$ on the number of points. On the other hand, the term corresponding to the integration does not depend on the number of points. Besides, it is important to note that $N_{\rm eval}$ is the only factor in \cref{FLOPs_num} that is going to be different each time the integration is done for a different value of $\boldsymbol{\theta}$. This becomes especially relevant in cases like the one we mentioned in Sec.~\ref{f_R_loss} where because of a singularity in the parameter space of the differential system, there is a region of that space, close to the singularity, where the numerical solver requires more function evaluations to obtain a solution.

For the ANN method we first only count the number of FLOPs that it takes for the trained network to give the output we want, and later we include the cost of training the network, which is where most of the computational cost is.

To estimate the amount of FLOPs that it takes to calculate the $k$ outputs of such an ANN given $n$ inputs (called a forward pass), which we denote $\mathcal{F}_{\rm forw}$, we focus on the output of an ANN of size $\left(n,m,m,k\right)$
\footnote{In \cref{fig:workflow} we show an example of an ANN with a certain amount of inputs, hidden layers (each with $m$ units), and a single output. A bit more generally, we can think of an ANN with an input layer with $n$ units, two hidden layers with $m$ units each, and an output layer with $k$ units, which we can denote as $\left(n,m, m, k\right)$.} [see \cref{hid_out}]. If we ignore the FLOPs necessary to calculate the activation function, all of the operations inside the ANN are just additions and multiplications. We count one FLOP per multiplication or addition. First there is the calculation of the $m$ outputs of the first hidden layer. Looking at \cref{hid_out}, we see that this requires $n$ additions and $n$ multiplications, which makes for a total of $2nm$ FLOPs to calculate all the outputs of the first hidden layer. Following the same procedure for each layer until reaching the output layer we obtain
\begin{equation}
\mathcal{F}_{\rm forw}\simeq2mn + 2m^2 + 2mk=2m\left(n+m+k\right).
\end{equation}

 Regarding the cost of the neural network method specifically, we recall from Sec.~\ref{bundles} that the inputs of the networks used correspond to the independent variable and parameters of the differential system. In addition, there is the process of passing the outputs of the ANN through the reparametrization, which entails some additional cost that depends on the reparametrization itself. Therefore, the final amount of FLOPs, $\mathcal{F}_{\rm ANN}$, that it takes for an ANN to output the values of a dependent variable at $N_{\rm out}$ different points can be estimated as follows:
\begin{equation}
\label{ANN_flops}
    \mathcal{F}_{\rm ANN} \simeq N_{\rm out}\left(\mathcal{F}_{\rm forw} + \mathcal{F}_{\rm rep}\right),
\end{equation}
where $\mathcal{F}_{\rm rep}$ is the amount FLOPs that it takes to calculate the reparametrization. Besides, because we associate each dependent variable with an individual ANN, as mentioned in Sec.~\ref{methods}, the formula in \cref{ANN_flops} is valid for each dependent variable. Therefore if for example one needs only two of the $M$ dependent variables, the cost is just $2 \mathcal{F}_{\rm ANN}$, and there is no need to calculate the $M-2$ remaining.
To close on this part of the analysis, the training cost must be addressed. The training of the ANNs constitutes an iterative process where in each iteration the loss $L$ must be computed at $N_{\rm train}$ training points. To calculate the value of $L$ at each of these points, the outputs of the $M$ ANNs need to be computed, which constitutes $M$ forward passes. Next, the outputs of the ANNs must be reparametrized. Once this has been achieved, the reparametrized outputs are used in \cref{loss_res_reparam_bundle} to calculate the value of $L$.\footnote{We ignore the contribution from the FLOPs necessary to calculate this last step.} Then, the internal parameters of the ANNs are optimized to minimize the value of $L$, called a backward pass. The cost of this last procedure can be estimated to be at most 3 times the cost of a forward pass \cite{back_forw_ratio}. Therefore, the amount of FLOPs, $\mathcal{F}_{\rm train}$, required for the training made up of $N_{\rm iter}$ iterations can be estimated to be
\begin{equation}
    \label{ANN_flops_training}
    \mathcal{F}_{\rm train}\simeq N_{\rm iter} N_{\rm train} M\left(4 \mathcal{F}_{\rm forw} + \mathcal{F}_{\rm rep}\right).
\end{equation}
It is evident from comparing \cref{ANN_flops,ANN_flops_training} that in the majority of cases the training is where most of the computational cost is going to be. In this case we see that the cost scales with three main factors: i) the sizes of the ANNs, ii) the batch size, and iii) the amount of iterations. All of these can be related to the accuracy of the results from the method. For the case of the batch size and amount of iterations, because there is no risk of overfitting, a higher number of these two quantities tends to correlate with more accurate solutions. In the case of the network size, it is not necessarily always the case that a bigger network, meaning more hidden layers and/or more units per hidden layer, is the route to achieve better results. Taking this into account, we can say that computational cost associated with a complex problem and/or a need for low-error solutions mostly translates into more computation resources in the training stage of the ANN method.
\subsection{Parallelization}
While the preceding calculations show a high amount of FLOPs in the ANN method, we have to take into account the viable hardware options to start translating this into time spent computing. This is where the highly parallelizable nature of ANNs comes into play. The usual choice of hardware to accomplish this task is GPUs. During training, these make it possible to calculate the outputs of the forward and backward passes for the $N_{\rm train}$ points in parallel, although this is limited by the memory of the GPU. This also applies for a trained ANN, where the $N_{\rm out}$ points can also be computed in parallel. Both of these uses result in a significant speed up when compared to using CPUs when $N_{\rm out}$ (or in training $N_{\rm train}$) is large enough. An example of this last case can be seen in Ref.~\cite{Cosmopower}, where the process of MCMC with trained ANNs was done on a GPU in an inference pipeline.
\subsection{Storage}
An additional capability to consider is the portability of the solutions. In the case of numerical methods, the option to store the solutions often comes in the form of storing a mesh of solutions in the parameter space to do interpolation afterwards. Such a method has several disadvantages such as the storage size of the mesh growing exponentially with the amount of parameters, and the interpolation adding a new source of error that can only be mitigated through making a denser mesh. Therefore, in many cases it is often more efficient to run the numerical solver each time a solution is needed, instead of storing the solutions. Conversely, neural networks are lightweight and, because of the very nature of the method being used in this work, the only things one needs are the trained ANNs and their associated reparametrizations to the dependent variables. To give a precise example of the storage needed for this method, each individual ANN used in this work takes $\sim 10\; \mathrm{kB}$ to store.
\subsection{When to use ANN}
Combining what we described throughout this appendix, we can now make a more educated selection of the use cases where one might find either numerical methods or ANN-based methods more appealing for solving an initial value problem.

If for a moment we ignore the training, we can take \cref{FLOPs_num,ANN_flops} and get

\begin{equation}
    \label{RK_ANN_ratio}
    \dfrac{\mathcal{F}_{\rm RK5(4)}}{M\mathcal{F}_{\rm ANN}} \simeq \dfrac{9 N_{\rm out}\left(M+1\right)+N_{\rm eval}\left(\mathcal{F}_{\rm eval}+20  M\right)}{MN_{\rm out}\left(\mathcal{F}_{\rm forw} + \mathcal{F}_{\rm rep}\right)},
\end{equation}
where we multiplied the ANN cost by $M$ to account for the FLOPs necessary to calculate all of the dependent variables. From \cref{RK_ANN_ratio} we can evaluate two different scenarios. First is the case where $9 N_{\rm out}\left(M+1\right)\gg N_{\rm eval} \left(\mathcal{F}_{\rm eval} + 20M\right)$, which represents a situation where either the amount of points is rather large, or the problem at hand is easy to integrate. It is clear that the Runge-Kutta method is the less costly implementation for this case, as we can see if we approximate \cref{RK_ANN_ratio} as
\begin{equation}
    \dfrac{\mathcal{F}_{\rm RK5(4)}}{M\mathcal{F}_{\rm ANN}} \simeq 9\left(\mathcal{F}_{\rm forw} + \mathcal{F}_{\rm rep}\right)^{-1}.
    \label{RK_ANN_ratio_2}
\end{equation}
On the other hand, given the case where the relation is $9 N_{\rm out}\left(M+1\right)\ll N_{\rm eval} \left(\mathcal{F}_{\rm eval} + 20M\right)$, we are left with
\begin{equation}
    \dfrac{\mathcal{F}_{\rm RK5(4)}}{M\mathcal{F}_{\rm ANN}} \simeq \dfrac{N_{\rm eval}\left(\mathcal{F}_{\rm eval}+20 M\right)}{MN_{\rm out}\left(\mathcal{F}_{\rm forw} + \mathcal{F}_{\rm rep}\right)}.
    \label{RK_ANN_ratio_3}
\end{equation}
This approximation is applicable to situations where the amount of points is fairly low or the problem is difficult to solve. Therefore, in these situations we can use \cref{RK_ANN_ratio_3} as an approximate rule of thumb to determine which method is likely to be the most efficient.

Now it is important to recall the training. From what we discussed previously about \cref{ANN_flops_training}, we can see that even in the situations where \cref{RK_ANN_ratio_3} suggests the use of ANNs, the additional cost of training the networks can still make it the less efficient option. Although this may seem discouraging, we have to be reminded of two important things mentioned before: the fact that the training is done just once, and parallelization. First, because the training is performed just once, we can see that the cost from this process can be easily overlooked if the solutions are used several times. We can see that this can be the case in a scenario that involves using MCMC. For example, if in such a scenario the number of chains is $N_{\rm chains}$, the total cost from the use of the ANN solutions is $N_{\rm chains}M\mathcal{F}_{\rm ANN} + \mathcal{F}_{\rm train}$. Therefore, the larger $N_{\rm chains}$, the less significant the cost of training the ANN. Besides, the use of parallelization can make the cost of both training and using the ANNs much more efficient in terms of actual time spent executing the computations, thus making the scenario discussed before more favorable to ANNs when the hardware options are available.

To conclude this appendix, it is important to note that the analysis done here for the computational capabilities of the ANN method is valid for any kind of differential system. The only caveat to have in mind for the most general case would be that now the $n$ inputs of the network represent the parameters as well as the multiple independent variables, instead of just a single independent variable in the initial value case.


\begin{thebibliography}{77}%
\makeatletter
\providecommand \@ifxundefined [1]{%
 \@ifx{#1\undefined}
}%
\providecommand \@ifnum [1]{%
 \ifnum #1\expandafter \@firstoftwo
 \else \expandafter \@secondoftwo
 \fi
}%
\providecommand \@ifx [1]{%
 \ifx #1\expandafter \@firstoftwo
 \else \expandafter \@secondoftwo
 \fi
}%
\providecommand \natexlab [1]{#1}%
\providecommand \enquote  [1]{``#1''}%
\providecommand \bibnamefont  [1]{#1}%
\providecommand \bibfnamefont [1]{#1}%
\providecommand \citenamefont [1]{#1}%
\providecommand \href@noop [0]{\@secondoftwo}%
\providecommand \href [0]{\begingroup \@sanitize@url \@href}%
\providecommand \@href[1]{\@@startlink{#1}\@@href}%
\providecommand \@@href[1]{\endgroup#1\@@endlink}%
\providecommand \@sanitize@url [0]{\catcode `\\12\catcode `\$12\catcode
  `\&12\catcode `\#12\catcode `\^12\catcode `\_12\catcode `\%12\relax}%
\providecommand \@@startlink[1]{}%
\providecommand \@@endlink[0]{}%
\providecommand \url  [0]{\begingroup\@sanitize@url \@url }%
\providecommand \@url [1]{\endgroup\@href {#1}{\urlprefix }}%
\providecommand \urlprefix  [0]{URL }%
\providecommand \Eprint [0]{\href }%
\providecommand \doibase [0]{https://doi.org/}%
\providecommand \selectlanguage [0]{\@gobble}%
\providecommand \bibinfo  [0]{\@secondoftwo}%
\providecommand \bibfield  [0]{\@secondoftwo}%
\providecommand \translation [1]{[#1]}%
\providecommand \BibitemOpen [0]{}%
\providecommand \bibitemStop [0]{}%
\providecommand \bibitemNoStop [0]{.\EOS\space}%
\providecommand \EOS [0]{\spacefactor3000\relax}%
\providecommand \BibitemShut  [1]{\csname bibitem#1\endcsname}%
\let\auto@bib@innerbib\@empty
\bibitem [{\citenamefont {Raissi}\ \emph {et~al.}(2019)\citenamefont {Raissi},
  \citenamefont {Perdikaris},\ and\ \citenamefont {Karniadakis}}]{pinns}%
  \BibitemOpen
  \bibfield  {author} {\bibinfo {author} {\bibfnamefont {M.}~\bibnamefont
  {Raissi}}, \bibinfo {author} {\bibfnamefont {P.}~\bibnamefont {Perdikaris}},\
  and\ \bibinfo {author} {\bibfnamefont {G.~E.}\ \bibnamefont {Karniadakis}},\
  }\bibfield  {title} {\bibinfo {title} {Physics-informed neural networks: A
  deep learning framework for solving forward and inverse problems involving
  nonlinear partial differential equations},\ }\href
  {https://doi.org/https://doi.org/10.1016/j.jcp.2018.10.045} {\bibfield
  {journal} {\bibinfo  {journal} {Journal of Computational Physics}\ }\textbf
  {\bibinfo {volume} {378}},\ \bibinfo {pages} {686} (\bibinfo {year}
  {2019})}\BibitemShut {NoStop}%
\bibitem [{\citenamefont {Dialektopoulos}\ \emph {et~al.}(2022)\citenamefont
  {Dialektopoulos}, \citenamefont {Said}, \citenamefont {Mifsud}, \citenamefont
  {Sultana},\ and\ \citenamefont {Adami}}]{ANN_cosmo_reconstruction}%
  \BibitemOpen
  \bibfield  {author} {\bibinfo {author} {\bibfnamefont {K.}~\bibnamefont
  {Dialektopoulos}}, \bibinfo {author} {\bibfnamefont {J.~L.}\ \bibnamefont
  {Said}}, \bibinfo {author} {\bibfnamefont {J.}~\bibnamefont {Mifsud}},
  \bibinfo {author} {\bibfnamefont {J.}~\bibnamefont {Sultana}},\ and\ \bibinfo
  {author} {\bibfnamefont {K.~Z.}\ \bibnamefont {Adami}},\ }\bibfield  {title}
  {\bibinfo {title} {Neural network reconstruction of late-time cosmology and
  null tests},\ }\href {https://doi.org/10.1088/1475-7516/2022/02/023}
  {\bibfield  {journal} {\bibinfo  {journal} {Journal of Cosmology and
  Astroparticle Physics}\ }\textbf {\bibinfo {volume} {2022}}\bibinfo  {number}
  { (02)},\ \bibinfo {pages} {023}}\BibitemShut {NoStop}%
\bibitem [{\citenamefont {Escamilla-Rivera}\ \emph {et~al.}(2020)\citenamefont
  {Escamilla-Rivera}, \citenamefont {Quintero},\ and\ \citenamefont
  {Capozziello}}]{RNN+BNN_trained_on_Pantheon_to_generate_dl_at_high_z}%
  \BibitemOpen
\bibfield  {number} {  }\bibfield  {author} {\bibinfo {author} {\bibfnamefont
  {C.}~\bibnamefont {Escamilla-Rivera}}, \bibinfo {author} {\bibfnamefont
  {M.~A.~C.}\ \bibnamefont {Quintero}},\ and\ \bibinfo {author} {\bibfnamefont
  {S.}~\bibnamefont {Capozziello}},\ }\bibfield  {title} {\bibinfo {title} {A
  deep learning approach to cosmological dark energy models},\ }\href
  {https://doi.org/10.1088/1475-7516/2020/03/008} {\bibfield  {journal}
  {\bibinfo  {journal} {Journal of Cosmology and Astroparticle Physics}\
  }\textbf {\bibinfo {volume} {2020}}\bibinfo  {number} { (03)},\ \bibinfo
  {pages} {008}}\BibitemShut {NoStop}%
\bibitem [{\citenamefont {Spurio Mancini}\ \emph {et~al.}(2022)\citenamefont
  {Spurio Mancini}, \citenamefont {Piras}, \citenamefont {Alsing},
  \citenamefont {Joachimi},\ and\ \citenamefont {Hobson}}]{Cosmopower}%
  \BibitemOpen
\bibfield  {number} {  }\bibfield  {author} {\bibinfo {author} {\bibfnamefont
  {A.}~\bibnamefont {Spurio Mancini}}, \bibinfo {author} {\bibfnamefont
  {D.}~\bibnamefont {Piras}}, \bibinfo {author} {\bibfnamefont
  {J.}~\bibnamefont {Alsing}}, \bibinfo {author} {\bibfnamefont
  {B.}~\bibnamefont {Joachimi}},\ and\ \bibinfo {author} {\bibfnamefont
  {M.~P.}\ \bibnamefont {Hobson}},\ }\bibfield  {title} {\bibinfo {title}
  {{CosmoPower: Emulating cosmological power spectra for accelerated Bayesian
  inference from next-generation surveys}},\ }\href
  {https://doi.org/10.1093/mnras/stac064} {\bibfield  {journal} {\bibinfo
  {journal} {Monthly Notices of the Royal Astronomical Society}\ }\textbf
  {\bibinfo {volume} {511}},\ \bibinfo {pages} {1771} (\bibinfo {year}
  {2022})}\BibitemShut {NoStop}%
\bibitem [{\citenamefont {Piscopo}\ \emph {et~al.}(2019)\citenamefont
  {Piscopo}, \citenamefont {Spannowsky},\ and\ \citenamefont
  {Waite}}]{early_universe_model_lagaris_method}%
  \BibitemOpen
  \bibfield  {author} {\bibinfo {author} {\bibfnamefont {M.~L.}\ \bibnamefont
  {Piscopo}}, \bibinfo {author} {\bibfnamefont {M.}~\bibnamefont
  {Spannowsky}},\ and\ \bibinfo {author} {\bibfnamefont {P.}~\bibnamefont
  {Waite}},\ }\bibfield  {title} {\bibinfo {title} {Solving differential
  equations with neural networks: Applications to the calculation of
  cosmological phase transitions},\ }\href
  {https://doi.org/10.1103/PhysRevD.100.016002} {\bibfield  {journal} {\bibinfo
   {journal} {Phys. Rev. D}\ }\textbf {\bibinfo {volume} {100}},\ \bibinfo
  {pages} {016002} (\bibinfo {year} {2019})}\BibitemShut {NoStop}%
\bibitem [{\citenamefont {Lagaris}\ \emph {et~al.}(1998)\citenamefont
  {Lagaris}, \citenamefont {Likas},\ and\ \citenamefont
  {Fotiadis}}]{ANN_diff_eqs}%
  \BibitemOpen
  \bibfield  {author} {\bibinfo {author} {\bibfnamefont {I.}~\bibnamefont
  {Lagaris}}, \bibinfo {author} {\bibfnamefont {A.}~\bibnamefont {Likas}},\
  and\ \bibinfo {author} {\bibfnamefont {D.}~\bibnamefont {Fotiadis}},\
  }\bibfield  {title} {\bibinfo {title} {Artificial neural networks for solving
  ordinary and partial differential equations},\ }\href
  {https://doi.org/10.1109/72.712178} {\bibfield  {journal} {\bibinfo
  {journal} {IEEE Transactions on Neural Networks}\ }\textbf {\bibinfo {volume}
  {9}},\ \bibinfo {pages} {987} (\bibinfo {year} {1998})}\BibitemShut {NoStop}%
\bibitem [{\citenamefont {Goodfellow}\ \emph {et~al.}(2016)\citenamefont
  {Goodfellow}, \citenamefont {Bengio},\ and\ \citenamefont
  {Courville}}]{goodfellow2016deep}%
  \BibitemOpen
  \bibfield  {author} {\bibinfo {author} {\bibfnamefont {I.}~\bibnamefont
  {Goodfellow}}, \bibinfo {author} {\bibfnamefont {Y.}~\bibnamefont {Bengio}},\
  and\ \bibinfo {author} {\bibfnamefont {A.}~\bibnamefont {Courville}},\
  }\href@noop {} {\emph {\bibinfo {title} {Deep Learning}}}\ (\bibinfo
  {publisher} {MIT press},\ \bibinfo {year} {2016})\BibitemShut {NoStop}%
\bibitem [{\citenamefont {Flamant}\ \emph {et~al.}(2020)\citenamefont
  {Flamant}, \citenamefont {Protopapas},\ and\ \citenamefont
  {Sondak}}]{bundlesolutions}%
  \BibitemOpen
  \bibfield  {author} {\bibinfo {author} {\bibfnamefont {C.}~\bibnamefont
  {Flamant}}, \bibinfo {author} {\bibfnamefont {P.}~\bibnamefont
  {Protopapas}},\ and\ \bibinfo {author} {\bibfnamefont {D.}~\bibnamefont
  {Sondak}},\ }\href@noop {} {\bibinfo {title} {Solving differential equations
  using neural network solution bundles}} (\bibinfo {year} {2020}),\ \Eprint
  {https://arxiv.org/abs/2006.14372} {arXiv:2006.14372} \BibitemShut {NoStop}%
\bibitem [{\citenamefont {Chen}\ \emph {et~al.}(2020)\citenamefont {Chen},
  \citenamefont {Sondak}, \citenamefont {Protopapas}, \citenamefont
  {Mattheakis}, \citenamefont {Liu}, \citenamefont {Agarwal},\ and\
  \citenamefont {Giovanni}}]{outdated_neurodiff_ref}%
  \BibitemOpen
  \bibfield  {author} {\bibinfo {author} {\bibfnamefont {F.}~\bibnamefont
  {Chen}}, \bibinfo {author} {\bibfnamefont {D.}~\bibnamefont {Sondak}},
  \bibinfo {author} {\bibfnamefont {P.}~\bibnamefont {Protopapas}}, \bibinfo
  {author} {\bibfnamefont {M.}~\bibnamefont {Mattheakis}}, \bibinfo {author}
  {\bibfnamefont {S.}~\bibnamefont {Liu}}, \bibinfo {author} {\bibfnamefont
  {D.}~\bibnamefont {Agarwal}},\ and\ \bibinfo {author} {\bibfnamefont {M.~D.}\
  \bibnamefont {Giovanni}},\ }\bibfield  {title} {\bibinfo {title}
  {Neurodiffeq: A python package for solving differential equations with neural
  networks},\ }\href {https://doi.org/10.21105/joss.01931} {\bibfield
  {journal} {\bibinfo  {journal} {Journal of Open Source Software}\ }\textbf
  {\bibinfo {volume} {5}},\ \bibinfo {pages} {1931} (\bibinfo {year}
  {2020})}\BibitemShut {NoStop}%
\bibitem [{\citenamefont {Riess}\ \emph {et~al.}(1998)\citenamefont {Riess},
  \citenamefont {Filippenko}, \citenamefont {Challis}, \citenamefont
  {Clocchiatti}, \citenamefont {Diercks}, \citenamefont {Garnavich},
  \citenamefont {Gilliland}, \citenamefont {Hogan}, \citenamefont {Jha},
  \citenamefont {Kirshner} \emph {et~al.}}]{Riess_1998}%
  \BibitemOpen
  \bibfield  {author} {\bibinfo {author} {\bibfnamefont {A.~G.}\ \bibnamefont
  {Riess}}, \bibinfo {author} {\bibfnamefont {A.~V.}\ \bibnamefont
  {Filippenko}}, \bibinfo {author} {\bibfnamefont {P.}~\bibnamefont {Challis}},
  \bibinfo {author} {\bibfnamefont {A.}~\bibnamefont {Clocchiatti}}, \bibinfo
  {author} {\bibfnamefont {A.}~\bibnamefont {Diercks}}, \bibinfo {author}
  {\bibfnamefont {P.~M.}\ \bibnamefont {Garnavich}}, \bibinfo {author}
  {\bibfnamefont {R.~L.}\ \bibnamefont {Gilliland}}, \bibinfo {author}
  {\bibfnamefont {C.~J.}\ \bibnamefont {Hogan}}, \bibinfo {author}
  {\bibfnamefont {S.}~\bibnamefont {Jha}}, \bibinfo {author} {\bibfnamefont
  {R.~P.}\ \bibnamefont {Kirshner}}, \emph {et~al.},\ }\bibfield  {title}
  {\bibinfo {title} {Observational evidence from supernovae for an accelerating
  universe and a cosmological constant},\ }\href
  {https://doi.org/10.1086/300499} {\bibfield  {journal} {\bibinfo  {journal}
  {The Astronomical Journal}\ }\textbf {\bibinfo {volume} {116}},\ \bibinfo
  {pages} {1009} (\bibinfo {year} {1998})}\BibitemShut {NoStop}%
\bibitem [{\citenamefont {Perlmutter}\ \emph {et~al.}(1999)\citenamefont
  {Perlmutter}, \citenamefont {Aldering}, \citenamefont {Goldhaber},
  \citenamefont {Knop}, \citenamefont {Nugent}, \citenamefont {Castro},
  \citenamefont {Deustua}, \citenamefont {Fabbro}, \citenamefont {Goobar},
  \citenamefont {Groom} \emph {et~al.}}]{Perlmutter}%
  \BibitemOpen
  \bibfield  {author} {\bibinfo {author} {\bibfnamefont {S.}~\bibnamefont
  {Perlmutter}}, \bibinfo {author} {\bibfnamefont {G.}~\bibnamefont
  {Aldering}}, \bibinfo {author} {\bibfnamefont {G.}~\bibnamefont {Goldhaber}},
  \bibinfo {author} {\bibfnamefont {R.~A.}\ \bibnamefont {Knop}}, \bibinfo
  {author} {\bibfnamefont {P.}~\bibnamefont {Nugent}}, \bibinfo {author}
  {\bibfnamefont {P.~G.}\ \bibnamefont {Castro}}, \bibinfo {author}
  {\bibfnamefont {S.}~\bibnamefont {Deustua}}, \bibinfo {author} {\bibfnamefont
  {S.}~\bibnamefont {Fabbro}}, \bibinfo {author} {\bibfnamefont
  {A.}~\bibnamefont {Goobar}}, \bibinfo {author} {\bibfnamefont {D.~E.}\
  \bibnamefont {Groom}}, \emph {et~al.},\ }\bibfield  {title} {\bibinfo {title}
  {Measurements of {$\Omega$} and {$\Lambda$} from 42 high-redshift
  supernovae},\ }\href {https://doi.org/10.1086/307221} {\bibfield  {journal}
  {\bibinfo  {journal} {The Astrophysical Journal}\ }\textbf {\bibinfo {volume}
  {517}},\ \bibinfo {pages} {565} (\bibinfo {year} {1999})}\BibitemShut
  {NoStop}%
\bibitem [{\citenamefont {Riess}\ \emph {et~al.}(2022)\citenamefont {Riess},
  \citenamefont {Yuan}, \citenamefont {Macri}, \citenamefont {Scolnic},
  \citenamefont {Brout}, \citenamefont {Casertano}, \citenamefont {Jones},
  \citenamefont {Murakami}, \citenamefont {Anand}, \citenamefont {Breuval},
  \citenamefont {Brink} \emph {et~al.}}]{riess2022comprehensive}%
  \BibitemOpen
  \bibfield  {author} {\bibinfo {author} {\bibfnamefont {A.~G.}\ \bibnamefont
  {Riess}}, \bibinfo {author} {\bibfnamefont {W.}~\bibnamefont {Yuan}},
  \bibinfo {author} {\bibfnamefont {L.~M.}\ \bibnamefont {Macri}}, \bibinfo
  {author} {\bibfnamefont {D.}~\bibnamefont {Scolnic}}, \bibinfo {author}
  {\bibfnamefont {D.}~\bibnamefont {Brout}}, \bibinfo {author} {\bibfnamefont
  {S.}~\bibnamefont {Casertano}}, \bibinfo {author} {\bibfnamefont {D.~O.}\
  \bibnamefont {Jones}}, \bibinfo {author} {\bibfnamefont {Y.}~\bibnamefont
  {Murakami}}, \bibinfo {author} {\bibfnamefont {G.~S.}\ \bibnamefont {Anand}},
  \bibinfo {author} {\bibfnamefont {L.}~\bibnamefont {Breuval}}, \bibinfo
  {author} {\bibfnamefont {T.~G.}\ \bibnamefont {Brink}}, \emph {et~al.},\
  }\bibfield  {title} {\bibinfo {title} {A comprehensive measurement of the
  local value of the {Hubble} constant with 1 km s$^{-1}$ mpc$^{-1}$
  uncertainty from the {Hubble Space Telescope} and the {SH0ES} team},\ }\href
  {https://doi.org/10.3847/2041-8213/ac5c5b} {\bibfield  {journal} {\bibinfo
  {journal} {The Astrophysical Journal Letters}\ }\textbf {\bibinfo {volume}
  {934}},\ \bibinfo {pages} {L7} (\bibinfo {year} {2022})}\BibitemShut
  {NoStop}%
\bibitem [{\citenamefont {{Planck Collaboration}}\ \emph
  {et~al.}(2020)\citenamefont {{Planck Collaboration}}, \citenamefont {{N.
  Aghanim}}, \citenamefont {{Y. Akrami}}, \citenamefont {{M. Ashdown}},
  \citenamefont {{J. Aumont}}, \citenamefont {{C. Baccigalupi}}, \citenamefont
  {{M. Ballardini}}, \citenamefont {{A. J. Banday}}, \citenamefont {{R. B.
  Barreiro}}, \citenamefont {{N. Bartolo}}, \citenamefont {{S. Basak}} \emph
  {et~al.}}]{Planck}%
  \BibitemOpen
  \bibfield  {author} {\bibinfo {author} {\bibnamefont {{Planck
  Collaboration}}}, \bibinfo {author} {\bibnamefont {{N. Aghanim}}}, \bibinfo
  {author} {\bibnamefont {{Y. Akrami}}}, \bibinfo {author} {\bibnamefont {{M.
  Ashdown}}}, \bibinfo {author} {\bibnamefont {{J. Aumont}}}, \bibinfo {author}
  {\bibnamefont {{C. Baccigalupi}}}, \bibinfo {author} {\bibnamefont {{M.
  Ballardini}}}, \bibinfo {author} {\bibnamefont {{A. J. Banday}}}, \bibinfo
  {author} {\bibnamefont {{R. B. Barreiro}}}, \bibinfo {author} {\bibnamefont
  {{N. Bartolo}}}, \bibinfo {author} {\bibnamefont {{S. Basak}}}, \emph
  {et~al.},\ }\bibfield  {title} {\bibinfo {title} {Planck 2018 results - {VI}.
  {Cosmological} parameters},\ }\href
  {https://doi.org/10.1051/0004-6361/201833910} {\bibfield  {journal} {\bibinfo
   {journal} {A\&A}\ }\textbf {\bibinfo {volume} {641}},\ \bibinfo {pages} {A6}
  (\bibinfo {year} {2020})}\BibitemShut {NoStop}%
\bibitem [{\citenamefont {{Shah}}\ \emph {et~al.}(2021)\citenamefont {{Shah}},
  \citenamefont {{Lemos}},\ and\ \citenamefont {{Lahav}}}]{H_0_tension_review}%
  \BibitemOpen
  \bibfield  {author} {\bibinfo {author} {\bibfnamefont {P.}~\bibnamefont
  {{Shah}}}, \bibinfo {author} {\bibfnamefont {P.}~\bibnamefont {{Lemos}}},\
  and\ \bibinfo {author} {\bibfnamefont {O.}~\bibnamefont {{Lahav}}},\
  }\bibfield  {title} {\bibinfo {title} {{A buyer's guide to the Hubble
  constant}},\ }\href {https://doi.org/10.1007/s00159-021-00137-4} {\bibfield
  {journal} {\bibinfo  {journal} {The Astronomy and Astrophysics Review}\
  }\textbf {\bibinfo {volume} {29}},\ \bibinfo {eid} {9} (\bibinfo {year}
  {2021})}\BibitemShut {NoStop}%
\bibitem [{\citenamefont {Abdalla}\ \emph {et~al.}(2022)\citenamefont
  {Abdalla}, \citenamefont {Abellán}, \citenamefont {Aboubrahim},
  \citenamefont {Agnello}, \citenamefont {Özgür Akarsu}, \citenamefont
  {Akrami}, \citenamefont {Alestas}, \citenamefont {Aloni}, \citenamefont
  {Amendola}, \citenamefont {Anchordoqui} \emph {et~al.}}]{cosmo_white_paper}%
  \BibitemOpen
  \bibfield  {author} {\bibinfo {author} {\bibfnamefont {E.}~\bibnamefont
  {Abdalla}}, \bibinfo {author} {\bibfnamefont {G.~F.}\ \bibnamefont
  {Abellán}}, \bibinfo {author} {\bibfnamefont {A.}~\bibnamefont
  {Aboubrahim}}, \bibinfo {author} {\bibfnamefont {A.}~\bibnamefont {Agnello}},
  \bibinfo {author} {\bibnamefont {Özgür Akarsu}}, \bibinfo {author}
  {\bibfnamefont {Y.}~\bibnamefont {Akrami}}, \bibinfo {author} {\bibfnamefont
  {G.}~\bibnamefont {Alestas}}, \bibinfo {author} {\bibfnamefont
  {D.}~\bibnamefont {Aloni}}, \bibinfo {author} {\bibfnamefont
  {L.}~\bibnamefont {Amendola}}, \bibinfo {author} {\bibfnamefont {L.~A.}\
  \bibnamefont {Anchordoqui}}, \emph {et~al.},\ }\bibfield  {title} {\bibinfo
  {title} {Cosmology intertwined: A review of the particle physics,
  astrophysics, and cosmology associated with the cosmological tensions and
  anomalies},\ }\href
  {https://doi.org/https://doi.org/10.1016/j.jheap.2022.04.002} {\bibfield
  {journal} {\bibinfo  {journal} {Journal of High Energy Astrophysics}\
  }\textbf {\bibinfo {volume} {34}},\ \bibinfo {pages} {49} (\bibinfo {year}
  {2022})}\BibitemShut {NoStop}%
\bibitem [{\citenamefont {Chevallier}\ and\ \citenamefont
  {Polarski}(2001)}]{darkenergy1}%
  \BibitemOpen
  \bibfield  {author} {\bibinfo {author} {\bibfnamefont {M.}~\bibnamefont
  {Chevallier}}\ and\ \bibinfo {author} {\bibfnamefont {D.}~\bibnamefont
  {Polarski}},\ }\bibfield  {title} {\bibinfo {title} {Accelerating universes
  with scaling dark matter},\ }\href
  {https://doi.org/10.1142/S0218271801000822} {\bibfield  {journal} {\bibinfo
  {journal} {International Journal of Modern Physics D}\ }\textbf {\bibinfo
  {volume} {10}},\ \bibinfo {pages} {213} (\bibinfo {year} {2001})}\BibitemShut
  {NoStop}%
\bibitem [{\citenamefont {Linder}(2003)}]{darkenergy2}%
  \BibitemOpen
  \bibfield  {author} {\bibinfo {author} {\bibfnamefont {E.~V.}\ \bibnamefont
  {Linder}},\ }\bibfield  {title} {\bibinfo {title} {Exploring the expansion
  history of the {Universe}},\ }\href
  {https://doi.org/10.1103/PhysRevLett.90.091301} {\bibfield  {journal}
  {\bibinfo  {journal} {Phys. Rev. Lett.}\ }\textbf {\bibinfo {volume} {90}},\
  \bibinfo {pages} {091301} (\bibinfo {year} {2003})}\BibitemShut {NoStop}%
\bibitem [{\citenamefont {Wetterich}(1988)}]{quintessence_exp_pot1}%
  \BibitemOpen
  \bibfield  {author} {\bibinfo {author} {\bibfnamefont {C.}~\bibnamefont
  {Wetterich}},\ }\bibfield  {title} {\bibinfo {title} {Cosmology and the fate
  of dilatation symmetry},\ }\href
  {https://doi.org/https://doi.org/10.1016/0550-3213(88)90193-9} {\bibfield
  {journal} {\bibinfo  {journal} {Nuclear Physics B}\ }\textbf {\bibinfo
  {volume} {302}},\ \bibinfo {pages} {668} (\bibinfo {year}
  {1988})}\BibitemShut {NoStop}%
\bibitem [{\citenamefont {Ratra}\ and\ \citenamefont
  {Peebles}(1988)}]{quintessence_exp_pot2}%
  \BibitemOpen
  \bibfield  {author} {\bibinfo {author} {\bibfnamefont {B.}~\bibnamefont
  {Ratra}}\ and\ \bibinfo {author} {\bibfnamefont {P.~J.~E.}\ \bibnamefont
  {Peebles}},\ }\bibfield  {title} {\bibinfo {title} {Cosmological consequences
  of a rolling homogeneous scalar field},\ }\href
  {https://doi.org/10.1103/PhysRevD.37.3406} {\bibfield  {journal} {\bibinfo
  {journal} {Phys. Rev. D}\ }\textbf {\bibinfo {volume} {37}},\ \bibinfo
  {pages} {3406} (\bibinfo {year} {1988})}\BibitemShut {NoStop}%
\bibitem [{\citenamefont {Hu}\ and\ \citenamefont
  {Sawicki}(2007)}]{Hu-Sawicki}%
  \BibitemOpen
  \bibfield  {author} {\bibinfo {author} {\bibfnamefont {W.}~\bibnamefont
  {Hu}}\ and\ \bibinfo {author} {\bibfnamefont {I.}~\bibnamefont {Sawicki}},\
  }\bibfield  {title} {\bibinfo {title} {Models of {$f(R)$} cosmic acceleration
  that evade solar system tests},\ }\href
  {https://doi.org/10.1103/PhysRevD.76.064004} {\bibfield  {journal} {\bibinfo
  {journal} {Phys. Rev. D}\ }\textbf {\bibinfo {volume} {76}},\ \bibinfo
  {pages} {064004} (\bibinfo {year} {2007})}\BibitemShut {NoStop}%
\bibitem [{\citenamefont {Simon}\ \emph {et~al.}(2005)\citenamefont {Simon},
  \citenamefont {Verde},\ and\ \citenamefont {Jimenez}}]{CC1}%
  \BibitemOpen
  \bibfield  {author} {\bibinfo {author} {\bibfnamefont {J.}~\bibnamefont
  {Simon}}, \bibinfo {author} {\bibfnamefont {L.}~\bibnamefont {Verde}},\ and\
  \bibinfo {author} {\bibfnamefont {R.}~\bibnamefont {Jimenez}},\ }\bibfield
  {title} {\bibinfo {title} {Constraints on the redshift dependence of the dark
  energy potential},\ }\href {https://doi.org/10.1103/PhysRevD.71.123001}
  {\bibfield  {journal} {\bibinfo  {journal} {Phys. Rev. D}\ }\textbf {\bibinfo
  {volume} {71}},\ \bibinfo {pages} {123001} (\bibinfo {year}
  {2005})}\BibitemShut {NoStop}%
\bibitem [{\citenamefont {Stern}\ \emph {et~al.}(2010)\citenamefont {Stern},
  \citenamefont {Jimenez}, \citenamefont {Verde}, \citenamefont
  {Kamionkowski},\ and\ \citenamefont {Stanford}}]{CC2}%
  \BibitemOpen
  \bibfield  {author} {\bibinfo {author} {\bibfnamefont {D.}~\bibnamefont
  {Stern}}, \bibinfo {author} {\bibfnamefont {R.}~\bibnamefont {Jimenez}},
  \bibinfo {author} {\bibfnamefont {L.}~\bibnamefont {Verde}}, \bibinfo
  {author} {\bibfnamefont {M.}~\bibnamefont {Kamionkowski}},\ and\ \bibinfo
  {author} {\bibfnamefont {S.~A.}\ \bibnamefont {Stanford}},\ }\bibfield
  {title} {\bibinfo {title} {Cosmic chronometers: constraining the equation of
  state of dark energy. {I}:{$H\left(z\right)$} measurements},\ }\href
  {https://doi.org/10.1088/1475-7516/2010/02/008} {\bibfield  {journal}
  {\bibinfo  {journal} {Journal of Cosmology and Astroparticle Physics}\
  }\textbf {\bibinfo {volume} {2010}}\bibinfo  {number} { (02)},\ \bibinfo
  {pages} {008}}\BibitemShut {NoStop}%
\bibitem [{\citenamefont {{Moresco}}\ \emph {et~al.}(2012)\citenamefont
  {{Moresco}}, \citenamefont {{Cimatti}}, \citenamefont {{Jimenez}},
  \citenamefont {{Pozzetti}}, \citenamefont {{Zamorani}}, \citenamefont
  {{Bolzonella}}, \citenamefont {{Dunlop}}, \citenamefont {{Lamareille}},
  \citenamefont {{Mignoli}}, \citenamefont {{Pearce}} \emph {et~al.}}]{CC3}%
  \BibitemOpen
\bibfield  {number} {  }\bibfield  {author} {\bibinfo {author} {\bibfnamefont
  {M.}~\bibnamefont {{Moresco}}}, \bibinfo {author} {\bibfnamefont
  {A.}~\bibnamefont {{Cimatti}}}, \bibinfo {author} {\bibfnamefont
  {R.}~\bibnamefont {{Jimenez}}}, \bibinfo {author} {\bibfnamefont
  {L.}~\bibnamefont {{Pozzetti}}}, \bibinfo {author} {\bibfnamefont
  {G.}~\bibnamefont {{Zamorani}}}, \bibinfo {author} {\bibfnamefont
  {M.}~\bibnamefont {{Bolzonella}}}, \bibinfo {author} {\bibfnamefont
  {J.}~\bibnamefont {{Dunlop}}}, \bibinfo {author} {\bibfnamefont
  {F.}~\bibnamefont {{Lamareille}}}, \bibinfo {author} {\bibfnamefont
  {M.}~\bibnamefont {{Mignoli}}}, \bibinfo {author} {\bibfnamefont
  {H.}~\bibnamefont {{Pearce}}}, \emph {et~al.},\ }\bibfield  {title} {\bibinfo
  {title} {{Improved constraints on the expansion rate of the Universe up to $z
  \sim 1.1$ from the spectroscopic evolution of cosmic chronometers}},\ }\href
  {https://doi.org/10.1088/1475-7516/2012/08/006} {\bibfield  {journal}
  {\bibinfo  {journal} {Journal of Cosmology and Astroparticle Physics}\
  }\textbf {\bibinfo {volume} {2012}}\bibinfo  {number} { (8)},\ \bibinfo {eid}
  {006}}\BibitemShut {NoStop}%
\bibitem [{\citenamefont {{Zhang}}\ \emph {et~al.}(2014)\citenamefont
  {{Zhang}}, \citenamefont {{Zhang}}, \citenamefont {{Yuan}}, \citenamefont
  {{Liu}}, \citenamefont {{Zhang}},\ and\ \citenamefont {{Sun}}}]{CC4}%
  \BibitemOpen
\bibfield  {number} {  }\bibfield  {author} {\bibinfo {author} {\bibfnamefont
  {C.}~\bibnamefont {{Zhang}}}, \bibinfo {author} {\bibfnamefont
  {H.}~\bibnamefont {{Zhang}}}, \bibinfo {author} {\bibfnamefont
  {S.}~\bibnamefont {{Yuan}}}, \bibinfo {author} {\bibfnamefont
  {S.}~\bibnamefont {{Liu}}}, \bibinfo {author} {\bibfnamefont {T.-J.}\
  \bibnamefont {{Zhang}}},\ and\ \bibinfo {author} {\bibfnamefont {Y.-C.}\
  \bibnamefont {{Sun}}},\ }\bibfield  {title} {\bibinfo {title} {{Four new
  observational $H\left(z\right)$ data from luminous red galaxies in the Sloan
  digital sky survey data release seven}},\ }\href
  {https://doi.org/10.1088/1674-4527/14/10/002} {\bibfield  {journal} {\bibinfo
   {journal} {Research in Astronomy and Astrophysics}\ }\textbf {\bibinfo
  {volume} {14}},\ \bibinfo {eid} {1221-1233} (\bibinfo {year}
  {2014})}\BibitemShut {NoStop}%
\bibitem [{\citenamefont {{Moresco}}(2015)}]{CC5}%
  \BibitemOpen
  \bibfield  {author} {\bibinfo {author} {\bibfnamefont {M.}~\bibnamefont
  {{Moresco}}},\ }\bibfield  {title} {\bibinfo {title} {{Raising the bar: new
  constraints on the {Hubble} parameter with cosmic chronometers at $z \sim
  2$.}},\ }\href {https://doi.org/10.1093/mnrasl/slv037} {\bibfield  {journal}
  {\bibinfo  {journal} {Monthly Notices Royal Astronomial Society}\ }\textbf
  {\bibinfo {volume} {450}},\ \bibinfo {pages} {L16} (\bibinfo {year}
  {2015})}\BibitemShut {NoStop}%
\bibitem [{\citenamefont {{Moresco}}\ \emph {et~al.}(2016)\citenamefont
  {{Moresco}}, \citenamefont {{Pozzetti}}, \citenamefont {{Cimatti}},
  \citenamefont {{Jimenez}}, \citenamefont {{Maraston}}, \citenamefont
  {{Verde}}, \citenamefont {{Thomas}}, \citenamefont {{Citro}}, \citenamefont
  {{Tojeiro}},\ and\ \citenamefont {{Wilkinson}}}]{CC6}%
  \BibitemOpen
  \bibfield  {author} {\bibinfo {author} {\bibfnamefont {M.}~\bibnamefont
  {{Moresco}}}, \bibinfo {author} {\bibfnamefont {L.}~\bibnamefont
  {{Pozzetti}}}, \bibinfo {author} {\bibfnamefont {A.}~\bibnamefont
  {{Cimatti}}}, \bibinfo {author} {\bibfnamefont {R.}~\bibnamefont
  {{Jimenez}}}, \bibinfo {author} {\bibfnamefont {C.}~\bibnamefont
  {{Maraston}}}, \bibinfo {author} {\bibfnamefont {L.}~\bibnamefont {{Verde}}},
  \bibinfo {author} {\bibfnamefont {D.}~\bibnamefont {{Thomas}}}, \bibinfo
  {author} {\bibfnamefont {A.}~\bibnamefont {{Citro}}}, \bibinfo {author}
  {\bibfnamefont {R.}~\bibnamefont {{Tojeiro}}},\ and\ \bibinfo {author}
  {\bibfnamefont {D.}~\bibnamefont {{Wilkinson}}},\ }\bibfield  {title}
  {\bibinfo {title} {{A $6\%$ measurement of the Hubble parameter at
  $z\sim0.45$: Direct evidence of the epoch of cosmic re-acceleration}},\
  }\href {https://doi.org/10.1088/1475-7516/2016/05/014} {\bibfield  {journal}
  {\bibinfo  {journal} {Journal of Cosmology and Astroparticle Physics}\
  }\textbf {\bibinfo {volume} {2016}}\bibinfo  {number} { (5)},\ \bibinfo {eid}
  {014}}\BibitemShut {NoStop}%
\bibitem [{\citenamefont {Scolnic}\ \emph {et~al.}(2018)\citenamefont
  {Scolnic}, \citenamefont {Jones}, \citenamefont {Rest}, \citenamefont {Pan},
  \citenamefont {Chornock}, \citenamefont {Foley}, \citenamefont {Huber},
  \citenamefont {Kessler}, \citenamefont {Narayan}, \citenamefont {Riess} \emph
  {et~al.}}]{SnIa_pantheon}%
  \BibitemOpen
\bibfield  {number} {  }\bibfield  {author} {\bibinfo {author} {\bibfnamefont
  {D.~M.}\ \bibnamefont {Scolnic}}, \bibinfo {author} {\bibfnamefont {D.~O.}\
  \bibnamefont {Jones}}, \bibinfo {author} {\bibfnamefont {A.}~\bibnamefont
  {Rest}}, \bibinfo {author} {\bibfnamefont {Y.~C.}\ \bibnamefont {Pan}},
  \bibinfo {author} {\bibfnamefont {R.}~\bibnamefont {Chornock}}, \bibinfo
  {author} {\bibfnamefont {R.~J.}\ \bibnamefont {Foley}}, \bibinfo {author}
  {\bibfnamefont {M.~E.}\ \bibnamefont {Huber}}, \bibinfo {author}
  {\bibfnamefont {R.}~\bibnamefont {Kessler}}, \bibinfo {author} {\bibfnamefont
  {G.}~\bibnamefont {Narayan}}, \bibinfo {author} {\bibfnamefont {A.~G.}\
  \bibnamefont {Riess}}, \emph {et~al.},\ }\bibfield  {title} {\bibinfo {title}
  {The complete light-curve sample of spectroscopically confirmed {SNe} {Ia}
  from {Pan}-{STARRS}1 and cosmological constraints from the combined pantheon
  sample},\ }\href {https://doi.org/10.3847/1538-4357/aab9bb} {\bibfield
  {journal} {\bibinfo  {journal} {The Astrophysical Journal}\ }\textbf
  {\bibinfo {volume} {859}},\ \bibinfo {pages} {101} (\bibinfo {year}
  {2018})}\BibitemShut {NoStop}%
\bibitem [{\citenamefont {Ross}\ \emph {et~al.}(2015)\citenamefont {Ross},
  \citenamefont {Samushia}, \citenamefont {Howlett}, \citenamefont {Percival},
  \citenamefont {Burden},\ and\ \citenamefont {Manera}}]{SDSS_DR7}%
  \BibitemOpen
  \bibfield  {author} {\bibinfo {author} {\bibfnamefont {A.~J.}\ \bibnamefont
  {Ross}}, \bibinfo {author} {\bibfnamefont {L.}~\bibnamefont {Samushia}},
  \bibinfo {author} {\bibfnamefont {C.}~\bibnamefont {Howlett}}, \bibinfo
  {author} {\bibfnamefont {W.~J.}\ \bibnamefont {Percival}}, \bibinfo {author}
  {\bibfnamefont {A.}~\bibnamefont {Burden}},\ and\ \bibinfo {author}
  {\bibfnamefont {M.}~\bibnamefont {Manera}},\ }\bibfield  {title} {\bibinfo
  {title} {{The clustering of the SDSS DR7 main Galaxy sample – I. A 4 per
  cent distance measure at $z=0.15$}},\ }\href
  {https://doi.org/10.1093/mnras/stv154} {\bibfield  {journal} {\bibinfo
  {journal} {Monthly Notices of the Royal Astronomical Society}\ }\textbf
  {\bibinfo {volume} {449}},\ \bibinfo {pages} {835} (\bibinfo {year}
  {2015})}\BibitemShut {NoStop}%
\bibitem [{\citenamefont {Kazin}\ \emph {et~al.}(2014)\citenamefont {Kazin},
  \citenamefont {Koda}, \citenamefont {Blake}, \citenamefont {Padmanabhan},
  \citenamefont {Brough}, \citenamefont {Colless}, \citenamefont {Contreras},
  \citenamefont {Couch}, \citenamefont {Croom}, \citenamefont {Croton} \emph
  {et~al.}}]{WiggleZ}%
  \BibitemOpen
  \bibfield  {author} {\bibinfo {author} {\bibfnamefont {E.~A.}\ \bibnamefont
  {Kazin}}, \bibinfo {author} {\bibfnamefont {J.}~\bibnamefont {Koda}},
  \bibinfo {author} {\bibfnamefont {C.}~\bibnamefont {Blake}}, \bibinfo
  {author} {\bibfnamefont {N.}~\bibnamefont {Padmanabhan}}, \bibinfo {author}
  {\bibfnamefont {S.}~\bibnamefont {Brough}}, \bibinfo {author} {\bibfnamefont
  {M.}~\bibnamefont {Colless}}, \bibinfo {author} {\bibfnamefont
  {C.}~\bibnamefont {Contreras}}, \bibinfo {author} {\bibfnamefont
  {W.}~\bibnamefont {Couch}}, \bibinfo {author} {\bibfnamefont
  {S.}~\bibnamefont {Croom}}, \bibinfo {author} {\bibfnamefont {D.~J.}\
  \bibnamefont {Croton}}, \emph {et~al.},\ }\bibfield  {title} {\bibinfo
  {title} {{The WiggleZ dark energy survey: improved distance measurements to
  $z=1$ with reconstruction of the baryonic acoustic feature}},\ }\href
  {https://doi.org/10.1093/mnras/stu778} {\bibfield  {journal} {\bibinfo
  {journal} {Monthly Notices of the Royal Astronomical Society}\ }\textbf
  {\bibinfo {volume} {441}},\ \bibinfo {pages} {3524} (\bibinfo {year}
  {2014})}\BibitemShut {NoStop}%
\bibitem [{\citenamefont {Ata}\ \emph {et~al.}(2017)\citenamefont {Ata},
  \citenamefont {Baumgarten}, \citenamefont {Bautista}, \citenamefont
  {Beutler}, \citenamefont {Bizyaev}, \citenamefont {Blanton}, \citenamefont
  {Blazek}, \citenamefont {Bolton}, \citenamefont {Brinkmann}, \citenamefont
  {Brownstein} \emph {et~al.}}]{SDSS-IV_quasars}%
  \BibitemOpen
  \bibfield  {author} {\bibinfo {author} {\bibfnamefont {M.}~\bibnamefont
  {Ata}}, \bibinfo {author} {\bibfnamefont {F.}~\bibnamefont {Baumgarten}},
  \bibinfo {author} {\bibfnamefont {J.}~\bibnamefont {Bautista}}, \bibinfo
  {author} {\bibfnamefont {F.}~\bibnamefont {Beutler}}, \bibinfo {author}
  {\bibfnamefont {D.}~\bibnamefont {Bizyaev}}, \bibinfo {author} {\bibfnamefont
  {M.~R.}\ \bibnamefont {Blanton}}, \bibinfo {author} {\bibfnamefont {J.~A.}\
  \bibnamefont {Blazek}}, \bibinfo {author} {\bibfnamefont {A.~S.}\
  \bibnamefont {Bolton}}, \bibinfo {author} {\bibfnamefont {J.}~\bibnamefont
  {Brinkmann}}, \bibinfo {author} {\bibfnamefont {J.~R.}\ \bibnamefont
  {Brownstein}}, \emph {et~al.},\ }\bibfield  {title} {\bibinfo {title} {{The
  clustering of the SDSS-IV extended Baryon Oscillation Spectroscopic Survey
  DR14 quasar sample: First measurement of baryon acoustic oscillations between
  redshift 0.8 and 2.2}},\ }\href {https://doi.org/10.1093/mnras/stx2630}
  {\bibfield  {journal} {\bibinfo  {journal} {Monthly Notices of the Royal
  Astronomical Society}\ }\textbf {\bibinfo {volume} {473}},\ \bibinfo {pages}
  {4773} (\bibinfo {year} {2017})}\BibitemShut {NoStop}%
\bibitem [{\citenamefont {Abbott}\ \emph {et~al.}(2018)\citenamefont {Abbott},
  \citenamefont {Abdalla}, \citenamefont {Alarcon}, \citenamefont {Allam},
  \citenamefont {Andrade-Oliveira}, \citenamefont {Annis}, \citenamefont
  {Avila}, \citenamefont {Banerji}, \citenamefont {Banik}, \citenamefont
  {Bechtol} \emph {et~al.}}]{DES_Y1}%
  \BibitemOpen
  \bibfield  {author} {\bibinfo {author} {\bibfnamefont {T.~M.~C.}\
  \bibnamefont {Abbott}}, \bibinfo {author} {\bibfnamefont {F.~B.}\
  \bibnamefont {Abdalla}}, \bibinfo {author} {\bibfnamefont {A.}~\bibnamefont
  {Alarcon}}, \bibinfo {author} {\bibfnamefont {S.}~\bibnamefont {Allam}},
  \bibinfo {author} {\bibfnamefont {F.}~\bibnamefont {Andrade-Oliveira}},
  \bibinfo {author} {\bibfnamefont {J.}~\bibnamefont {Annis}}, \bibinfo
  {author} {\bibfnamefont {S.}~\bibnamefont {Avila}}, \bibinfo {author}
  {\bibfnamefont {M.}~\bibnamefont {Banerji}}, \bibinfo {author} {\bibfnamefont
  {N.}~\bibnamefont {Banik}}, \bibinfo {author} {\bibfnamefont
  {K.}~\bibnamefont {Bechtol}}, \emph {et~al.},\ }\bibfield  {title} {\bibinfo
  {title} {{Dark energy survey year 1 results: measurement of the baryon
  acoustic oscillation scale in the distribution of galaxies to redshift 1}},\
  }\href {https://doi.org/10.1093/mnras/sty3351} {\bibfield  {journal}
  {\bibinfo  {journal} {Monthly Notices of the Royal Astronomical Society}\
  }\textbf {\bibinfo {volume} {483}},\ \bibinfo {pages} {4866} (\bibinfo {year}
  {2018})}\BibitemShut {NoStop}%
\bibitem [{\citenamefont {Alam}\ \emph {et~al.}(2017)\citenamefont {Alam},
  \citenamefont {Ata}, \citenamefont {Bailey}, \citenamefont {Beutler},
  \citenamefont {Bizyaev}, \citenamefont {Blazek}, \citenamefont {Bolton},
  \citenamefont {Brownstein}, \citenamefont {Burden}, \citenamefont {Chuang}
  \emph {et~al.}}]{BOSS}%
  \BibitemOpen
  \bibfield  {author} {\bibinfo {author} {\bibfnamefont {S.}~\bibnamefont
  {Alam}}, \bibinfo {author} {\bibfnamefont {M.}~\bibnamefont {Ata}}, \bibinfo
  {author} {\bibfnamefont {S.}~\bibnamefont {Bailey}}, \bibinfo {author}
  {\bibfnamefont {F.}~\bibnamefont {Beutler}}, \bibinfo {author} {\bibfnamefont
  {D.}~\bibnamefont {Bizyaev}}, \bibinfo {author} {\bibfnamefont {J.~A.}\
  \bibnamefont {Blazek}}, \bibinfo {author} {\bibfnamefont {A.~S.}\
  \bibnamefont {Bolton}}, \bibinfo {author} {\bibfnamefont {J.~R.}\
  \bibnamefont {Brownstein}}, \bibinfo {author} {\bibfnamefont
  {A.}~\bibnamefont {Burden}}, \bibinfo {author} {\bibfnamefont {C.-H.}\
  \bibnamefont {Chuang}}, \emph {et~al.},\ }\bibfield  {title} {\bibinfo
  {title} {{The clustering of galaxies in the completed SDSS-III Baryon
  Oscillation Spectroscopic Survey: Cosmological analysis of the DR12 galaxy
  sample}},\ }\href {https://doi.org/10.1093/mnras/stx721} {\bibfield
  {journal} {\bibinfo  {journal} {Monthly Notices of the Royal Astronomical
  Society}\ }\textbf {\bibinfo {volume} {470}},\ \bibinfo {pages} {2617}
  (\bibinfo {year} {2017})}\BibitemShut {NoStop}%
\bibitem [{\citenamefont {Bautista}\ \emph {et~al.}(2020)\citenamefont
  {Bautista}, \citenamefont {Paviot}, \citenamefont {Vargas Magaña},
  \citenamefont {de la Torre}, \citenamefont {Fromenteau}, \citenamefont
  {Gil-Marín}, \citenamefont {Ross}, \citenamefont {Burtin}, \citenamefont
  {Dawson}, \citenamefont {Hou} \emph {et~al.}}]{SDSS_IV_LRG_anisotropicCF}%
  \BibitemOpen
  \bibfield  {author} {\bibinfo {author} {\bibfnamefont {J.~E.}\ \bibnamefont
  {Bautista}}, \bibinfo {author} {\bibfnamefont {R.}~\bibnamefont {Paviot}},
  \bibinfo {author} {\bibfnamefont {M.}~\bibnamefont {Vargas Magaña}},
  \bibinfo {author} {\bibfnamefont {S.}~\bibnamefont {de la Torre}}, \bibinfo
  {author} {\bibfnamefont {S.}~\bibnamefont {Fromenteau}}, \bibinfo {author}
  {\bibfnamefont {H.}~\bibnamefont {Gil-Marín}}, \bibinfo {author}
  {\bibfnamefont {A.~J.}\ \bibnamefont {Ross}}, \bibinfo {author}
  {\bibfnamefont {E.}~\bibnamefont {Burtin}}, \bibinfo {author} {\bibfnamefont
  {K.~S.}\ \bibnamefont {Dawson}}, \bibinfo {author} {\bibfnamefont
  {J.}~\bibnamefont {Hou}}, \emph {et~al.},\ }\bibfield  {title} {\bibinfo
  {title} {{The completed SDSS-IV extended Baryon Oscillation Spectroscopic
  Survey: measurement of the BAO and growth rate of structure of the luminous
  red galaxy sample from the anisotropic correlation function between redshifts
  0.6 and 1}},\ }\href {https://doi.org/10.1093/mnras/staa2800} {\bibfield
  {journal} {\bibinfo  {journal} {Monthly Notices of the Royal Astronomical
  Society}\ }\textbf {\bibinfo {volume} {500}},\ \bibinfo {pages} {736}
  (\bibinfo {year} {2020})}\BibitemShut {NoStop}%
\bibitem [{\citenamefont {Neveux}\ \emph {et~al.}(2020)\citenamefont {Neveux},
  \citenamefont {Burtin}, \citenamefont {de Mattia}, \citenamefont {Smith},
  \citenamefont {Ross}, \citenamefont {Hou}, \citenamefont {Bautista},
  \citenamefont {Brinkmann}, \citenamefont {Chuang}, \citenamefont {Dawson}
  \emph {et~al.}}]{SDSS_IV_QSO_anisotropicPS}%
  \BibitemOpen
  \bibfield  {author} {\bibinfo {author} {\bibfnamefont {R.}~\bibnamefont
  {Neveux}}, \bibinfo {author} {\bibfnamefont {E.}~\bibnamefont {Burtin}},
  \bibinfo {author} {\bibfnamefont {A.}~\bibnamefont {de Mattia}}, \bibinfo
  {author} {\bibfnamefont {A.}~\bibnamefont {Smith}}, \bibinfo {author}
  {\bibfnamefont {A.~J.}\ \bibnamefont {Ross}}, \bibinfo {author}
  {\bibfnamefont {J.}~\bibnamefont {Hou}}, \bibinfo {author} {\bibfnamefont
  {J.}~\bibnamefont {Bautista}}, \bibinfo {author} {\bibfnamefont
  {J.}~\bibnamefont {Brinkmann}}, \bibinfo {author} {\bibfnamefont {C.-H.}\
  \bibnamefont {Chuang}}, \bibinfo {author} {\bibfnamefont {K.~S.}\
  \bibnamefont {Dawson}}, \emph {et~al.},\ }\bibfield  {title} {\bibinfo
  {title} {{The completed SDSS-IV extended Baryon Oscillation Spectroscopic
  Survey: BAO and RSD measurements from the anisotropic power spectrum of the
  quasar sample between redshift 0.8 and 2.2}},\ }\href
  {https://doi.org/10.1093/mnras/staa2780} {\bibfield  {journal} {\bibinfo
  {journal} {Monthly Notices of the Royal Astronomical Society}\ }\textbf
  {\bibinfo {volume} {499}},\ \bibinfo {pages} {210} (\bibinfo {year}
  {2020})}\BibitemShut {NoStop}%
\bibitem [{\citenamefont {{Julian E. Bautista,}}\ \emph
  {et~al.}(2017)\citenamefont {{Julian E. Bautista,}}, \citenamefont
  {{Nicol\'as G. Busca,}}, \citenamefont {{Julien Guy,}}, \citenamefont {{James
  Rich,}}, \citenamefont {{Michael Blomqvist,}}, \citenamefont {{H\'elion du
  Mas des Bourboux,}}, \citenamefont {{Matthew M. Pieri,}}, \citenamefont
  {{Andreu Font-Ribera,}}, \citenamefont {{Stephen Bailey,}}, \citenamefont
  {{Timoth\'ee Delubac}} \emph {et~al.}}]{SDSS-III_La_forests}%
  \BibitemOpen
  \bibfield  {author} {\bibinfo {author} {\bibnamefont {{Julian E.
  Bautista,}}}, \bibinfo {author} {\bibnamefont {{Nicol\'as G. Busca,}}},
  \bibinfo {author} {\bibnamefont {{Julien Guy,}}}, \bibinfo {author}
  {\bibnamefont {{James Rich,}}}, \bibinfo {author} {\bibnamefont {{Michael
  Blomqvist,}}}, \bibinfo {author} {\bibnamefont {{H\'elion du Mas des
  Bourboux,}}}, \bibinfo {author} {\bibnamefont {{Matthew M. Pieri,}}},
  \bibinfo {author} {\bibnamefont {{Andreu Font-Ribera,}}}, \bibinfo {author}
  {\bibnamefont {{Stephen Bailey,}}}, \bibinfo {author} {\bibnamefont
  {{Timoth\'ee Delubac}}}, \emph {et~al.},\ }\bibfield  {title} {\bibinfo
  {title} {Measurement of baryon acoustic oscillation correlations at $z$ with
  {SDSS} {DR12} {Lyα-Forests}},\ }\href
  {https://doi.org/10.1051/0004-6361/201730533} {\bibfield  {journal} {\bibinfo
   {journal} {A\&A}\ }\textbf {\bibinfo {volume} {603}},\ \bibinfo {pages}
  {A12} (\bibinfo {year} {2017})}\BibitemShut {NoStop}%
\bibitem [{\citenamefont {{H\'elion du Mas des Bourboux,}}\ \emph
  {et~al.}(2017)\citenamefont {{H\'elion du Mas des Bourboux,}}, \citenamefont
  {{Jean-Marc Le Goff,}}, \citenamefont {{Michael Blomqvist,}}, \citenamefont
  {{Nicol\'as G. Busca,}}, \citenamefont {{Julien Guy,}}, \citenamefont {{James
  Rich,}}, \citenamefont {{Christophe Y\`eche,}}, \citenamefont {{Julian E.
  Bautista,}}, \citenamefont {{\'Etienne Burtin,}}, \citenamefont {{Kyle S.
  Dawson}} \emph {et~al.}}]{La_forests_quasars_cross}%
  \BibitemOpen
  \bibfield  {author} {\bibinfo {author} {\bibnamefont {{H\'elion du Mas des
  Bourboux,}}}, \bibinfo {author} {\bibnamefont {{Jean-Marc Le Goff,}}},
  \bibinfo {author} {\bibnamefont {{Michael Blomqvist,}}}, \bibinfo {author}
  {\bibnamefont {{Nicol\'as G. Busca,}}}, \bibinfo {author} {\bibnamefont
  {{Julien Guy,}}}, \bibinfo {author} {\bibnamefont {{James Rich,}}}, \bibinfo
  {author} {\bibnamefont {{Christophe Y\`eche,}}}, \bibinfo {author}
  {\bibnamefont {{Julian E. Bautista,}}}, \bibinfo {author} {\bibnamefont
  {{\'Etienne Burtin,}}}, \bibinfo {author} {\bibnamefont {{Kyle S. Dawson}}},
  \emph {et~al.},\ }\bibfield  {title} {\bibinfo {title} {Baryon acoustic
  oscillations from the complete {SDSS}-{III} {Lyα}-quasar cross-correlation
  function at $z = 2.4$},\ }\href {https://doi.org/10.1051/0004-6361/201731731}
  {\bibfield  {journal} {\bibinfo  {journal} {A\&A}\ }\textbf {\bibinfo
  {volume} {608}},\ \bibinfo {pages} {A130} (\bibinfo {year}
  {2017})}\BibitemShut {NoStop}%
\bibitem [{\citenamefont {Caldwell}\ \emph {et~al.}(1998)\citenamefont
  {Caldwell}, \citenamefont {Dave},\ and\ \citenamefont
  {Steinhardt}}]{quintessence1}%
  \BibitemOpen
  \bibfield  {author} {\bibinfo {author} {\bibfnamefont {R.~R.}\ \bibnamefont
  {Caldwell}}, \bibinfo {author} {\bibfnamefont {R.}~\bibnamefont {Dave}},\
  and\ \bibinfo {author} {\bibfnamefont {P.~J.}\ \bibnamefont {Steinhardt}},\
  }\bibfield  {title} {\bibinfo {title} {Cosmological imprint of an energy
  component with general equation of state},\ }\href
  {https://doi.org/10.1103/PhysRevLett.80.1582} {\bibfield  {journal} {\bibinfo
   {journal} {Phys. Rev. Lett.}\ }\textbf {\bibinfo {volume} {80}},\ \bibinfo
  {pages} {1582} (\bibinfo {year} {1998})}\BibitemShut {NoStop}%
\bibitem [{\citenamefont {Zlatev}\ \emph {et~al.}(1999)\citenamefont {Zlatev},
  \citenamefont {Wang},\ and\ \citenamefont {Steinhardt}}]{quintessence2}%
  \BibitemOpen
  \bibfield  {author} {\bibinfo {author} {\bibfnamefont {I.}~\bibnamefont
  {Zlatev}}, \bibinfo {author} {\bibfnamefont {L.}~\bibnamefont {Wang}},\ and\
  \bibinfo {author} {\bibfnamefont {P.~J.}\ \bibnamefont {Steinhardt}},\
  }\bibfield  {title} {\bibinfo {title} {Quintessence, cosmic coincidence, and
  the cosmological constant},\ }\href
  {https://doi.org/10.1103/PhysRevLett.82.896} {\bibfield  {journal} {\bibinfo
  {journal} {Phys. Rev. Lett.}\ }\textbf {\bibinfo {volume} {82}},\ \bibinfo
  {pages} {896} (\bibinfo {year} {1999})}\BibitemShut {NoStop}%
\bibitem [{\citenamefont {Armendariz-Picon}\ \emph {et~al.}(2000)\citenamefont
  {Armendariz-Picon}, \citenamefont {Mukhanov},\ and\ \citenamefont
  {Steinhardt}}]{quintessence3}%
  \BibitemOpen
  \bibfield  {author} {\bibinfo {author} {\bibfnamefont {C.}~\bibnamefont
  {Armendariz-Picon}}, \bibinfo {author} {\bibfnamefont {V.}~\bibnamefont
  {Mukhanov}},\ and\ \bibinfo {author} {\bibfnamefont {P.~J.}\ \bibnamefont
  {Steinhardt}},\ }\bibfield  {title} {\bibinfo {title} {Dynamical solution to
  the problem of a small cosmological constant and late-time cosmic
  acceleration},\ }\href {https://doi.org/10.1103/PhysRevLett.85.4438}
  {\bibfield  {journal} {\bibinfo  {journal} {Phys. Rev. Lett.}\ }\textbf
  {\bibinfo {volume} {85}},\ \bibinfo {pages} {4438} (\bibinfo {year}
  {2000})}\BibitemShut {NoStop}%
\bibitem [{\citenamefont {Copeland}\ \emph {et~al.}(1998)\citenamefont
  {Copeland}, \citenamefont {Liddle},\ and\ \citenamefont
  {Wands}}]{var_change_quintessence}%
  \BibitemOpen
  \bibfield  {author} {\bibinfo {author} {\bibfnamefont {E.~J.}\ \bibnamefont
  {Copeland}}, \bibinfo {author} {\bibfnamefont {A.~R.}\ \bibnamefont
  {Liddle}},\ and\ \bibinfo {author} {\bibfnamefont {D.}~\bibnamefont
  {Wands}},\ }\bibfield  {title} {\bibinfo {title} {Exponential potentials and
  cosmological scaling solutions},\ }\href
  {https://doi.org/10.1103/PhysRevD.57.4686} {\bibfield  {journal} {\bibinfo
  {journal} {Phys. Rev. D}\ }\textbf {\bibinfo {volume} {57}},\ \bibinfo
  {pages} {4686} (\bibinfo {year} {1998})}\BibitemShut {NoStop}%
\bibitem [{\citenamefont {Clifton}\ \emph {et~al.}(2012)\citenamefont
  {Clifton}, \citenamefont {Ferreira}, \citenamefont {Padilla},\ and\
  \citenamefont {Skordis}}]{ReviewMOG}%
  \BibitemOpen
  \bibfield  {author} {\bibinfo {author} {\bibfnamefont {T.}~\bibnamefont
  {Clifton}}, \bibinfo {author} {\bibfnamefont {P.~G.}\ \bibnamefont
  {Ferreira}}, \bibinfo {author} {\bibfnamefont {A.}~\bibnamefont {Padilla}},\
  and\ \bibinfo {author} {\bibfnamefont {C.}~\bibnamefont {Skordis}},\
  }\bibfield  {title} {\bibinfo {title} {Modified gravity and cosmology},\
  }\href {https://doi.org/https://doi.org/10.1016/j.physrep.2012.01.001}
  {\bibfield  {journal} {\bibinfo  {journal} {Physics Reports}\ }\textbf
  {\bibinfo {volume} {513}},\ \bibinfo {pages} {1} (\bibinfo {year}
  {2012})}\BibitemShut {NoStop}%
\bibitem [{\citenamefont {{Buchdahl}}(1970)}]{first_f_R_paper}%
  \BibitemOpen
  \bibfield  {author} {\bibinfo {author} {\bibfnamefont {H.~A.}\ \bibnamefont
  {{Buchdahl}}},\ }\bibfield  {title} {\bibinfo {title} {{Non-linear
  Lagrangians and cosmological theory}},\ }\href
  {https://doi.org/10.1093/mnras/150.1.1} {\bibfield  {journal} {\bibinfo
  {journal} {Mon. Not. R. Astron. Soc.}\ }\textbf {\bibinfo {volume} {150}},\
  \bibinfo {pages} {1} (\bibinfo {year} {1970})}\BibitemShut {NoStop}%
\bibitem [{\citenamefont {Brax}\ \emph {et~al.}(2008)\citenamefont {Brax},
  \citenamefont {van~de Bruck}, \citenamefont {Davis},\ and\ \citenamefont
  {Shaw}}]{Brax2008}%
  \BibitemOpen
  \bibfield  {author} {\bibinfo {author} {\bibfnamefont {P.}~\bibnamefont
  {Brax}}, \bibinfo {author} {\bibfnamefont {C.}~\bibnamefont {van~de Bruck}},
  \bibinfo {author} {\bibfnamefont {A.-C.}\ \bibnamefont {Davis}},\ and\
  \bibinfo {author} {\bibfnamefont {D.~J.}\ \bibnamefont {Shaw}},\ }\bibfield
  {title} {\bibinfo {title} {{$f(R)$} gravity and chameleon theories},\ }\href
  {https://doi.org/10.1103/PhysRevD.78.104021} {\bibfield  {journal} {\bibinfo
  {journal} {Phys. Rev. D}\ }\textbf {\bibinfo {volume} {78}},\ \bibinfo
  {pages} {104021} (\bibinfo {year} {2008})}\BibitemShut {NoStop}%
\bibitem [{\citenamefont {Hui}\ \emph {et~al.}(2009)\citenamefont {Hui},
  \citenamefont {Nicolis},\ and\ \citenamefont {Stubbs}}]{Hui2009}%
  \BibitemOpen
  \bibfield  {author} {\bibinfo {author} {\bibfnamefont {L.}~\bibnamefont
  {Hui}}, \bibinfo {author} {\bibfnamefont {A.}~\bibnamefont {Nicolis}},\ and\
  \bibinfo {author} {\bibfnamefont {C.~W.}\ \bibnamefont {Stubbs}},\ }\bibfield
   {title} {\bibinfo {title} {Equivalence principle implications of modified
  gravity models},\ }\href {https://doi.org/10.1103/PhysRevD.80.104002}
  {\bibfield  {journal} {\bibinfo  {journal} {Phys. Rev. D}\ }\textbf {\bibinfo
  {volume} {80}},\ \bibinfo {pages} {104002} (\bibinfo {year}
  {2009})}\BibitemShut {NoStop}%
\bibitem [{\citenamefont {Kandhai}\ and\ \citenamefont
  {Dunsby}(2015)}]{f_R_var}%
  \BibitemOpen
  \bibfield  {author} {\bibinfo {author} {\bibfnamefont {S.}~\bibnamefont
  {Kandhai}}\ and\ \bibinfo {author} {\bibfnamefont {P.~K.~S.}\ \bibnamefont
  {Dunsby}},\ }\href@noop {} {\bibinfo {title} {Cosmological dynamics of viable
  {$f(R)$} theories of gravity}} (\bibinfo {year} {2015}),\ \Eprint
  {https://arxiv.org/abs/1511.00101} {arXiv:1511.00101} \BibitemShut {NoStop}%
\bibitem [{\citenamefont {Liu}\ \emph {et~al.}(2022)\citenamefont {Liu},
  \citenamefont {Huang},\ and\ \citenamefont {Protopapas}}]{liu2022evaluating}%
  \BibitemOpen
  \bibfield  {author} {\bibinfo {author} {\bibfnamefont {S.}~\bibnamefont
  {Liu}}, \bibinfo {author} {\bibfnamefont {X.}~\bibnamefont {Huang}},\ and\
  \bibinfo {author} {\bibfnamefont {P.}~\bibnamefont {Protopapas}},\
  }\href@noop {} {\bibinfo {title} {Evaluating error bound for physics-informed
  neural networks on linear dynamical systems}} (\bibinfo {year} {2022}),\
  \Eprint {https://arxiv.org/abs/2207.01114} {arXiv:2207.01114} \BibitemShut
  {NoStop}%
\bibitem [{\citenamefont {Mattheakis}\ \emph {et~al.}(2022)\citenamefont
  {Mattheakis}, \citenamefont {Sondak}, \citenamefont {Dogra},\ and\
  \citenamefont {Protopapas}}]{marios_exp_reparam}%
  \BibitemOpen
  \bibfield  {author} {\bibinfo {author} {\bibfnamefont {M.}~\bibnamefont
  {Mattheakis}}, \bibinfo {author} {\bibfnamefont {D.}~\bibnamefont {Sondak}},
  \bibinfo {author} {\bibfnamefont {A.~S.}\ \bibnamefont {Dogra}},\ and\
  \bibinfo {author} {\bibfnamefont {P.}~\bibnamefont {Protopapas}},\ }\bibfield
   {title} {\bibinfo {title} {Hamiltonian neural networks for solving equations
  of motion},\ }\href {https://doi.org/10.1103/PhysRevE.105.065305} {\bibfield
  {journal} {\bibinfo  {journal} {Phys. Rev. E}\ }\textbf {\bibinfo {volume}
  {105}},\ \bibinfo {pages} {065305} (\bibinfo {year} {2022})}\BibitemShut
  {NoStop}%
\bibitem [{\citenamefont {Basilakos}\ \emph {et~al.}(2013)\citenamefont
  {Basilakos}, \citenamefont {Nesseris},\ and\ \citenamefont
  {Perivolaropoulos}}]{Basilakos}%
  \BibitemOpen
  \bibfield  {author} {\bibinfo {author} {\bibfnamefont {S.}~\bibnamefont
  {Basilakos}}, \bibinfo {author} {\bibfnamefont {S.}~\bibnamefont
  {Nesseris}},\ and\ \bibinfo {author} {\bibfnamefont {L.}~\bibnamefont
  {Perivolaropoulos}},\ }\bibfield  {title} {\bibinfo {title} {Observational
  constraints on viable {$f(R)$} parametrizations with geometrical and
  dynamical probes},\ }\href {https://doi.org/10.1103/PhysRevD.87.123529}
  {\bibfield  {journal} {\bibinfo  {journal} {Phys. Rev. D}\ }\textbf {\bibinfo
  {volume} {87}},\ \bibinfo {pages} {123529} (\bibinfo {year}
  {2013})}\BibitemShut {NoStop}%
\bibitem [{\citenamefont {Leizerovich}\ \emph {et~al.}(2022)\citenamefont
  {Leizerovich}, \citenamefont {Kraiselburd}, \citenamefont {Landau},\ and\
  \citenamefont {Sc\'occola}}]{Updated_Matias_paper}%
  \BibitemOpen
  \bibfield  {author} {\bibinfo {author} {\bibfnamefont {M.}~\bibnamefont
  {Leizerovich}}, \bibinfo {author} {\bibfnamefont {L.}~\bibnamefont
  {Kraiselburd}}, \bibinfo {author} {\bibfnamefont {S.}~\bibnamefont
  {Landau}},\ and\ \bibinfo {author} {\bibfnamefont {C.~G.}\ \bibnamefont
  {Sc\'occola}},\ }\bibfield  {title} {\bibinfo {title} {Testing {$f(R)$}
  gravity models with quasar x-ray and {UV} fluxes},\ }\href
  {https://doi.org/10.1103/PhysRevD.105.103526} {\bibfield  {journal} {\bibinfo
   {journal} {Phys. Rev. D}\ }\textbf {\bibinfo {volume} {105}},\ \bibinfo
  {pages} {103526} (\bibinfo {year} {2022})}\BibitemShut {NoStop}%
\bibitem [{\citenamefont {D'Agostino}\ and\ \citenamefont
  {Nunes}(2020)}]{D'agostino_Nunes}%
  \BibitemOpen
  \bibfield  {author} {\bibinfo {author} {\bibfnamefont {R.}~\bibnamefont
  {D'Agostino}}\ and\ \bibinfo {author} {\bibfnamefont {R.~C.}\ \bibnamefont
  {Nunes}},\ }\bibfield  {title} {\bibinfo {title} {Measurements of ${H}_{0}$
  in modified gravity theories: The role of lensed quasars in the late-time
  universe},\ }\href {https://doi.org/10.1103/PhysRevD.101.103505} {\bibfield
  {journal} {\bibinfo  {journal} {Phys. Rev. D}\ }\textbf {\bibinfo {volume}
  {101}},\ \bibinfo {pages} {103505} (\bibinfo {year} {2020})}\BibitemShut
  {NoStop}%
\bibitem [{\citenamefont {Farrugia}\ \emph {et~al.}(2021)\citenamefont
  {Farrugia}, \citenamefont {Sultana},\ and\ \citenamefont
  {Mifsud}}]{Farrugia2021}%
  \BibitemOpen
  \bibfield  {author} {\bibinfo {author} {\bibfnamefont {C.~R.}\ \bibnamefont
  {Farrugia}}, \bibinfo {author} {\bibfnamefont {J.}~\bibnamefont {Sultana}},\
  and\ \bibinfo {author} {\bibfnamefont {J.}~\bibnamefont {Mifsud}},\
  }\bibfield  {title} {\bibinfo {title} {Spatial curvature in {$f(R)$}
  gravity},\ }\href {https://doi.org/10.1103/PhysRevD.104.123503} {\bibfield
  {journal} {\bibinfo  {journal} {Phys. Rev. D}\ }\textbf {\bibinfo {volume}
  {104}},\ \bibinfo {pages} {123503} (\bibinfo {year} {2021})}\BibitemShut
  {NoStop}%
\bibitem [{\citenamefont {Graf}\ \emph {et~al.}(2021)\citenamefont {Graf},
  \citenamefont {Flores}, \citenamefont {Protopapas},\ and\ \citenamefont
  {Pichara}}]{graf2021uncertainty}%
  \BibitemOpen
  \bibfield  {author} {\bibinfo {author} {\bibfnamefont {O.}~\bibnamefont
  {Graf}}, \bibinfo {author} {\bibfnamefont {P.}~\bibnamefont {Flores}},
  \bibinfo {author} {\bibfnamefont {P.}~\bibnamefont {Protopapas}},\ and\
  \bibinfo {author} {\bibfnamefont {K.}~\bibnamefont {Pichara}},\ }\href@noop
  {} {\bibinfo {title} {Uncertainty quantification in neural differential
  equations}} (\bibinfo {year} {2021}),\ \Eprint
  {https://arxiv.org/abs/2111.04207} {arXiv:2111.04207} \BibitemShut {NoStop}%
\bibitem [{sup()}]{supp_material}%
  \BibitemOpen
  \href@noop {} {}\bibinfo {note} {See Supplemental Material at
  \url{https://github.com/at-chantada/Supplemental-Materials} for a description
  of the Runge-Kutta 5(4) algorithm and an estimation of its computational cost
  measured in floating point operations.}\BibitemShut {Stop}%
\bibitem [{\citenamefont {Jimenez}\ and\ \citenamefont
  {Loeb}(2002)}]{CC_technique}%
  \BibitemOpen
  \bibfield  {author} {\bibinfo {author} {\bibfnamefont {R.}~\bibnamefont
  {Jimenez}}\ and\ \bibinfo {author} {\bibfnamefont {A.}~\bibnamefont {Loeb}},\
  }\bibfield  {title} {\bibinfo {title} {Constraining cosmological parameters
  based on relative galaxy ages},\ }\href {https://doi.org/10.1086/340549}
  {\bibfield  {journal} {\bibinfo  {journal} {The Astrophysical Journal}\
  }\textbf {\bibinfo {volume} {573}},\ \bibinfo {pages} {37} (\bibinfo {year}
  {2002})}\BibitemShut {NoStop}%
\bibitem [{\citenamefont {{Tripp}}(1998)}]{Tripp}%
  \BibitemOpen
  \bibfield  {author} {\bibinfo {author} {\bibfnamefont {R.}~\bibnamefont
  {{Tripp}}},\ }\bibfield  {title} {\bibinfo {title} {{A two-parameter
  luminosity correction for Type IA supernovae}},\ }\href
  {https://ui.adsabs.harvard.edu/abs/1998A&A...331..815T} {\bibfield  {journal}
  {\bibinfo  {journal} {\aap}\ }\textbf {\bibinfo {volume} {331}},\ \bibinfo
  {pages} {815} (\bibinfo {year} {1998})}\BibitemShut {NoStop}%
\bibitem [{\citenamefont {{M. Betoule,}}\ \emph {et~al.}(2014)\citenamefont
  {{M. Betoule,}}, \citenamefont {{R. Kessler,}}, \citenamefont {{J. Guy,}},
  \citenamefont {{J. Mosher,}}, \citenamefont {{D. Hardin,}}, \citenamefont
  {{R. Biswas,}}, \citenamefont {{P. Astier,}}, \citenamefont {{P. El-Hage,}},
  \citenamefont {{M. Konig,}}, \citenamefont {{S. Kuhlmann}} \emph
  {et~al.}}]{SALT2}%
  \BibitemOpen
  \bibfield  {author} {\bibinfo {author} {\bibnamefont {{M. Betoule,}}},
  \bibinfo {author} {\bibnamefont {{R. Kessler,}}}, \bibinfo {author}
  {\bibnamefont {{J. Guy,}}}, \bibinfo {author} {\bibnamefont {{J. Mosher,}}},
  \bibinfo {author} {\bibnamefont {{D. Hardin,}}}, \bibinfo {author}
  {\bibnamefont {{R. Biswas,}}}, \bibinfo {author} {\bibnamefont {{P.
  Astier,}}}, \bibinfo {author} {\bibnamefont {{P. El-Hage,}}}, \bibinfo
  {author} {\bibnamefont {{M. Konig,}}}, \bibinfo {author} {\bibnamefont {{S.
  Kuhlmann}}}, \emph {et~al.},\ }\bibfield  {title} {\bibinfo {title} {Improved
  cosmological constraints from a joint analysis of the {SDSS}-{II} and {SNLS}
  supernova samples},\ }\href {https://doi.org/10.1051/0004-6361/201423413}
  {\bibfield  {journal} {\bibinfo  {journal} {A\&A}\ }\textbf {\bibinfo
  {volume} {568}},\ \bibinfo {pages} {A22} (\bibinfo {year}
  {2014})}\BibitemShut {NoStop}%
\bibitem [{\citenamefont {Kessler}\ \emph {et~al.}(2009)\citenamefont
  {Kessler}, \citenamefont {Bernstein}, \citenamefont {Cinabro}, \citenamefont
  {Dilday}, \citenamefont {Frieman}, \citenamefont {Jha}, \citenamefont
  {Kuhlmann}, \citenamefont {Miknaitis}, \citenamefont {Sako}, \citenamefont
  {Taylor},\ and\ \citenamefont {Vanderplas}}]{SNANA}%
  \BibitemOpen
  \bibfield  {author} {\bibinfo {author} {\bibfnamefont {R.}~\bibnamefont
  {Kessler}}, \bibinfo {author} {\bibfnamefont {J.~P.}\ \bibnamefont
  {Bernstein}}, \bibinfo {author} {\bibfnamefont {D.}~\bibnamefont {Cinabro}},
  \bibinfo {author} {\bibfnamefont {B.}~\bibnamefont {Dilday}}, \bibinfo
  {author} {\bibfnamefont {J.~A.}\ \bibnamefont {Frieman}}, \bibinfo {author}
  {\bibfnamefont {S.}~\bibnamefont {Jha}}, \bibinfo {author} {\bibfnamefont
  {S.}~\bibnamefont {Kuhlmann}}, \bibinfo {author} {\bibfnamefont
  {G.}~\bibnamefont {Miknaitis}}, \bibinfo {author} {\bibfnamefont
  {M.}~\bibnamefont {Sako}}, \bibinfo {author} {\bibfnamefont {M.}~\bibnamefont
  {Taylor}},\ and\ \bibinfo {author} {\bibfnamefont {J.}~\bibnamefont
  {Vanderplas}},\ }\bibfield  {title} {\bibinfo {title} {{SNANA}: A public
  software package for supernova analysis},\ }\href
  {https://doi.org/10.1086/605984} {\bibfield  {journal} {\bibinfo  {journal}
  {Publications of the Astronomical Society of the Pacific}\ }\textbf {\bibinfo
  {volume} {121}},\ \bibinfo {pages} {1028} (\bibinfo {year}
  {2009})}\BibitemShut {NoStop}%
\bibitem [{\citenamefont {Negrelli}\ \emph {et~al.}(2020)\citenamefont
  {Negrelli}, \citenamefont {Kraiselburd}, \citenamefont {Landau},\ and\
  \citenamefont {Sc{\'{o}}ccola}}]{Negrelli_2020}%
  \BibitemOpen
  \bibfield  {author} {\bibinfo {author} {\bibfnamefont {C.}~\bibnamefont
  {Negrelli}}, \bibinfo {author} {\bibfnamefont {L.}~\bibnamefont
  {Kraiselburd}}, \bibinfo {author} {\bibfnamefont {S.}~\bibnamefont
  {Landau}},\ and\ \bibinfo {author} {\bibfnamefont {C.~G.}\ \bibnamefont
  {Sc{\'{o}}ccola}},\ }\bibfield  {title} {\bibinfo {title} {Testing modified
  gravity theory ({MOG}) with {Type} {Ia} supernovae, cosmic chronometers and
  baryon acoustic oscillations},\ }\href
  {https://doi.org/10.1088/1475-7516/2020/07/015} {\bibfield  {journal}
  {\bibinfo  {journal} {Journal of Cosmology and Astroparticle Physics}\
  }\textbf {\bibinfo {volume} {2020}}\bibinfo  {number} { (07)},\ \bibinfo
  {pages} {015}}\BibitemShut {NoStop}%
\bibitem [{\citenamefont {{Silk}}(1968)}]{BAO_theory_1}%
  \BibitemOpen
\bibfield  {number} {  }\bibfield  {author} {\bibinfo {author} {\bibfnamefont
  {J.}~\bibnamefont {{Silk}}},\ }\bibfield  {title} {\bibinfo {title} {{Cosmic
  black-body radiation and galaxy formation}},\ }\href
  {https://doi.org/10.1086/149449} {\bibfield  {journal} {\bibinfo  {journal}
  {The Astrophysical Journal}\ }\textbf {\bibinfo {volume} {151}},\ \bibinfo
  {pages} {459} (\bibinfo {year} {1968})}\BibitemShut {NoStop}%
\bibitem [{\citenamefont {{Peebles}}\ and\ \citenamefont
  {{Yu}}(1970)}]{BAO_theory_2}%
  \BibitemOpen
  \bibfield  {author} {\bibinfo {author} {\bibfnamefont {P.~J.~E.}\
  \bibnamefont {{Peebles}}}\ and\ \bibinfo {author} {\bibfnamefont {J.~T.}\
  \bibnamefont {{Yu}}},\ }\bibfield  {title} {\bibinfo {title} {{Primeval
  adiabatic perturbation in an expanding universe}},\ }\href
  {https://doi.org/10.1086/150713} {\bibfield  {journal} {\bibinfo  {journal}
  {The Astrophysical Journal}\ }\textbf {\bibinfo {volume} {162}},\ \bibinfo
  {pages} {815} (\bibinfo {year} {1970})}\BibitemShut {NoStop}%
\bibitem [{\citenamefont {{Sunyaev}}\ and\ \citenamefont
  {{Zeldovich}}(1970)}]{BAO_theory_3}%
  \BibitemOpen
  \bibfield  {author} {\bibinfo {author} {\bibfnamefont {R.~A.}\ \bibnamefont
  {{Sunyaev}}}\ and\ \bibinfo {author} {\bibfnamefont {Y.~B.}\ \bibnamefont
  {{Zeldovich}}},\ }\bibfield  {title} {\bibinfo {title} {{Small-scale
  fluctuations of relic radiation}},\ }\href
  {https://ui.adsabs.harvard.edu/abs/1970Ap&SS...7....3S} {\bibfield  {journal}
  {\bibinfo  {journal} {Astrophysics and Space Science}\ }\textbf {\bibinfo
  {volume} {7}},\ \bibinfo {pages} {3} (\bibinfo {year} {1970})}\BibitemShut
  {NoStop}%
\bibitem [{\citenamefont {Eisenstein}\ \emph {et~al.}(2005)\citenamefont
  {Eisenstein}, \citenamefont {Zehavi}, \citenamefont {Hogg}, \citenamefont
  {Scoccimarro}, \citenamefont {Blanton}, \citenamefont {Nichol}, \citenamefont
  {Scranton}, \citenamefont {Seo}, \citenamefont {Tegmark}, \citenamefont
  {Zheng} \emph {et~al.}}]{D_V}%
  \BibitemOpen
  \bibfield  {author} {\bibinfo {author} {\bibfnamefont {D.~J.}\ \bibnamefont
  {Eisenstein}}, \bibinfo {author} {\bibfnamefont {I.}~\bibnamefont {Zehavi}},
  \bibinfo {author} {\bibfnamefont {D.~W.}\ \bibnamefont {Hogg}}, \bibinfo
  {author} {\bibfnamefont {R.}~\bibnamefont {Scoccimarro}}, \bibinfo {author}
  {\bibfnamefont {M.~R.}\ \bibnamefont {Blanton}}, \bibinfo {author}
  {\bibfnamefont {R.~C.}\ \bibnamefont {Nichol}}, \bibinfo {author}
  {\bibfnamefont {R.}~\bibnamefont {Scranton}}, \bibinfo {author}
  {\bibfnamefont {H.-J.}\ \bibnamefont {Seo}}, \bibinfo {author} {\bibfnamefont
  {M.}~\bibnamefont {Tegmark}}, \bibinfo {author} {\bibfnamefont
  {Z.}~\bibnamefont {Zheng}}, \emph {et~al.},\ }\bibfield  {title} {\bibinfo
  {title} {Detection of the baryon acoustic peak in the large-scale correlation
  function of {SDSS} luminous red galaxies},\ }\href
  {https://doi.org/10.1086/466512} {\bibfield  {journal} {\bibinfo  {journal}
  {The Astrophysical Journal}\ }\textbf {\bibinfo {volume} {633}},\ \bibinfo
  {pages} {560} (\bibinfo {year} {2005})}\BibitemShut {NoStop}%
\bibitem [{\citenamefont {Lazkoz}\ \emph {et~al.}(2018)\citenamefont {Lazkoz},
  \citenamefont {Ortiz-Ba{\~{n}}os},\ and\ \citenamefont {Salzano}}]{r_s}%
  \BibitemOpen
  \bibfield  {author} {\bibinfo {author} {\bibfnamefont {R.}~\bibnamefont
  {Lazkoz}}, \bibinfo {author} {\bibfnamefont {M.}~\bibnamefont
  {Ortiz-Ba{\~{n}}os}},\ and\ \bibinfo {author} {\bibfnamefont
  {V.}~\bibnamefont {Salzano}},\ }\bibfield  {title} {\bibinfo {title}
  {{$f(R)$} gravity modifications: {from} the action to the data},\ }\href
  {https://doi.org/10.1140/epjc/s10052-018-5711-6} {\bibfield  {journal}
  {\bibinfo  {journal} {The European Physical Journal C}\ }\textbf {\bibinfo
  {volume} {78}},\ \bibinfo {pages} {213} (\bibinfo {year} {2018})}\BibitemShut
  {NoStop}%
\bibitem [{\citenamefont {Eisenstein}\ and\ \citenamefont
  {Hu}(1998)}]{z_drag_formula}%
  \BibitemOpen
  \bibfield  {author} {\bibinfo {author} {\bibfnamefont {D.~J.}\ \bibnamefont
  {Eisenstein}}\ and\ \bibinfo {author} {\bibfnamefont {W.}~\bibnamefont
  {Hu}},\ }\bibfield  {title} {\bibinfo {title} {Baryonic features in the
  matter transfer function},\ }\href {https://doi.org/10.1086/305424}
  {\bibfield  {journal} {\bibinfo  {journal} {The Astrophysical Journal}\
  }\textbf {\bibinfo {volume} {496}},\ \bibinfo {pages} {605} (\bibinfo {year}
  {1998})}\BibitemShut {NoStop}%
\bibitem [{\citenamefont {{Foreman-Mackey}}\ \emph {et~al.}(2013)\citenamefont
  {{Foreman-Mackey}}, \citenamefont {{Hogg}}, \citenamefont {{Lang}},\ and\
  \citenamefont {{Goodman}}}]{emcee}%
  \BibitemOpen
  \bibfield  {author} {\bibinfo {author} {\bibfnamefont {D.}~\bibnamefont
  {{Foreman-Mackey}}}, \bibinfo {author} {\bibfnamefont {D.~W.}\ \bibnamefont
  {{Hogg}}}, \bibinfo {author} {\bibfnamefont {D.}~\bibnamefont {{Lang}}},\
  and\ \bibinfo {author} {\bibfnamefont {J.}~\bibnamefont {{Goodman}}},\
  }\bibfield  {title} {\bibinfo {title} {emcee: The {MCMC} {Hammer}},\ }\href
  {https://doi.org/10.1086/670067} {\bibfield  {journal} {\bibinfo  {journal}
  {PASP}\ }\textbf {\bibinfo {volume} {125}},\ \bibinfo {pages} {306} (\bibinfo
  {year} {2013})},\ \Eprint {https://arxiv.org/abs/1202.3665} {1202.3665}
  \BibitemShut {NoStop}%
\bibitem [{\citenamefont {{Goodman}}\ and\ \citenamefont
  {{Weare}}(2010)}]{Affine_invariance_sampler}%
  \BibitemOpen
  \bibfield  {author} {\bibinfo {author} {\bibfnamefont {J.}~\bibnamefont
  {{Goodman}}}\ and\ \bibinfo {author} {\bibfnamefont {J.}~\bibnamefont
  {{Weare}}},\ }\bibfield  {title} {\bibinfo {title} {{Ensemble samplers with
  affine invariance}},\ }\href {https://doi.org/10.2140/camcos.2010.5.65}
  {\bibfield  {journal} {\bibinfo  {journal} {Communications in Applied
  Mathematics and Computational Science}\ }\textbf {\bibinfo {volume} {5}},\
  \bibinfo {pages} {65} (\bibinfo {year} {2010})}\BibitemShut {NoStop}%
\bibitem [{\citenamefont {Lewis}(2019)}]{getdist}%
  \BibitemOpen
  \bibfield  {author} {\bibinfo {author} {\bibfnamefont {A.}~\bibnamefont
  {Lewis}},\ }\href@noop {} {\bibinfo {title} {Getdist: A {Python} package for
  analysing {Monte} {Carlo} samples}} (\bibinfo {year} {2019}),\ \Eprint
  {https://arxiv.org/abs/1910.13970} {arXiv:1910.13970} \BibitemShut {NoStop}%
\bibitem [{\citenamefont {{Motta}}\ \emph {et~al.}(2021)\citenamefont
  {{Motta}}, \citenamefont {{Garc{\'\i}a-Aspeitia}}, \citenamefont
  {{Hern{\'a}ndez-Almada}}, \citenamefont {{Maga{\~n}a}},\ and\ \citenamefont
  {{Verdugo}}}]{2021Univ....7..163M}%
  \BibitemOpen
  \bibfield  {author} {\bibinfo {author} {\bibfnamefont {V.}~\bibnamefont
  {{Motta}}}, \bibinfo {author} {\bibfnamefont {M.~A.}\ \bibnamefont
  {{Garc{\'\i}a-Aspeitia}}}, \bibinfo {author} {\bibfnamefont {A.}~\bibnamefont
  {{Hern{\'a}ndez-Almada}}}, \bibinfo {author} {\bibfnamefont {J.}~\bibnamefont
  {{Maga{\~n}a}}},\ and\ \bibinfo {author} {\bibfnamefont {T.}~\bibnamefont
  {{Verdugo}}},\ }\bibfield  {title} {\bibinfo {title} {{Taxonomy of Dark
  Energy Models}},\ }\href {https://doi.org/10.3390/universe7060163} {\bibfield
   {journal} {\bibinfo  {journal} {Universe}\ }\textbf {\bibinfo {volume}
  {7}},\ \bibinfo {pages} {163} (\bibinfo {year} {2021})}\BibitemShut {NoStop}%
\bibitem [{\citenamefont {{Akrami}}\ \emph {et~al.}(2019)\citenamefont
  {{Akrami}}, \citenamefont {{Kallosh}}, \citenamefont {{Linde}},\ and\
  \citenamefont {{Vardanyan}}}]{2019ForPh..6700075A}%
  \BibitemOpen
  \bibfield  {author} {\bibinfo {author} {\bibfnamefont {Y.}~\bibnamefont
  {{Akrami}}}, \bibinfo {author} {\bibfnamefont {R.}~\bibnamefont {{Kallosh}}},
  \bibinfo {author} {\bibfnamefont {A.}~\bibnamefont {{Linde}}},\ and\ \bibinfo
  {author} {\bibfnamefont {V.}~\bibnamefont {{Vardanyan}}},\ }\bibfield
  {title} {\bibinfo {title} {{The landscape, the swampland and the era of
  precision cosmology}},\ }\href {https://doi.org/10.1002/prop.201800075}
  {\bibfield  {journal} {\bibinfo  {journal} {Fortschritte der Physik}\
  }\textbf {\bibinfo {volume} {67}},\ \bibinfo {pages} {1800075} (\bibinfo
  {year} {2019})}\BibitemShut {NoStop}%
\bibitem [{\citenamefont {{Tosone}}\ \emph {et~al.}(2019)\citenamefont
  {{Tosone}}, \citenamefont {{Haridasu}}, \citenamefont {{Lukovi{\'c}}},\ and\
  \citenamefont {{Vittorio}}}]{2019PhRvD..99d3503T}%
  \BibitemOpen
  \bibfield  {author} {\bibinfo {author} {\bibfnamefont {F.}~\bibnamefont
  {{Tosone}}}, \bibinfo {author} {\bibfnamefont {B.~S.}\ \bibnamefont
  {{Haridasu}}}, \bibinfo {author} {\bibfnamefont {V.~V.}\ \bibnamefont
  {{Lukovi{\'c}}}},\ and\ \bibinfo {author} {\bibfnamefont {N.}~\bibnamefont
  {{Vittorio}}},\ }\bibfield  {title} {\bibinfo {title} {Constraints on field
  flows of quintessence dark energy},\ }\href
  {https://doi.org/10.1103/PhysRevD.99.043503} {\bibfield  {journal} {\bibinfo
  {journal} {Phys. Rev. D}\ }\textbf {\bibinfo {volume} {99}},\ \bibinfo
  {pages} {043503} (\bibinfo {year} {2019})}\BibitemShut {NoStop}%
\bibitem [{\citenamefont {Riess}\ \emph {et~al.}(2018)\citenamefont {Riess},
  \citenamefont {Casertano}, \citenamefont {Yuan}, \citenamefont {Macri},
  \citenamefont {Anderson}, \citenamefont {MacKenty}, \citenamefont {Bowers},
  \citenamefont {Clubb}, \citenamefont {Filippenko}, \citenamefont {Jones},\
  and\ \citenamefont {Tucker}}]{Riess_2018}%
  \BibitemOpen
  \bibfield  {author} {\bibinfo {author} {\bibfnamefont {A.~G.}\ \bibnamefont
  {Riess}}, \bibinfo {author} {\bibfnamefont {S.}~\bibnamefont {Casertano}},
  \bibinfo {author} {\bibfnamefont {W.}~\bibnamefont {Yuan}}, \bibinfo {author}
  {\bibfnamefont {L.}~\bibnamefont {Macri}}, \bibinfo {author} {\bibfnamefont
  {J.}~\bibnamefont {Anderson}}, \bibinfo {author} {\bibfnamefont {J.~W.}\
  \bibnamefont {MacKenty}}, \bibinfo {author} {\bibfnamefont {J.~B.}\
  \bibnamefont {Bowers}}, \bibinfo {author} {\bibfnamefont {K.~I.}\
  \bibnamefont {Clubb}}, \bibinfo {author} {\bibfnamefont {A.~V.}\ \bibnamefont
  {Filippenko}}, \bibinfo {author} {\bibfnamefont {D.~O.}\ \bibnamefont
  {Jones}},\ and\ \bibinfo {author} {\bibfnamefont {B.~E.}\ \bibnamefont
  {Tucker}},\ }\bibfield  {title} {\bibinfo {title} {New parallaxes of galactic
  cepheids from spatially scanning the {Hubble} space telescope: Implications
  for the {Hubble} constant},\ }\href
  {https://doi.org/10.3847/1538-4357/aaadb7} {\bibfield  {journal} {\bibinfo
  {journal} {The Astrophysical Journal}\ }\textbf {\bibinfo {volume} {855}},\
  \bibinfo {pages} {136} (\bibinfo {year} {2018})}\BibitemShut {NoStop}%
\bibitem [{\citenamefont {Wong}\ \emph {et~al.}(2019)\citenamefont {Wong},
  \citenamefont {Suyu}, \citenamefont {Chen}, \citenamefont {Rusu},
  \citenamefont {Millon}, \citenamefont {Sluse}, \citenamefont {Bonvin},
  \citenamefont {Fassnacht}, \citenamefont {Taubenberger}, \citenamefont
  {Auger} \emph {et~al.}}]{HOLICOW2019}%
  \BibitemOpen
  \bibfield  {author} {\bibinfo {author} {\bibfnamefont {K.~C.}\ \bibnamefont
  {Wong}}, \bibinfo {author} {\bibfnamefont {S.~H.}\ \bibnamefont {Suyu}},
  \bibinfo {author} {\bibfnamefont {G.~C.-F.}\ \bibnamefont {Chen}}, \bibinfo
  {author} {\bibfnamefont {C.~E.}\ \bibnamefont {Rusu}}, \bibinfo {author}
  {\bibfnamefont {M.}~\bibnamefont {Millon}}, \bibinfo {author} {\bibfnamefont
  {D.}~\bibnamefont {Sluse}}, \bibinfo {author} {\bibfnamefont
  {V.}~\bibnamefont {Bonvin}}, \bibinfo {author} {\bibfnamefont {C.~D.}\
  \bibnamefont {Fassnacht}}, \bibinfo {author} {\bibfnamefont {S.}~\bibnamefont
  {Taubenberger}}, \bibinfo {author} {\bibfnamefont {M.~W.}\ \bibnamefont
  {Auger}}, \emph {et~al.},\ }\bibfield  {title} {\bibinfo {title} {{H0LiCOW
  – XIII. A 2.4 per cent measurement of $H_0$ from lensed quasars: 5.3σ
  tension between early- and late-universe probes}},\ }\href
  {https://doi.org/10.1093/mnras/stz3094} {\bibfield  {journal} {\bibinfo
  {journal} {Monthly Notices of the Royal Astronomical Society}\ }\textbf
  {\bibinfo {volume} {498}},\ \bibinfo {pages} {1420} (\bibinfo {year}
  {2019})}\BibitemShut {NoStop}%
\bibitem [{\citenamefont {Dormand}\ and\ \citenamefont
  {Prince}(1980)}]{RK5(4)_formulae}%
  \BibitemOpen
  \bibfield  {author} {\bibinfo {author} {\bibfnamefont {J.}~\bibnamefont
  {Dormand}}\ and\ \bibinfo {author} {\bibfnamefont {P.}~\bibnamefont
  {Prince}},\ }\bibfield  {title} {\bibinfo {title} {A family of embedded
  {Runge}-{Kutta} formulae},\ }\href
  {https://doi.org/https://doi.org/10.1016/0771-050X(80)90013-3} {\bibfield
  {journal} {\bibinfo  {journal} {Journal of Computational and Applied
  Mathematics}\ }\textbf {\bibinfo {volume} {6}},\ \bibinfo {pages} {19}
  (\bibinfo {year} {1980})}\BibitemShut {NoStop}%
\bibitem [{\citenamefont {Shampine}(1986)}]{RK_interpolation}%
  \BibitemOpen
  \bibfield  {author} {\bibinfo {author} {\bibfnamefont {L.~F.}\ \bibnamefont
  {Shampine}},\ }\bibfield  {title} {\bibinfo {title} {Some practical
  {Runge}-{Kutta} formulas},\ }\href
  {https://doi.org/https://doi.org/10.1090/S0025-5718-1986-0815836-3}
  {\bibfield  {journal} {\bibinfo  {journal} {Mathematics of Computation}\
  }\textbf {\bibinfo {volume} {46}},\ \bibinfo {pages} {135} (\bibinfo {year}
  {1986})}\BibitemShut {NoStop}%
\bibitem [{\citenamefont {Kingma}\ and\ \citenamefont {Ba}(2017)}]{Adam}%
  \BibitemOpen
  \bibfield  {author} {\bibinfo {author} {\bibfnamefont {D.~P.}\ \bibnamefont
  {Kingma}}\ and\ \bibinfo {author} {\bibfnamefont {J.}~\bibnamefont {Ba}},\
  }\href@noop {} {\bibinfo {title} {Adam: A method for stochastic
  optimization}} (\bibinfo {year} {2017}),\ \Eprint
  {https://arxiv.org/abs/1412.6980} {arXiv:1412.6980} \BibitemShut {NoStop}%
\bibitem [{\citenamefont {Virtanen}\ \emph {et~al.}(2020)\citenamefont
  {Virtanen}, \citenamefont {Gommers}, \citenamefont {Oliphant}, \citenamefont
  {Haberland}, \citenamefont {Reddy}, \citenamefont {Cournapeau}, \citenamefont
  {Burovski}, \citenamefont {Peterson}, \citenamefont {Weckesser},
  \citenamefont {Bright} \emph {et~al.}}]{scipy}%
  \BibitemOpen
  \bibfield  {author} {\bibinfo {author} {\bibfnamefont {P.}~\bibnamefont
  {Virtanen}}, \bibinfo {author} {\bibfnamefont {R.}~\bibnamefont {Gommers}},
  \bibinfo {author} {\bibfnamefont {T.~E.}\ \bibnamefont {Oliphant}}, \bibinfo
  {author} {\bibfnamefont {M.}~\bibnamefont {Haberland}}, \bibinfo {author}
  {\bibfnamefont {T.}~\bibnamefont {Reddy}}, \bibinfo {author} {\bibfnamefont
  {D.}~\bibnamefont {Cournapeau}}, \bibinfo {author} {\bibfnamefont
  {E.}~\bibnamefont {Burovski}}, \bibinfo {author} {\bibfnamefont
  {P.}~\bibnamefont {Peterson}}, \bibinfo {author} {\bibfnamefont
  {W.}~\bibnamefont {Weckesser}}, \bibinfo {author} {\bibfnamefont
  {J.}~\bibnamefont {Bright}}, \emph {et~al.},\ }\bibfield  {title} {\bibinfo
  {title} {{{SciPy} 1.0: Fundamental algorithms for scientific computing in
  Python}},\ }\href {https://doi.org/10.1038/s41592-019-0686-2} {\bibfield
  {journal} {\bibinfo  {journal} {Nature Methods}\ }\textbf {\bibinfo {volume}
  {17}},\ \bibinfo {pages} {261} (\bibinfo {year} {2020})}\BibitemShut
  {NoStop}%
\bibitem [{\citenamefont {Møller}(1993)}]{back_forw_ratio}%
  \BibitemOpen
  \bibfield  {author} {\bibinfo {author} {\bibfnamefont {M.~F.}\ \bibnamefont
  {Møller}},\ }\bibfield  {title} {\bibinfo {title} {A scaled conjugate
  gradient algorithm for fast supervised learning},\ }\href
  {https://doi.org/https://doi.org/10.1016/S0893-6080(05)80056-5} {\bibfield
  {journal} {\bibinfo  {journal} {Neural Networks}\ }\textbf {\bibinfo {volume}
  {6}},\ \bibinfo {pages} {525} (\bibinfo {year} {1993})}\BibitemShut {NoStop}%
\end{thebibliography}
\end{document}